\documentclass{aa}

\usepackage{hyperref}
\usepackage{graphicx}
\usepackage{txfonts}
\usepackage[svgnames]{xcolor}
\usepackage{natbib}

\bibpunct{(}{)}{;}{a}{}{,}

\newcommand{\red}[1]{\textcolor{red}{#1}}
\newcommand{\violet}[1]{\textcolor{violet}{#1}}
\newcommand{\blue}[1]{\textcolor{blue}{#1}}

\hypersetup{
 colorlinks=true,
 allcolors=blue
}

\begin{document}

\date{Received 20 July 2023 / Accepted 11 June 2024}

\title{Very high energy afterglow of structured jets:\\ GW~170817 and prospects for future detections}

\titlerunning{VHE afterglow of structured jets: GW~170817 and prospects for future detections}

\author{Cl\'ement~Pellouin\inst{\ref{iap}}$^,$\inst{\ref{tau}}
\and Fr\'ed\'eric~Daigne\inst{\ref{iap}}$^,$\inst{\ref{iuf}}}

\institute{Sorbonne Universit\'e, CNRS, UMR 7095, Institut d’Astrophysique de Paris (IAP), 98 bis boulevard Arago, 75014 Paris,
France\label{iap}; e-mails: \texttt{pellouin@iap.fr}; \texttt{daigne@iap.fr}
\and 
{School of Physics and Astronomy, Tel Aviv University, Tel Aviv 6997801, Israel\label{tau}}
\and 
Institut Universitaire de France\label{iuf}
}

\authorrunning{Pellouin \& Daigne}

\abstract{
We present a complete numerical model of the afterglow of a laterally structured relativistic ejecta from the radio to very high energies (VHE). This includes a self-consistent calculation of the synchrotron radiation, with its maximum frequency, and of synchrotron self-Compton (SSC) scattering that takes the Klein-Nishina regime into account. Attenuation due to pair production is also included. This model is computationally efficient and allows multi-wavelength data fitting. 
As a validation test, the radiative model was used to fit the broad-band spectrum of GRB~190114C at 90~s up to the TeV range. The full model was then used to fit the afterglow of GW~170817 and predict its VHE emission. We find that the SSC flux at the peak was much dimmer than the upper limit from H.E.S.S. observations. However, we show that either a smaller viewing angle or a higher external density would make similar off-axis events detectable in the future at VHE, even above 100 Mpc with the sensitivity of the Cherenkov telescope array. High external densities are expected in the case of fast mergers, but the existence of a formation channel for these binary neutron stars is still uncertain. We highlight that VHE afterglow detections would help to efficiently probe systems like this. 
}
  

\maketitle


\section{Introduction}\label{sec:introduction}

The first detection of gravitational waves (GWs) from a binary neutron star (BNS) merger, GW~170817 \citep{2017PhRvL.119p1101A}, was followed by the detection of several electromagnetic counterparts (\citealt{2017ApJ...848L..12A} and references therein). The short gamma-ray burst GRB~170817A, was detected $\sim 1.7~\mathrm{s}$ after the GW signal \citep{2017ApJ...848L..14G,2017ApJ...848L..15S}; a fast-decaying thermal transient in the visible and infrared range, a kilonova, was observed for $\sim 10$ days after the merger \citep[see e.g.][]{2017ApJ...851L..21V,2017ApJ...848L..27T,2017ApJ...848L..17C}; and a non-thermal afterglow was observed from radio to X-rays for more than three years \citep[see e.g.][]{2021ApJ...914L..20B,2020MNRAS.498.5643T,2019ApJ...886L..17H}. Among the many advances made possible by this exceptional multi-messenger event, GW~170817/GRB~170817A is in particular the first direct association between a BNS merger and a short gamma-ray burst (GRB) and it has helped us to better understand the relativistic ejection associated with events like this. Indeed, the study of the relativistic ejecta benefited not only from an exceptional multi-wavelength follow-up, but also from observing conditions that were very different from those for other short GRBs: a much smaller distance (GW~170817 was hosted in NGC~4993 at  $\sim 40~\mathrm{Mpc}$; \citealt{2017ApJ...849L..34P}; \citealt{2018ApJ...854L..31C}), and a significantly off-axis observation ($32^{+10}_{-13}\pm 1.7$~deg derived from the GW signal using the accurate localisation and distance of NGC~4993, \citealt{2018ApJ...860L...2F}).

The prompt short GRB is puzzling: it is extremely weak despite the short distance, but its peak energy is above $150\, \mathrm{keV}$ \citep{2017ApJ...848L..14G}. It is very unlikely that it is produced by internal dissipation in the ultra-relativistic core jet like in other GRBs at cosmological distance \citep{2019MNRAS.483.1247M}. GRB~170817A was rather emitted in mildly relativistic and mildly energetic material on the line of sight: The shock-breakout emission when the relativistic core jet emerges from the kilonova ejecta is a promising mechanism \citep{2018MNRAS.475.2971B,2018MNRAS.479..588G}.

The interpretation of the afterglow is better understood. The slow rise of the light curve until its peak after $\sim 120-160$ days indicates that this was an off-axis observation of a decelerating jet surrounded by a lateral structure \citep[see e.g.][]{davanzo18} that may have been inherited from the early interaction with the kilonova ejecta. This was also invoked for the weak prompt emission \citep{2018MNRAS.479..588G}. The relativistic ejecta seen off-axis is also confirmed by the compactness of the source and its apparent superluminal motion as measured by very long baseline interferometry (VLBI) imagery \citep{2018Natur.561..355M, 2019Sci...363..968G, 2022Natur.610..273M}. The light curve rise is dominated by the lateral structure of the relativistic ejecta, while its peak and decay are dominated by the deceleration of the ultra-relativistic jet core, which requires a kinetic energy comparable to values commonly found in short GRBs \citep[see e.g.][]{2019Sci...363..968G}.

A lateral structure like this may be a common feature in GRBs due to the early propagation of the relativistic ejecta, either through the infalling envelope of the stellar progenitor (long GRBs), or through the post-merger ejecta that produces the kilonova emission (short GRBs) \citep{2011ApJ...740..100B}. Possible signatures of this lateral structure in  GRBs at cosmological distance viewed slightly off-axis were recently discussed, for instance to explain the complex phenomenology observed in the early afterglow \citep{2020MNRAS.492.2847B, 2020ApJ...893...88O,2020A&A...641A..61A,2022MNRAS.513..951D} or the non-standard decay of the afterglow of the extremely bright GRB~221009A \citep{2023SciA....9I1405O, 2023MNRAS.524L..78G}. Accounting for the lateral structure of the jet in afterglow models has thus become necessary not only for GW~170817, but for other cosmic GRBs as well.

Most models of the off-axis afterglow of GW~170817 only considered synchrotron emission and are therefore limited to the spectral range of observations, from radio to X-rays. However, upper limits on the flux at very high energy (VHE) were obtained by the high energy stereoscopic system telescopes (H.E.S.S.) at two epochs, a few days after the merger \citep{2017ApJ...850L..22A} and around its peak \citep{2020ApJ...894L..16A}. Less constraining upper limits were also obtained by the high-altitude water Cherenkov gamma-ray observatory (HAWC) at early times \citep{2019ICRC...36..681G} and by the major atmospheric gamma-ray imaging Cherenkov telescope (MAGIC) around the afterglow peak \citep{2022icrc.confE.944S}. A discussion of the constraints associated with these upper limits requires that the synchrotron self-Compton (SSC) emission is included in afterglow models, that is, the inverse Compton (IC) scattering of the synchrotron photons by the relativistic electrons that emit them. Recent detections of the VHE afterglow of several long GRBs\footnote{Candidate VHE photons are associated at a lower level of confidence with several other bursts by MAGIC, including the short GRB~160821B \citep{2021ApJ...908...90A}.} by H.E.S.S. (GRB~180720B; \citealt{2019Natur.575..464A}; GRB~190829A; \citealt{2021Sci...372.1081H}), MAGIC (GRB~190114C; \citealt{2019Natur.575..455M, 2019Natur.575..459M}; GRB~201216C; \citealt{2020GCN.29075....1B}), and the large high altitude air shower observatory (LHAASO) (GRB~221009A; \citealt{LHAASO}) challenge purely synchrotron afterglow models \citep[see however][]{2021Sci...372.1081H} and suggest the dominant contribution of a new emission process at VHE \citep[see the discussion of  possible processes at VHE by][]{2022Galax..10...74G}, most probably, SSC emission, as suggested for instance by the modelling of the afterglows of GRB~190114C \citep{2019Natur.575..459M,2019ApJ...884..117W,2021ApJ...923..135D}, GRB~190829A \citep{2022ApJ...931L..19S}, or GRB~221009A \citep{2023MNRAS.522L..56S}.

While SSC emission can be efficiently computed analytically in the Thomson regime \citep{2000ApJ...543...66P,2001ApJ...548..787S}, several studies have pointed out that the Klein-Nishina (KN) attenuation at high energy also needs to be accounted for \citep[see e.g.][]{2009ApJ...703..675N, 2011ApJ...732...77M, 2015MNRAS.454.1073B, 2021MNRAS.504..528J,2022MNRAS.512.2142Y}. In the KN regime, the IC power of an electron strongly depends on its Lorentz factor. This affects the cooling of the electron distribution, and it therefore also impacts the synchrotron spectrum \citep{2001A&A...372.1071D,2009ApJ...703..675N,2009A&A...498..677B}, which can significantly differ from the standard prediction  given by \citet{1998ApJ...497L..17S} for the purely synchrotron case.

Motivated by these recent advances in GRB afterglow studies, we propose in this paper a consistent model of GRB afterglow emission from a decelerating laterally structured jet where electrons accelerated at the external forward shock radiate at all wavelengths by synchrotron emission and SSC scattering. The treatment of SSC in the KN regime follows the approach proposed by \citet{2009ApJ...703..675N}, extended to account for the maximum Lorentz factor of the accelerated electrons and the attenuation at high energy due to pair creation. The numerical implementation of the model was optimised to be computationally efficient. This allowed to use Bayesian statistics to infer the parameters. We then applied this model to the multi-wavelength observations of the afterglow of GW~170817 to understand the physical constraints brought by the H.E.S.S. upper limits and to discuss whether the detection of post-merger VHE afterglows may become possible in the future, especially in the coming era of the Cherenkov telescope array (CTA; see \citealt{2019scta.book.....C}).

The model and its assumptions are detailed in Sect.~\ref{sec:vhe_model} and are then tested and compared to other afterglow models in Sect.~\ref{sec:comparison_other_models}. In Sect.~\ref{sec:170817} we show the results obtained when fitting the afterglow of GW~170817. We discuss the predicted VHE emission in the light of the H.E.S.S. upper limit. We study the conditions for future post-merger detections of VHE afterglows in Sect.~\ref{sec:conditions_tev} and highlight that VHE emission is favoured by a higher external density compared to GW~170817, which may help to probe the population of fast-merging binaries, if it exists. Our conclusions are summarised in Sect.~\ref{sec:conclusion}.

\section{Modelling the VHE afterglow of structured jets}\label{sec:vhe_model}

Our aim is to model the GRB afterglow at all wavelengths from radio bands to the TeV range. We limit ourselves to the contribution of the forward external shock propagating in the external medium and leave the extension of the model to a future work that includes the reverse-shock contribution at early times. Motivated by the observations of GW~170817, we rather focus here on two other aspects: the lateral structure of the jet, and the detailed calculation of the IC spectral component, including KN effects. 

In this section, we describe our model: The assumptions for the structure of the initial relativistic outflow (Sect.~\ref{sec:geometry_structure}), the dynamics of its deceleration (Sect.~\ref{sec:dynamics}), the assumptions for the acceleration of electrons and the amplification of the magnetic field at the shock front (Sect.~\ref{sec:acc_el_b_field}), the detailed calculation of the emission in the comoving frame of the shocked external medium (including synchrotron and IC components; Sect.~\ref{sec:emicom}), and finally, the integration over equal-arrival time surfaces to compute the observed flux in the observer frame (light curves and spectra; Sect.~\ref{sec:eatsurfaces}). Our numerical implementation of this model optimises the computation time (typically a few seconds for computing light curves and/or spectra at different frequencies on a laptop) so that we can explore the parameter space with a Bayesian approach for the data fitting, as described in Sect.~\ref{sec:170817}. In case of ambiguity, a physical quantity $q$ is written $q$ without a prime in the fixed source frame, and $q'$ with a prime in the comoving frame of the emitting material.

\subsection{Relativistic outflow: Geometry and structure}\label{sec:geometry_structure}

We consider a laterally structured jet. The initial energy per solid angle $\epsilon_0(\theta)$ and the initial Lorentz factor $\Gamma_0(\theta)$ decrease with $\theta$, which is the angle from the jet axis. We define $\theta_{\mathrm{c}}$ as the opening angle of the core, that is, the ultra-relativistic/ultra-energetic central part of the jet. We then write
\begin{equation}\label{eq:epsilon_0}
    \epsilon_0(\theta) = \epsilon_0^{\mathrm{c}} \times f_a\left(\frac{\theta}{\theta_{\mathrm{c}}}\right)
\end{equation}
and
\begin{equation}
    \Gamma_0(\theta) = 1 + (\Gamma_0^{\rm c} - 1) \times g_b\left(\frac{\theta}{\theta_{\rm c}}\right)\, ,
\end{equation}
where $\epsilon_0^{\mathrm{c}}$ is the initial energy per solid angle, and $\Gamma_0^{\rm c}$ is the initial Lorentz factor, both on the jet axis ($\theta=0$). $f_a$ and $g_b$ are the normalised profiles for the energy and Lorentz factor in the lateral structure. 

Different lateral structures have been suggested in the literature, following the photometric and VLBI observations of the GRB afterglow of GW~170817. In this paper, we model this afterglow assuming the power-law structure used by \cite{2019A&A...631A..39D},
\begin{eqnarray}
        f_a\left(x\right) & = & \left\lbrace\begin{array}{ccc}
    1 & \mathrm{if} & x < 1\\
    x^{-a} & \mathrm{if} & x \geq 1\\
    \end{array}\right.\label{eq:PLjet_epsilon}\, ,\\
        g_b\left(x\right) & = & \left\lbrace\begin{array}{ccc}
    1 & \mathrm{if} & x < 1\\
    x^{-b} & \mathrm{if} & x \geq 1\\
    \end{array}\right.\label{eq:PLjet_gamma}
    \, .
\end{eqnarray}
This prescription allows  a direct comparison with a top-hat jet ($f_a(x)=g_b(x)=1$ for $x\le 1$ and $0$ otherwise). For comparison with other studies (see Sect.~\ref{sec:comparison_other_models}), we also considered the following possible structures:
\begin{itemize}
    \item A power-law jet from \citet{2018MNRAS.478.4128G},
        $f_a(x)=X^{-a}$ and
        $g_b(x)=X^{-b}$, with $X=\sqrt{1+x^2}$.
    \item A power-law jet as defined in \texttt{afterglowpy}, from \citet{2020ApJ...896..166R},
    $       f_a\left(x\right) = \left(\sqrt{1+ {x^2}/{a}}\right)^{-a}$.
        The model used in \texttt{afterglowpy} is limited to the self-similar evolution of the jet, which is independent of the initial value of the Lorentz factor. Therefore, $g_b$ is not specified in this case.
    \item A Gaussian jet from \citet{2018MNRAS.478.4128G},
    $f_a(x)=g_b(x) = \max{\left(
    e^{-{x^2}/{2}}\, ;
    e^{-{x_{\mathrm{max}}^2}/{2}}
    \right)}\, ,$
        where $x_{\mathrm{max}}=\theta_{\mathrm{max}}/\theta_{\mathrm{c}}$ and
        $\theta_{\rm max}$ is defined as the angle where $\beta_{0,\mathrm{min}}\left(\theta_{\rm max}\right) = 0.01$ (by default) or any other specific value. We note that \cite{2020ApJ...896..166R} used a similar parametrisation of the lateral structure for Gaussian jets in \texttt{afterglowpy}, with $\theta_{\rm max} = 90~\mathrm{deg}$.
\end{itemize}

The initial total energy of the jet and of its  core are given by
\begin{eqnarray}\label{eq:energy_0}
    E_{0} & = & 2 \int_0^{\pi/2} \epsilon_0(\theta) \, 2\pi \sin{\theta}\, \mathrm{d}\theta\, ,\\
    E_{0}^{\mathrm{c}} & = & 2 \int_0^{\theta_{\mathrm{c}}} \epsilon_0(\theta) \, 2\pi \sin{\theta}\, \mathrm{d}\theta\, .
\end{eqnarray}
The factor of $2$ accounts for the counter-jet. The usual isotropic-equivalent energy of the core jet equals
\begin{equation}
    E_{0,\mathrm{iso}}^{\mathrm{c}}=\frac{E_{0}^{\mathrm{c}}}{1-\cos{\theta_{\mathrm{c}}}}\, .
\end{equation}
For the power-law structure considered here (Eq.~(\ref{eq:PLjet_epsilon})), we have $E_{0}^{\mathrm{c}}=4\pi \epsilon_{0}^{\mathrm{c}} \left(1-\cos{\theta_{\mathrm{c}}}\right)$ and $E_{0,\mathrm{iso}}^{\mathrm{c}}=4\pi \epsilon_{0}^{\mathrm{c}}$.

We did not consider a possible radial structure of the outflow that may also very well be present due to the variability of the central engine. This radial structure may be at least partially smoothed out during the early propagation, for example via internal shocks \citep{1997ApJ...490...92K,1998MNRAS.296..275D,2000A&A...358.1157D}, and is expected to have more impact on the reverse shock \citep[see e.g.][]{2007ApJ...665L..93U,2007MNRAS.381..732G}, which is not included in the current version of the model. Therefore, $E_{0 \rm, iso}^{\rm c}$, $\Gamma_0^{\rm c}$, $\theta_{\rm c}$, $a$, and $b$ (or $\beta_{0, \rm min}$ for the Gaussian jet) are the only four free parameters needed to fully describe the initial structure of the jet before the deceleration starts.

In our numerical implementation of the model, we suppressed the lateral structure above the maximum angle $\theta_{\mathrm{max}}$. This angle is taken as the maximum between $\theta_\epsilon$, defined as the angle up to which a fraction $(1-\epsilon)$ of the jet energy is contained, and the viewing angle $\theta_{\mathrm{v}}$ (see Sect.~\ref{sec:eatsurfaces}). We therefore solved for $\theta_\epsilon$ the equation
\begin{equation}
(1 - \epsilon)E_0 = 2 \int_0^{\theta_\epsilon} \epsilon_0(\theta) \, 2\pi \sin{\theta}\, \mathrm{d}\theta \, .
\end{equation}
We use $\epsilon = 0.01$ in the following. \cite{2020ApJ...896..166R} introduced a similar parameter $\theta_\mathrm{W}$ to minimise computation time. Taking $\theta_{\mathrm{max}}=\max{\left(\theta_\epsilon;\theta_{\mathrm{v}}\right)}$ allowed us to keep a precise calculation at early times even for very large viewing angles.

In practice, the lateral structure is discretised into $N+1$ components, $i=0$ being the core jet and $i=1\to N$ being rings at increasing angles in the lateral structure. Each component is defined by $\theta_{\mathrm{min},i}\le\theta\le\theta_{\mathrm{max},i}$ with $\theta_{\mathrm{min},0}=0$ and $\theta_{\mathrm{max},0}=\theta_\mathrm{c}$ for the core jet; $\theta_{\mathrm{min},i}=\theta_{\mathrm{max},i-1}$ for $i=1\to N$, and $\theta_{\mathrm{max},N}=\theta_\mathrm{max}$. Each component is treated independently with fixed limits in latitude (no lateral spreading). The dynamics is computed such that $R_i(t)=R(\theta_i;t)$ and $\Gamma_i(t)=\Gamma(\theta_i;t)$, where $R(\theta,t)$ and $\Gamma(\theta,t)$ are given by the solution for the dynamics of the deceleration discussed below, and where $\theta_i= (\theta_{\mathrm{min}, i} + \theta_{\mathrm{max}, i}) / 2$ for $i \geq 1$ and $\theta_0 = 0$. We define the successive $\theta_{\mathrm{min}, i}$ and $\theta_{\mathrm{max}, i}$ such that each component carries an equal amount of energy (except for the core, $i=0$), although a linear increase can also be chosen. Finally, we typically choose a discretisation comprised of $N = 15$ components, which we find to be a good compromise between model accuracy and computational efficiency.

\subsection{Dynamics of the deceleration}
\label{sec:dynamics}

The structured outflow decelerates in an external medium, with an assumed density profile as a function of radius $R$,
\begin{equation}
\rho_{\rm ext}(R) = \frac{A}{R^{s}}\, ,
\end{equation}
with $s=0$ for a constant-density external medium as considered in the following to model the afterglow of GW~170817. In this case, we write $A =  n_{\rm ext} m_{\rm p}$, with $n_{\rm ext}$ the external medium density. We focus on the dynamics of the shocked external medium at the forward shock. It is computed assuming that (i) the dynamics of each ring of material at angle $\theta$ is independent of other angles, and that (ii) the lateral expansion of the outflow is negligible. These two assumptions are questionable at late times, close to the transition to the Newtonian regime \citep{1997ApJ...487L...1R}.

The Lorentz factor $\Gamma(\theta; R)$ of the shocked material at angle $\theta$ and radius $R$ is computed in a simplified way, using energy conservation \citep[see e.g.][]{2000ApJ...543...66P,2018MNRAS.478.4128G},
\begin{eqnarray}
    \Gamma(\theta;R) & \!\!\!=\!\!\! & \frac{\Gamma_0(\theta)}{2m(\theta;R)}\times\nonumber\\
    & &  \left[ -1 + \sqrt{1 + 4m(\theta;R) + 4 \left( \frac{m(\theta;R)}{\Gamma_0(\theta)} \right)^{2}} \right]\, ,\label{eq:gamma_shock}
\end{eqnarray}
where  
\begin{equation}
    m(\theta;R) = \frac{M_{\mathrm{ext}}(R)}{M_{\mathrm{ej}}(\theta)/\Gamma_0(\theta)}=\left(\frac{R}{R_{\mathrm{dec}}(\theta)}\right)^{3-s}\, ,
\end{equation}
with $M_{\mathrm{ext}}(R)=\int_0^R \rho_\mathrm{ext}(R)\, R^2\mathrm{d}R$ the swept-up mass per unit solid angle, and $M_{\mathrm{ej}}(\theta)=\epsilon_0(\theta)/\Gamma_0(\theta) c^2$ the ejected mass per unit solid angle. The deceleration radius of the material at angle $\theta$ is defined by
\begin{equation}
R_{\mathrm{dec}}(\theta) = \left( \frac{(3-s)\epsilon_0(\theta)}{ A \Gamma_0^2(\theta)c^2} \right)^{1/(3-s)}\, .
\end{equation}
For a power-law structure as defined by Eqs.~(\ref{eq:PLjet_epsilon}) and (\ref{eq:PLjet_gamma}), the deceleration radius scales as $R_{\mathrm{dec}}(\theta)\propto \theta^{\frac{2b-a}{3-s}}$. Best-fit models of the afterglow of GW~170817 usually have $2b > a$ so that the deceleration radius is larger at high latitude (see Sect.~\ref{sec:170817_fits}). More generally, \citet{2020MNRAS.493.3521B} derived the conditions for the observed afterglow light curve to have a single peak, as in the case of GW~170817. For an off-axis observation with a large viewing angle, these conditions are $\Gamma_0^\mathrm{c}\theta_\mathrm{c}>1$; $2b>a/(4-s)$ (i.e. $8b>a$ for a uniform external medium), $b>-\ln{\Gamma_0^\mathrm{c}}/\ln{\theta_\mathrm{c}}$, and $\theta_\mathrm{v} > \theta_\mathrm{c} \left( \Gamma_0^\mathrm{c}\theta_\mathrm{c} \right)^{\frac{1}{b-1}}$.

This description allows us to characterise the early-time dynamics in the coasting phase $R\ll R_{\mathrm{dec}}(\theta)$ where the Lorentz factor remains constant. It continuously branches out to the relativistic self-similar evolution \citep{1976PhFl...19.1130B} for $R_{\mathrm{dec}}(\theta) \ll R\ll R_{\mathrm{N}}(\theta)=\Gamma_0^{\frac{2}{3-s}}(\theta)R_{\mathrm{dec}}(\theta)$, and finally, to the Sedov-Taylor phase \citep{1946JApMM..10..241S,1950RSPSA.201..159T} in the non-relativistic regime for $R\gg R_{\mathrm{N}}(\theta)$. For $R\gg R_{\mathrm{dec}}(\theta)$, the self-similar evolution becomes independent of the value of the initial Lorentz factor $\Gamma_0(\theta)$.

When the Lorentz factor $\Gamma(\theta;R)$ and the velocity $\beta(\theta;R)=\frac{\sqrt{\Gamma^2\left(\theta;R\right)-1}}{\Gamma(\theta;R)}$ are known, the corresponding time $t$ in the source frame is given by
\begin{equation}
    t(\theta;R) = \int_0^R \frac{\mathrm{d}r}{\beta(\theta;r) c}\, .
\end{equation}
This leads to the solution $R(\theta,t)$ and $\Gamma(\theta,t)$ that we use for the dynamics of each component of the structured jet.

The physical conditions in the shocked medium are easily deduced from the shock-jump conditions in the strong-shock regime \citep{1976PhFl...19.1130B}.
In the comoving frame, the mass density equals 
\begin{equation}
    \rho_*(\theta;R)=\left(4 \Gamma(\theta;R)+3\right) \rho_{\mathrm{ext}}(R)
    \label{eq:rhostarFS}
\end{equation}
and the internal energy per unit mass is given by
\begin{equation}
    \epsilon_*(\theta;R)=\left(\Gamma(\theta;R)-1\right) c^2
    \label{eq:epsilonstarFS}\, .
\end{equation}
We assumed an adiabatic index of $4/3$, which is valid as long as $\epsilon_*\gg c^2$. Finally, the timescale of the adiabatic cooling of the shocked region due to the spherical expansion is given in the comoving frame by
\begin{equation}
  t'_{\mathrm{dyn}}(\theta;R)
  =\frac{R}{\Gamma(\theta;R)\beta(\theta;R)\, c}
  =\frac{R}{c\, \sqrt{\Gamma^2(\theta;R) -1}}\, .
   \label{eq:tdyn}
\end{equation}

\subsection{Accelerated electrons and amplified magnetic field}\label{sec:acc_el_b_field}

We consider a shocked region where the physical conditions in the comoving frame are given by the mass density $\rho_*$, the internal energy per unit mass $\epsilon_*$, and the dynamical timescale $t'_{\mathrm{dyn}}$ (i.e. the characteristic timescale of the adiabatic cooling due to the spherical expansion), and we assume that the emission is produced by non-thermal shock-accelerated electrons that radiate in a local turbulent magnetic field amplified at the shock. In practice, we only consider the forward external shock, so that in this study, $\rho_*$, $\epsilon_*$, and $t'_{\mathrm{dyn}}$ are given by Eqs.~(\ref{eq:rhostarFS})--(\ref{eq:tdyn}).

We use the following standard parametrisation of the microphysics at the shock: (i) a fraction $\epsilon_{\mathrm{B}}$ of the internal energy is injected in the magnetic field, 
\begin{equation}
    u_{\mathrm{B}}=\frac{{B'}^2}{8\pi}=\epsilon_{\mathrm{B}}\, \rho_*\epsilon_*\, ,
\end{equation}
leading to
\begin{equation}
    B' = \sqrt{8\pi \epsilon_{\rm B} \rho_* \epsilon_*}\, ;
\end{equation}
(ii) a fraction $\epsilon_{\mathrm{e}}$ of the internal energy is injected into non-thermal electrons that represent a fraction $\zeta$ of all available electrons. Their number density ($\mathrm{cm}^{-3}$) and energy density ($\mathrm{erg \cdot cm^{-3}}$) in the comoving frame are therefore given by
\begin{equation}
    n_{\mathrm{e}}^{\mathrm{acc}}=\zeta \frac{\rho_*}{m_{\mathrm{p}}}
\end{equation}
and
\begin{equation}
    u_{\mathrm{e}}=\epsilon_{\mathrm{e}}\, \rho_*\epsilon_*\, .
\end{equation}
We assume a power-law distribution at injection with an index $-p$ and $2<p<3$. This leads to the following distribution of accelerated electrons ($\mathrm{cm}^{-3}$):
\begin{equation}
    n(\gamma) = 
    (p-1)
    \frac{n_{\mathrm{e}}^{\mathrm{acc}}}{\gamma_{\mathrm{m}}}
    \left(\frac{\gamma}{\gamma_{\mathrm{m}}}\right)^{-p}\,\,\,\,
    \mathrm{for}\,\,\,\, \gamma_{\mathrm{m}}\le \gamma\le \gamma_{\mathrm{max}}\, ,
    \label{eq:ngamma_acc}
\end{equation}
where the minimum Lorentz factor at injection equals
\begin{equation}
    \gamma_{\mathrm{m}}=\frac{p-2}{p-1} \frac{\epsilon_{\mathrm{e}}}{\zeta} \frac{m_{\mathrm{p}}}{m_{\mathrm{e}}}\frac{\epsilon_*}{c^2}\, .
    \label{eq:gm}
\end{equation}

As discussed below, it is important to take a realistic estimate of the maximum Lorentz factor $\gamma_{\mathrm{max}}$ up to which electrons can be accelerated at the shock into account for a discussion of the GeV-TeV afterglow emission. This was not included in the study by \citet{2009ApJ...703..675N}, on which we base our calculation of the emission. Eqs.~(\ref{eq:ngamma_acc}) and (\ref{eq:gm}) above were obtained by assuming that the maximum Lorentz factor $\gamma_{\mathrm{max}}$ is much higher than $\gamma_{\mathrm{m}}$. This is fully justified as we have typically $\gamma_\mathrm{max}/\gamma_\mathrm{m}> 10^6$ in the best-fit models of GW~170817 at the peak (Sect.~\ref{sec:170817}). The maximum electron Lorentz factor $\gamma_{\mathrm{max}}$ is evaluated by imposing that the acceleration timescale always remains shorter than the radiative and dynamical timescales,
\begin{equation}\label{eq:gamma_max_condition}
t'_{\mathrm{acc}}(\gamma) \le \min{\left(t'_{\mathrm{rad}}(\gamma);t'_{\mathrm{dyn}}\right)}\, .
\end{equation}

\begin{table*}[t]
        \caption{Values of the dimensionless parameters $K_P$; $K_\nu$ and $K_{P_\mathrm{max}}$ for different afterglow models used in each reference.}
    \centering
 \begin{tabular}{l|ccc}
 \hline\hline
        Reference 		& $K_P$ 			& $K_\nu$ 			& $K_{P_{\mathrm{max}}}$ 	 \\
        \hline
        \citet{1998ApJ...497L..17S} 				& $1$			& $1$				& $1$ 					\\
        \citet{2000ApJ...543...66P}			& $0.30$			& $0.78$				& $0.39$ 					 \\
        \citet{2001ApJ...548..787S,2018MNRAS.478.4128G} & $3.52\frac{p-1}{3p-1}$ & $3\pi/8$ & $0.88\frac{32}{3\pi}\frac{p-1}{3p-1}$\\
        & $\simeq 0.745$ for $p=2.16$ & $\simeq 1.178$ & $\simeq 0.633$ for $p=2.16$\\
        \hline
        \end{tabular}
    \label{tab:KPKnu}
\end{table*}

The acceleration timescale is written as a function of the Larmor time $R'_{\mathrm{L}}/c$ following Bohm's scaling, 
\begin{equation}\label{eq:tacc_gamma}
    t'_{\mathrm{acc}}(\gamma) 
    = K_{\rm acc} \frac{R'_{\mathrm{L}}(\gamma)}{c}
    = K_{\rm acc} \frac{\gamma m_{\rm e} c}{eB'}\, ,
\end{equation}
where $K_{\rm acc}\ge 1$ is a dimensionless factor. In practice, electrons at $\gamma_{\mathrm{max}}$ usually cool fast, and the maximum electron Lorentz factor is determined by the radiative timescale, $t'_{\mathrm{rad}}(\gamma)$. Its evaluation is non-trivial when IC scattering is taken into account (see Sect.~\ref{sec:emicom}). The resulting detailed calculation of $\gamma_{\mathrm{max}}$ is explained in Appendix~\ref{ap:gamma_max}. In the following, we assume $K_{\mathrm{acc}}=1$. Our value of $\gamma_{\mathrm{max}}$ should therefore be considered as an upper limit for the true value of the maximum electron Lorentz factor.

\subsection{Emissivity in the comoving frame}
\label{sec:emicom}

Our calculation of the emission from non-thermal electrons is mostly taken from \citet{2009ApJ...703..675N}. The population of electrons cools down by adiabatic cooling, synchrotron radiation, and via SSC scattering. As we are interested in the VHE afterglow emission, we include the attenuation due to pair production in Sect.~\ref{sec:pair_production}, but do not include the emission of secondary leptons. We focus on the afterglow of GW~170817, for which the radio observations did not show any evidence for absorption, and we therefore do not include the effect of synchrotron self-absorption in the current version of the model. In this subsection, all quantities are written in the comoving frame of the considered shocked region. We therefore omit the prime to simplify the notations.

\subsubsection{Synchrotron and inverse-Compton radiation for a single electron}

\paragraph{Synchrotron radiation.} 
The synchrotron power of a single  electron ($\mathrm{erg}\cdot\mathrm{s}^{-1}$) is given by
\begin{equation}
    P^{\mathrm{syn}}(\gamma) = K_1 B^2 \gamma^2\, ,
\end{equation}
with $K_1=K_P \frac{\sigma_{\mathrm{T}}c}{6\pi}$ and $K_P$ a dimensionless parameter, and its synchrotron frequency equals
\begin{equation}
    \nu_{\mathrm{syn}}(\gamma)= K_2 B \gamma^2\, ,
\end{equation}
with $K_2 = K_{\nu} \frac{e}{2\pi m_{\mathrm{e}}c}$ and $K_\nu$ a dimensionless parameter. The dimensionless parameters $K_P$ and $K_\nu$ depend on the assumptions made on the pitch angle of electrons relative to the magnetic field lines and different values are found in afterglow models, as listed in Table~\ref{tab:KPKnu}. By default, we use the same values as \citet{2018MNRAS.478.4128G}, taken from \citet{1999ApJ...513..679G}, who obtained them by fitting the broken power-law approximation of  the synchrotron power of a power-law distribution of electrons to the exact calculation.

The corresponding synchrotron power at frequency $\nu$ ($\mathrm{erg \cdot s^{-1} \cdot Hz^{-1}}$) is given by
\begin{equation}
    P^{\mathrm{syn}}_\nu(\gamma) = P_{\mathrm{max}} \Phi\left(\frac{\nu}{\nu_{\mathrm{syn}}(\gamma)}\right)\, ,
    \label{eq:Psynelec}
\end{equation}
with
\begin{equation}
P_{\mathrm{max}} =
\frac{P^{\mathrm{syn}}(\gamma)}{\nu_{\mathrm{syn}}(\gamma)}
= \frac{K_1}{K_2} B
=K_{P_{\mathrm{max}}} \frac{\sigma_{\mathrm{T}}m_{\mathrm{e}}c^2}{3e}\, B\, ,
\end{equation}
$K_{P_{\mathrm{max}}}=K_{P}/K_{\nu}$ and $\int_0^\infty \Phi(x)\mathrm{d}x=1$. In practice, we adopt a simplified shape $\Phi(x) = \frac{4}{3} x^{1/3}$ for $x\le 1$ and $0$ otherwise. 

\paragraph{Inverse-Compton scattering.}
The spectral number density of seed photons is $n_\nu^{\mathrm{syn}}$ ($\mathrm{cm^{-3} \cdot Hz^{-1}}$), which is assumed to result only from the synchrotron radiation of the electrons on which they also scatter (SSC). The IC power of an electron ($\mathrm{erg\cdot s^{-1}}$) is then given by
\begin{equation}
   P^{\mathrm{SSC}}(\gamma) = \int_0^\infty
   \mathrm{d}\nu\, n_\nu^{\mathrm{syn}}\, \sigma_{\mathrm{T}} f_{\mathrm{KN}}(w) c\,\, K_\mathrm{IC}\,\gamma^2 g_{\mathrm{KN}}(w) h\nu\, ,
   \label{eq:PSSCfirst}
\end{equation}
where $w= \gamma h\nu / m_{\mathrm{e}}c^2$; and $f_{\mathrm{KN}}$ and $g_{\mathrm{KN}}$ are the KN corrections to the cross-section and to the mean energy of the scattered photon. In the Thomson regime ($w\ll w_\mathrm{KN}$), the cross-section is $\sigma_{\mathrm{T}}$ and the mean energy of scattered photons is $K_\mathrm{IC}\, \gamma^2 h \nu$ for a seed synchrotron photon with a frequency $\nu$, so that $f_{\mathrm{KN}}=g_{\mathrm{KN}}=1$ for $w\ll w_\mathrm{KN}$. By default, we use the standard values $K_\mathrm{IC} = 4/3$ and $w_\mathrm{KN} = 1$, as in \citet{2009ApJ...703..675N}. We note, however, that \citet{2022MNRAS.512.2142Y} recently suggested that a better agreement with a detailed numerical spectral calculation of the SSC component was achieved with $w_\mathrm{KN} = 0.2$, especially at the transition between the Thomson and the KN regime.
For simplicity, we do not write in Eq.~(\ref{eq:PSSCfirst}) the dependence on the angle between the upscattered photon and the scattering electron explicitly. This was discussed in \citet{2009ApJ...703..675N}. We assume an isotropic seed radiation field. We only account for single IC scattering as the effects of multiple scatterings are expected to be negligible because the afterglow is produced in the optically thin regime. The Compton parameter is defined by
\begin{equation}
Y(\gamma) = \frac{P^{\mathrm{SSC}}(\gamma)}{P^{\mathrm{syn}}(\gamma)}\, .
\end{equation}
This Compton parameter is constant only when SSC scattering occurs in the Thomson regime. We take the KN regime into account, which explains the dependence on the electron Lorentz factor. This affects not only the SSC emission, but also the synchrotron component. Following \citet{2009ApJ...703..675N} and keeping the same notations, we simplify the IC cross-section by assuming that scattering is entirely suppressed in the KN regime: $f_{\mathrm{KN}}(w)=0$ for $w\ge w_\mathrm{KN}$. Then, a synchrotron photon produced by an electron with Lorentz factor $\gamma$ can only be upscattered by electrons with Lorentz factors below a certain limit $\widehat{\gamma}$ to remain in the Thomson regime, 
\begin{equation}
    \widehat{\gamma} = \frac{w_\mathrm{KN}\, m_{\rm e}c^2}{h\nu_{\rm syn}(\gamma)} \propto \gamma^{-2}\, .
\end{equation}
We note that $\widehat{\gamma}$ is a decreasing function of $\gamma$: If the energy of the synchrotron photon is higher, the maximum energy of any electron on which it could be upscattered in the Thomson regime must be lower. In an equivalent way, electrons with Lorentz factor $\gamma$ only scatter photons below the energy 
\begin{equation}
	h\widetilde{\nu} = w_\mathrm{KN} m_{\mathrm{e}}c^2/\gamma\, .
\end{equation}
Then, it is convenient to define $\widetilde{\gamma}$ such that $\nu_\mathrm{syn}(\widetilde{\gamma})=\widetilde{\nu}$,
\begin{equation}\label{eq:gamma_tilde_def}
\widetilde{\gamma}=\sqrt{w_\mathrm{KN} \gamma m_{\rm e}c^2/h\nu_\mathrm{syn}(\gamma)}\propto \gamma^{-1/2}\, .
\end{equation}
Electrons with a Lorentz factor $\gamma$ can only scatter synchrotron photons produced by electrons with Lorentz factors below $\widetilde{\gamma}$. The Lorentz factor $\gamma_\mathrm{self}$ is defined by 
\begin{equation}\label{eq:gamma_self}
\gamma_\mathrm{self}=\widehat{\gamma}_\mathrm{self}=\widetilde{\gamma}_\mathrm{self}=\left(\frac{w_\mathrm{KN}m_\mathrm{e}c^2}{h K_2 B}\right)^{1/3}\, .
\end{equation}

The definition of $\widetilde{\nu}$ allows us to simplify the expression of the SSC power given by Eq.~(\ref{eq:PSSCfirst}):
\begin{equation}
  P^{\mathrm{SSC}}(\gamma) = K_\mathrm{IC} 
  \sigma_{\mathrm{T}}c\, \gamma^2 \int_0^{\widetilde{\nu}}\mathrm{d}\nu\, u_\nu^{\mathrm{syn}}\, ,
  \label{eq:PSSC}
\end{equation}
with $u_\nu^{\mathrm{syn}}=n_\nu^{\mathrm{syn}} \times h\nu$ the energy density of the seed synchrotron photons ($\mathrm{erg \cdot cm^{-3} \cdot Hz^{-1}}$). Then,
\begin{equation}
    Y(\gamma) = \frac{3}{4} \frac{K_\mathrm{IC}}{K_P}  \frac{\int_0^{\widetilde{\nu}}\mathrm{d}\nu\, u_\nu^{\mathrm{syn}}}{u_{\mathrm{B}}}\, .
    \label{eq:Ygamma}
\end{equation}
Assuming a strict suppression in the KN regime leads to a simple expression for the SSC power at frequency $\nu$ ($\mathrm{erg \cdot s^{-1} \cdot Hz^{-1}}$) for a single electron with a Lorentz factor $\gamma$,
\begin{equation}\label{eq:pnusscgamma}
    P^{\mathrm{SSC}}_\nu(\gamma) = \sigma_{\mathrm{T}}c \, u_{\nu_\mathrm{seed}=\nu/\left(K_\mathrm{IC}
    \gamma^2\right)}^{\mathrm{syn}}
\end{equation}
if $h\nu \le K_\mathrm{IC} \gamma m_{\mathrm{e}}c^2$ and $0$ otherwise.

\subsubsection{Radiative regime of a single electron}

Following \citet{1998ApJ...497L..17S}, the first expected break in the electron distribution occurs at the critical electron Lorentz factor $\gamma_{\mathrm{c}}$, defined by the condition $t_{\mathrm{rad}}(\gamma)=t_{\mathrm{dyn}}$, so that high-energy electrons with $\gamma\gg \gamma_{\mathrm{c}}$ are radiatively efficient (fast cooling), and low-energy electrons with  $\gamma\ll \gamma_{\mathrm{c}}$ mainly cool via the adiabatic expansion and are radiatively inefficient (slow cooling). The radiative timescale is given by
\begin{equation}\label{eq:trad_gamma}
    t_{\mathrm{rad}}(\gamma) = \frac{\gamma m_{\mathrm{e}}c^2}{P_{\mathrm{syn}}(\gamma)+P_{\mathrm{SSC}}(\gamma)}
= \frac{m_{\mathrm{e}}c^2}{\left[1+Y(\gamma)\right] K_1 B^2 \gamma}\, .
\end{equation}
Then, self-consistently computing $\gamma_{\mathrm{c}}$ requires us to solve the equation
\begin{equation}\label{eq:gamma_c_y_c}
    \gamma_{\mathrm{c}} \left[1 + Y(\gamma_{\mathrm{c}})\right] = \gamma_{\mathrm{c}}^{\mathrm{syn}} =
    \frac{m_{\mathrm{e}}c^2}{K_1 B^2\, t_{\mathrm{dyn}}}\, ,
\end{equation}
where $\gamma_{\rm c}^{\rm syn}$ is the value of the critical Lorentz factor obtained when only the synchrotron radiation is taken into account. Our procedure to compute $\gamma_\mathrm{c}$ and $Y(\gamma_\mathrm{c})$ in the most general case where the SSC cooling impacts the electron distribution is discussed below in Sect.~\ref{sec:fgh}.

\subsubsection{Electron distribution and associated emission}

\paragraph{Electron distribution.} Following \citet{2009ApJ...703..675N}, we compute the time-averaged distribution of electrons $\bar{n}(\gamma)$ over the dynamical timescale $t_\mathrm{dyn}$ by keeping only the dominant term for the electron cooling: The instantaneous power of an electron is either dominated by the synchrotron and SSC radiation $\left(1+Y(\gamma)\right)P^\mathrm{syn}(\gamma)$ for $\gamma>\gamma_\mathrm{c}$ or by the adiabatic cooling $\gamma\, m_\mathrm{e}c^2/t_\mathrm{dyn}$ if $\gamma < \gamma_\mathrm{c}$. This leads to
\begin{equation}
    \bar{n}(\gamma) = \frac{n_\mathrm{e}^\mathrm{acc}}{1+Y(\gamma)}\frac{\gamma_\mathrm{c}^\mathrm{syn}}{\gamma_\mathrm{m}^2}\times\left\lbrace\begin{array}{cl}
    \left(\gamma/\gamma_\mathrm{m}\right)^{-2} & \mathrm{if}\, \gamma_\mathrm{c} < \gamma < \gamma_\mathrm{m}\\
    \left(\gamma/\gamma_\mathrm{m}\right)^{-(p+1)} & \mathrm{if}\, \gamma_\mathrm{m} < \gamma < \gamma_\mathrm{max}\\
\end{array}\right.
\label{eq:barnFC}
\end{equation}
in fast-cooling regime ($\gamma_\mathrm{m}>\gamma_\mathrm{c}$) and to
\begin{eqnarray}
     \bar{n}(\gamma)  & \!\!\!\!\!=\!\!\!\!\! &  n_\mathrm{e}^\mathrm{acc}\frac{\gamma_\mathrm{c}^\mathrm{syn}}{\gamma_\mathrm{m}^2}\left(\frac{\gamma_\mathrm{c}}{\gamma_\mathrm{m}}\right)^{-(p+1)}\nonumber\\
    & & \!\!\!\!\!\!\times\left\lbrace\begin{array}{cl}
    \left(\gamma/\gamma_\mathrm{c}\right)^{-p}/\left(1+Y(\gamma_\mathrm{c})\right) & \mathrm{if}\, \gamma_\mathrm{m} < \gamma < \gamma_\mathrm{c}\\
    \left(\gamma/\gamma_\mathrm{c}\right)^{-(p+1)}/\left(1+Y(\gamma)\right) & \mathrm{if}\, \gamma_\mathrm{c} < \gamma < \gamma_\mathrm{max}\\
\end{array}\right.\nonumber\\
\label{eq:barnSC}
\end{eqnarray}
in slow-cooling regime ($\gamma_\mathrm{m}<\gamma_\mathrm{c}$). We define $\gamma_\mathrm{m,c}=\max{\left(\gamma_{\mathrm{m}};\gamma_{\mathrm{c}}\right)}$ and $\gamma_\mathrm{c,m}=\min{\left(\gamma_{\mathrm{m}};\gamma_{\mathrm{c}}\right)}$ so that $\gamma_\mathrm{m,c} = \gamma_\mathrm{m}$ and $\gamma_\mathrm{c,m} = \gamma_\mathrm{c}$ in fast cooling and $\gamma_\mathrm{m,c} = \gamma_\mathrm{c}$ and $\gamma_\mathrm{c,m} = \gamma_\mathrm{m}$ in slow cooling. The minimum electron Lorentz factor therefore is $\gamma_{\mathrm{min}}=\gamma_\mathrm{c,m}$. We define the corresponding synchrotron frequencies $\nu_\mathrm{m,c}=\nu_\mathrm{syn}(\gamma_\mathrm{m,c})$ and $\nu_\mathrm{c,m}=\nu_\mathrm{syn}(\gamma_\mathrm{c,m})$. We also always only keep the dominant term in the radiated power, that is, $1+Y(\gamma)\simeq 1$ if $Y(\gamma)<1$ and $Y(\gamma)$ otherwise. As described in detail in \citet{2009ApJ...703..675N}, the resulting distribution shows several breaks in addition to $\gamma_\mathrm{m}$ and $\gamma_\mathrm{c}$: A break is expected at $\gamma_0$ such that $Y(\gamma_0)=1$, and several additional breaks are possible and must be identified in an iterative way, as described in Sect.~\ref{sec:fgh}. 

Following \citet{1998ApJ...497L..17S} and \citet{2009ApJ...703..675N}, we approximate $\bar{n}(\gamma)$ and the associated synchrotron spectrum by broken power laws to allow for a semi-analytical calculation. Electrons Lorentz factors are normalised by $\gamma_\mathrm{m,c}$, that is, $x=\gamma/\gamma_\mathrm{m,c}$, and photon frequencies by $\nu_\mathrm{m,c}$, that is, $y=\nu/\nu_\mathrm{m,c}$. Hence, the normalised synchrotron frequency of an electron with normalised Lorentz factor $x$ is $y=x^2$. In the following, the notations introduced for characteristic electron Lorentz factors or photon frequencies are implicitly conserved for the normalised quantities. For instance, $x_\mathrm{self}=\gamma_\mathrm{self}/\gamma_\mathrm{m,c}$, $\widehat{x}=\widehat{\gamma}/\gamma_\mathrm{m,c}=x_\mathrm{self}^3/x^2$, and $\widetilde{y}=\widetilde{\nu}/\nu_\mathrm{m,c}=\widetilde{x}^2$.

\paragraph{Normalised electron distribution.}
The electron distribution is given by
\begin{equation}
    \bar{n}(\gamma) = \frac{1}{I_0} \frac{n_{\mathrm{e}}^{\mathrm{acc}}}{\gamma_{\mathrm{m,c}}}f(x)\, ,
\end{equation}
where $f(x)$ is the normalised broken power-law electron distribution with $N$ breaks and power-law segments,
\begin{equation}\label{eq:f_x}
f(x) = \left\lbrace\begin{array}{rl}
x^{-p_1} & \mathrm{if}\, x_1\le x \le x_2\\
x_2^{p_2-p_1}\, x^{-p_2} & \mathrm{if}\, x_2\le x \le x_3\\
x_2^{p_2-p_1} x_3^{p_3-p_2}\, x^{-p_3} & \mathrm{if}\, x_2\le x \le x_3\\
\vdots & \vdots\\
x_2^{p_2-p_1} x_3^{p_3-p_2} \cdots\,
x_N^{p_N-p_{N-1}}
\, x^{-p_N} & \mathrm{if}\, x_N\le x \le x_{N+1}\\
\end{array}\right.
\end{equation}

\vspace*{1ex}

\noindent with $x_1=\gamma_{\mathrm{min}}/\gamma_{\mathrm{m,c}}$ and $x_{N+1}=\gamma_{\mathrm{max}}/\gamma_{\mathrm{m,c}}$. The number of breaks in the purely synchrotron case or when all IC scatterings are in the Thomson regime is $N=2$, as discussed in Sect.~\ref{sec:pureSYN} and \ref{sec:SSCnoKN}. In the general case, where KN corrections cannot be neglected, more breaks appear in the electron distribution, as discussed in Sect.~\ref{sec:fgh}. All possible cases are provided in Appendix~\ref{sec:sol_fgh}. Typically, $N=2$ to $5$, except in some very specific regions of the parameter space (case III in \citealt{2009ApJ...703..675N}). This is also discussed in Appendix~\ref{sec:sol_fgh}. 

The dimensionless integral $I_0$ is given by
\begin{equation}\label{eq:I_0}
I_0 = \int_{x_1}^{x_{N+1}} f(x)\, \mathrm{d}x
\end{equation}
to conserve the number of electrons, $\int_{\gamma_\mathrm{min}}^{\gamma_\mathrm{max}} \bar{n}(\gamma)\, \mathrm{d}\gamma=n_\mathrm{e}^\mathrm{acc}$.

\paragraph{Normalised synchrotron spectrum.}
The synchrotron power per electron and per unit frequency $p_\nu^\mathrm{syn}$ ($\mathrm{erg \cdot s^{-1} \cdot Hz^{-1} \cdot electron^{-1}}$) averaged over the timescale $t_\mathrm{dyn}$ is deduced from Eq.~(\ref{eq:Psynelec}) using the simplified shape for $\Phi$. This leads to a broken power-law synchrotron spectrum,
\begin{eqnarray}
  p_\nu^\mathrm{syn} & = & \frac{1}{n_\mathrm{e}^\mathrm{acc}} \int_{\gamma_\mathrm{min}}^{\gamma_\mathrm{max}} \mathrm{d}\gamma\, \bar{n}(\gamma) P_\nu^\mathrm{syn}(\gamma)\\
  & = & \frac{\gamma_\mathrm{m,c}^2\, m_\mathrm{e}c^2}{\gamma_\mathrm{c}^\mathrm{syn} \nu_\mathrm{m,c}\, t_\mathrm{dyn}}\, \times \frac{4}{3} y^{1/3}\int_{\max{\left(x_1; y^{1/2}
  \right)}}^{x_{N+1}} \!\!\!\!\mathrm{d}x\,  \frac{f(x)}{x^{2/3}}\nonumber\\ 
  & \simeq & \frac{1}{I_0}\, \frac{I_2}{J_0}   \frac{\gamma_\mathrm{m,c}^2\, m_\mathrm{e}c^2}{\gamma_\mathrm{c}^\mathrm{syn} \nu_\mathrm{m,c}\, t_\mathrm{dyn}}\, \times g(y)\, ,
    \label{eq:pnusynpere}
\end{eqnarray}
where the normalised broken power-law spectral shape is obtained by keeping the dominant term in the integral of $f(x)/x^{2/3}$,
\begin{equation}
g(y)
=
\left\lbrace\begin{array}{rl}
x_1^{\frac{1}{3}- p_1} y^{\frac{1}{3}} & \!\mathrm{if}\, y\! < x_1^2\\
y^{-\frac{p_1-1}{2}} & \!\mathrm{if}\, x_1^2 < \!y\! < x_2^2\\
x_2^{p_2-p_1}\, y^{-\frac{p_2-1}{2}} & \!\mathrm{if}\, x_2^2 < \!y\! < x_3^2\\
x_2^{p_2-p_1} x_3^{p_3-p_2}\, y^{-\frac{p_3-1}{2}} & \!\mathrm{if}\, x_3^2 < \!y\! < x_4^2\\
\vdots &  \vdots \\
\!\!x_2^{p_2-p_1} x_3^{p_3-p_2}\, \cdots\,  x_N^{p_N-p_{N-1}}\, y^{-\frac{p_N-1}{2}} & \!\mathrm{if}\, x_N^2 < \!y\! < x_{N+1}^2\\
\end{array}\right..
\label{eq:defg}
\end{equation}
The dimensionless factors $I_2$ and $J_0$ are defined by
\begin{eqnarray}\label{eq:I_2}
   I_2 & = & \int_{x_1}^{x_{N+1}} \mathrm{d}x\, x^2\, f(x)\\
   J_0 & = & \int_0^{x_{N+1}^2} \mathrm{d}y\, g(y)\, \label{eq:J_0}
\end{eqnarray}
to ensure that the total synchrotron power per electron ($\mathrm{erg \cdot s^{-1} \cdot electron^{-1}}$) is conserved,
\begin{eqnarray}
    p^\mathrm{syn} & = &\int_0^\infty \mathrm{d}\nu\,  p_\nu^\mathrm{syn} = \frac{I_2}{I_0}
    \frac{\gamma_\mathrm{m,c}^2\, m_\mathrm{e}c^2}{\gamma_\mathrm{c}^\mathrm{syn} \, t_\mathrm{dyn}}\nonumber\\
    & = & \frac{1}{n_\mathrm{e}^\mathrm{acc}}\int_{\gamma_\mathrm{min}}^{\gamma_\mathrm{max}} \mathrm{d}\gamma\, \bar{n}(\gamma) P^\mathrm{syn}(\gamma)
    \, .
\end{eqnarray}

Finally, we define $\nu_\mathrm{p}$ as the peak frequency of the synchrotron spectrum, that is, the frequency $\nu$ where $\nu^2  p_\nu^\mathrm{syn}$ is maximum. This peak frequency corresponds to synchrotron photons emitted by electrons at a Lorentz factor $\gamma_\mathrm{p}$, that is, $\nu_\mathrm{p}=\nu_\mathrm{syn}(\gamma_\mathrm{p})$. We note $i_\mathrm{p}$ the corresponding index of the break in the normalised distributions: $x_\mathrm{p}=x_{i_\mathrm{p}}$ and $y_\mathrm{p}=y_{i_\mathrm{p}}=x_\mathrm{p}^2$.

\paragraph{Normalised SSC spectrum.}
The SSC power per electron and per unit frequency $p_\nu^\mathrm{SSC}$ ($\mathrm{erg \cdot s^{-1} \cdot Hz^{-1} \cdot electron^{-1}}$) emitted over the timescale $t_\mathrm{dyn}$ is deduced from Eq.~(\ref{eq:pnusscgamma}) with 
\begin{equation}
u_\nu^\mathrm{syn} = n_\mathrm{e}^\mathrm{acc}\,  p_\nu^\mathrm{syn}t_\mathrm{dyn}\, .
\label{eq:unusyn}
\end{equation}
This leads to
\begin{eqnarray}
   p_\nu^\mathrm{SSC} & = & \frac{1}{n_\mathrm{e}^\mathrm{acc}} \int_{\gamma_\mathrm{min}}^{\gamma_\mathrm{max}}\mathrm{d}\gamma\, \bar{n}(\gamma) P_\nu^\mathrm{SSC}(\gamma)\\
   & \simeq & \frac{1}{I_0}\frac{I_2}{J_0} \frac{\tau_\mathrm{T}}{I_0} \frac{\gamma_\mathrm{m,c}^2\, m_\mathrm{e}c^2}{\gamma_\mathrm{c}^\mathrm{syn} \nu_\mathrm{m,c}\, t_\mathrm{dyn}}\times G(y)\, ,
   \label{eq:pnuSSCpere}
\end{eqnarray}
where 
\begin{equation}\label{eq:tau_Thomson}
\tau_\mathrm{T}=n_\mathrm{e}^\mathrm{acc}\, \sigma_\mathrm{T}\, c t_\mathrm{dyn}
\end{equation}
is the Thomson optical depth, and where the normalised spectral shape is given by
\begin{equation}\label{eq:G_y_SSC}
    G(y) = \int_{\max{\left(x_1;\frac{1}{K_\mathrm{IC}}\frac{h\nu_\mathrm{m,c}}{\gamma_\mathrm{m,c} m_\mathrm{e}c^2}y\right)}}^{x_{N+1}} \mathrm{d}x\, f(x)\, g\left(y_{\mathrm{seed}}
    = \frac{1}{K_\mathrm{IC}}\frac{y}{\gamma_\mathrm{m,c}^2 x^2}\right)\, .
\end{equation}
The cutoff at $x_\mathrm{KN}=\gamma_\mathrm{KN}/\gamma_\mathrm{m,c}$ with $\gamma_\mathrm{KN}= \frac{1}{K_\mathrm{IC}}\frac{h\nu}{ m_\mathrm{e}c^2}$ is a direct consequence of the strict suppression assumed in the KN regime: The VHE flux due to the scattering in the KN regime is neglected. \cite{2022MNRAS.512.2142Y} introduced a factor $f_\mathrm{KN}$ in their calculations to compensate for this approximation, which we do not include here. The KN regime affects the high-energy part of the emitted spectrum: Above a photon energy $h\nu= K_\mathrm{IC}\gamma_\mathrm{min} m_\mathrm{e} c^2$, we have $\gamma_\mathrm{KN}>\gamma_\mathrm{min}$ and the SSC emission is reduced. The SSC emission is even entirely suppressed ($G(y)=0$) at very high photon energies $h\nu> K_\mathrm{IC}\gamma_\mathrm{max} m_\mathrm{e} c^2$, corresponding to $\gamma_\mathrm{KN}>\gamma_\mathrm{max}$.

In practice, $G(y)$ is computed exactly in our numerical implementation. We also note that $J_0\simeq 2 I_2$ when only the leading term is kept in the integrals. However, all dimensionless factors $I_0$, $I_2$, and $J_0$ are also computed exactly in our numerical implementation, which ensures the continuity of the electron distribution and emitted spectrum at the transition from the fast- to the slow-cooling regime, as well as between the different possible cases in either regime (see Sect.~\ref{sec:fgh}).

\subsubsection{Purely synchrotron case}
\label{sec:pureSYN}
The standard purely synchrotron case in which the SSC emission is neglected ($Y(\gamma)=0$ for all electrons) was fully described in \citet{1998ApJ...497L..17S}. We have $\gamma_\mathrm{c}=\gamma_\mathrm{c}^\mathrm{syn}$ in this case. In the fast-cooling regime ($\gamma_\mathrm{m} > \gamma_\mathrm{c}$), the normalised distributions are given by
\begin{eqnarray}
   f(x) & = & \left\lbrace\begin{array}{rl}
   x^{-2} & \mathrm{if}\, x_1  < x < x_2\\
   x_2^{p-1} x^{-(p+1)} & \mathrm{if}\, x_2 < x < x_\mathrm{3} \\
   \end{array}\right.\, ,\\
   g(y) & = &  \left\lbrace\begin{array}{rl}
   x_1^{-5/3} y^{1/3} & \mathrm{if}\, y < x_1^2\\
   y^{-1/2} & \mathrm{if}\, x_1^2 < y < x_2^2 \\
   x_2^{p-1} y^{-p/2} & \mathrm{if}\, x_2^2 < y < x_3^2 \\
   \end{array}\right.\, ,
\end{eqnarray}
with $x_1= {\gamma_\mathrm{c}}/{\gamma_\mathrm{m}}$, $x_2=1$ and $x_3={\gamma_\mathrm{max}}/{\gamma_\mathrm{m}}$. In the slow-cooling regime ($\gamma_\mathrm{m} < \gamma_\mathrm{c}$), they are given by
\begin{eqnarray}
   f(x) & = & \left\lbrace\begin{array}{rl}
   x^{-p} & \mathrm{if}\, x_1  < x < x_2\\
   x_2\, x^{-(p+1)} & \mathrm{if}\, x_2 < x < x_\mathrm{3} \\
   \end{array}\right.\, ,\\
   g(y) & = &  \left\lbrace\begin{array}{rl}
   x_1^{1/3-p} y^{1/3} & \mathrm{if}\, y < x_1^2\\
   y^{-(p-1)/2} & \mathrm{if}\, x_1^2 < y < x_2^2 \\
   x_2\, y^{-p/2} & \mathrm{if}\, x_2^2 < y < x_3^2 \\
   \end{array}\right.\, ,
\end{eqnarray}
with $x_1= {\gamma_\mathrm{m}}/{\gamma_\mathrm{c}}$, $x_2=1$ and $x_3={\gamma_\mathrm{max}}/{\gamma_\mathrm{c}}$. Calculations using this prescription are labelled "no SSC" in the following.

\subsubsection{Synchrotron self-Compton case in the Thomson regime}
\label{sec:SSCnoKN}
If the KN regime is neglected, all IC scatterings occur in the Thomson regime, and the Compton parameter is the same for all electrons: $Y(\gamma)=Y^\mathrm{no\, KN}=\mathrm{cst}$. The corresponding solution was given by \citet{2001ApJ...548..787S}: The normalised distributions $f(x)$ and $g(y)$ are the same as in the purely synchrotron case above, but the value of the critical Lorentz factor $\gamma_\mathrm{c}$ is decreased due to the IC cooling, $\gamma_\mathrm{c}=\gamma_\mathrm{c}^\mathrm{syn}/\left(1+Y^\mathrm{no\, KN}\right)$. Using Eqs.~(\ref{eq:Ygamma}), (\ref{eq:pnusynpere}), and~(\ref{eq:unusyn}), we obtain
\begin{equation}
    Y^\mathrm{no\, KN} \left(1+Y^\mathrm{no\, KN}\right) = \frac{3}{4}\frac{K_\mathrm{IC}}{K_P}\, \frac{p-2}{p-1}\, \frac{\epsilon_\mathrm{e}}{\epsilon_\mathrm{B}}\, \frac{\gamma_\mathrm{m,c}}{\gamma_\mathrm{c,m}}\, \frac{I_2}{I_0}\, .
    \label{eq:YSSCnoKN}
\end{equation}
As $I_2/I_0$ is a function of $\gamma_\mathrm{c}/\gamma_\mathrm{m}=\gamma_\mathrm{c}^\mathrm{syn}/\gamma_\mathrm{m}/(1+Y^\mathrm{no\, KN})$, this is an implicit equation to be solved numerically to obtain $Y^\mathrm{no\, KN}$ and $\gamma_\mathrm{c}$. The limits for $\left(1+Y^\mathrm{no\, KN}\right)Y^\mathrm{no\, KN}$ are $\frac{3}{4}\frac{K_\mathrm{IC}}{K_P}\,  \frac{\epsilon_\mathrm{e}}{\epsilon_\mathrm{B}}$ for $\gamma_\mathrm{m}\gg\gamma_\mathrm{c}$ and $\frac{3}{4}\frac{K_\mathrm{IC}}{K_P}\,  \frac{\epsilon_\mathrm{e}}{\epsilon_\mathrm{B}} \frac{1}{3-p}\left(\frac{\gamma_\mathrm{m}}{\gamma_\mathrm{c}^\mathrm{syn}}\right)^{p-2}$ for $\gamma_\mathrm{m}\ll\gamma_\mathrm{c}$. Calculations using this prescription are labelled "SSC (Thomson)" in the following.

\subsubsection{Self-consistent calculation of the electron distribution and the Compton parameter in the general case}
\label{sec:fgh}

\paragraph{Normalised Compton parameter.}
In the general case where the KN suppression at high energy is included, the Compton parameter $Y(\gamma)$ is a decreasing function of the electron Lorentz factor. From Eqs.~(\ref{eq:Ygamma}), (\ref{eq:pnusynpere}), and~(\ref{eq:unusyn}), we obtain
\begin{equation}
   Y(\gamma)\left(1+Y(\gamma_\mathrm{c})\right)
   = \frac{3}{4}\frac{K_\mathrm{IC}}{K_P}\, \frac{p-2}{p-1}\, \frac{\epsilon_\mathrm{e}}{\epsilon_\mathrm{B}}\, \frac{\gamma_\mathrm{m,c}}{\gamma_\mathrm{c,m}}\, \frac{I_2}{I_0}\, \frac{1}{J_0}\int_0^{\widetilde{y}}\!\!\mathrm{d}y\, g(y)\, .
	\label{eq:YfullSSC}
\end{equation}
For low Lorentz factors $\gamma$, $\widetilde{\nu}$ becomes very large, so that for $\widetilde{\nu}>\nu_\mathrm{syn}(\gamma_\mathrm{max})$, we recover the Thomson regime, where the right-hand side of Eq.~(\ref{eq:YfullSSC}) is formally the same as in Eq.~(\ref{eq:YSSCnoKN}), even if the integrals $I_0$ and $I_ 2$ can be different when the normalised electron distribution $f(x)$ is different. As described in \citet{2009ApJ...703..675N}, when only the leading term in the integral of $g(y)$ is kept, the corresponding scaling law for the Compton parameter is $Y(\gamma)=\mathrm{cst}$ if $\widetilde{\nu}$ is above the peak frequency $\nu_\mathrm{p}$ and $Y(\gamma)\propto \gamma^{-\frac{3-\widetilde{p}}{2}}$ otherwise, where $\widetilde{p}$ is the slope of electron distribution in the power-law segment of the electron distribution including $\widetilde{\gamma}$. We therefore introduce a normalised Compton parameter defined by
\begin{equation}
   Y(\gamma)\left(1+Y(\gamma_\mathrm{c})\right)
=
    \frac{3}{4}\frac{K_\mathrm{IC}}{K_P}\, \frac{p-2}{p-1}\, \frac{\epsilon_\mathrm{e}}{\epsilon_\mathrm{B}}\, \frac{\gamma_\mathrm{m,c}}{\gamma_\mathrm{c,m}}\, \frac{I_2}{I_0}\, 
   h(x)\, ,
   \label{eq:defh}
\end{equation}
where $h(x)$ follows this scaling and is normalised so that Eq.~(\ref{eq:defh}) has the exact limit in the Thomson regime. This leads to
\begin{equation}
h(x) 
=
\left\lbrace\begin{array}{rl}
1 & \!\mathrm{if}\, x\! < \widehat{x}_\mathrm{p}\\
\widehat{x}_\mathrm{p}^{\frac{3-p_{i_\mathrm{p}-1}}{2}}
x^{-\frac{3-p_{i_\mathrm{p}-1}}{2}}
& \!\mathrm{if}\, \widehat{x}_\mathrm{p} < \!x\! < \widehat{x}_{i_\mathrm{p}-1}\\
\!\!\!\widehat{x}_\mathrm{p}^{\frac{3-p_{i_\mathrm{p}-1}}{2}}
\widehat{x}_{i_\mathrm{p}-1}^{\frac{p_{i_\mathrm{p}-1}-p_{i_\mathrm{p}-2}}{2}}
x^{-\frac{3-p_{i_\mathrm{p}-2}}{2}}
& \!\mathrm{if}\, \widehat{x}_{i_\mathrm{p}-1} < \!x\! < \widehat{x}_{i_\mathrm{p}-2}\\
\vdots & \vdots \\
\widehat{x}_\mathrm{p}^{\frac{3-p_{i_\mathrm{p}-1}}{2}}
\widehat{x}_{i_\mathrm{p}-1}^{\frac{p_{i_\mathrm{p}-1}-p_{i_\mathrm{p}-2}}{2}}
\cdots & \\
\cdots\, \widehat{x}_2^{\frac{p_2-p_1}{2}}
x^{-\frac{3-p_1}{2}}
& \!\mathrm{if}\, \widehat{x}_{2} < \!x < \widehat{x}_{1}\\
\widehat{x}_\mathrm{p}^{\frac{3-p_{i_\mathrm{p}-1}}{2}}
\widehat{x}_{i_\mathrm{p}-1}^{\frac{p_{i_\mathrm{p}-1}-p_{i_\mathrm{p}-2}}{2}}
\cdots & \\
\cdots \,\!
\widehat{x}_2^{\frac{p_2-p_1}{2}}
\widehat{x}_1^{\frac{p_1}{2}-\frac{1}{6}}
x^{-4/3}
& \!\mathrm{if}\, x\! > \widehat{x}_1
\end{array}\right. .
\end{equation}
We recall that $i_\mathrm{p}$ is the index of the break corresponding to the peak frequency of the synchrotron spectrum. In most cases (see Appendix~\ref{sec:sol_fgh}), we have $i_\mathrm{p}=2$, leading to
\begin{equation}
h(x) = \left\lbrace\begin{array}{rl}
1 & \mathrm{if}\, x < \widehat{x}_2\\
\widehat{x}_2^{\frac{3-p_1}{2}}
x^{-\frac{3-p_1}{2}}
& \mathrm{if}\, \widehat{x}_2 < x < \widehat{x}_1\\
\widehat{x}_2^{\frac{3-p_2}{2}}
\widehat{x}_1^{\frac{p_1}{2}-\frac{1}{6}}
x^{-4/3}
& \mathrm{if}\, x > \widehat{x}_1
\end{array}\right.\, .
\end{equation}

\paragraph{Self-consistent solution for the normalised electron distribution.}
Following \citet{2009ApJ...703..675N}, we define\footnote{As $Y(\gamma)$ is a decreasing function with a constant first segment, if $Y(\gamma) < 1$ for $\gamma < \max{\left(\gamma_\mathrm{c};\widehat{\gamma}_\mathrm{p}\right)}$, then $Y(\gamma) < 1$ for all values of $\gamma$, and $\gamma_0$ is therefore undefined. In these cases, we retrieve the "no SSC" solution.} $\gamma_0$ by $Y(\gamma_0)=1$ and assume $1+Y(\gamma) = Y(\gamma)$ for $\gamma<\gamma_0$ and $1$ otherwise. From Eqs.~(\ref{eq:barnFC}) and~(\ref{eq:barnSC}), this shows that the IC cooling only affects the electron distribution in the interval
\begin{equation}
    \max{\left(\gamma_\mathrm{c};\widehat{\gamma}_\mathrm{p}\right)}<\gamma<\gamma_0\, .
    \label{eq:impactSSC}
\end{equation}
Therefore, the solution is entirely determined by the orderings of $\gamma_\mathrm{m}$, $\gamma_\mathrm{c}$, $\gamma_0$, $\widehat{\gamma}_\mathrm{m}$, and $\widehat{\gamma_\mathrm{c}}$. All possible cases and the corresponding solutions are listed in Appendix~\ref{sec:sol_fgh}. In some cases, subcases are introduced as breaks can appear at $\widehat{\gamma}_0$, $\widehat{\widehat{\gamma}}_\mathrm{m}$, $\widehat{\widehat{\gamma}}_\mathrm{c}$, etc. All the most relevant cases for GRB afterglows were described in \citet{2009ApJ...703..675N}, along with the detailed method for obtaining the corresponding solution for the electron distribution. For completeness, we list in Appendix~\ref{sec:sol_fgh} the additional cases that allow us to fully describe the parameter space, even in regions that are unlikely to be explored in GRBs.

\paragraph{Calculation of the critical Lorentz factor $\gamma_\mathrm{c}$.}
From Eq.~(\ref{eq:defh}), we obtain the following equation for $Y(\gamma_\mathrm{c})$:
\begin{equation}
Y(\gamma_\mathrm{c}) = 
\left\lbrace\begin{array}{cl}
    A\, \frac{\gamma_\mathrm{m}}{\gamma_\mathrm{c}^\mathrm{syn}}\, \frac{I_2}{I_0}\, 
   h(x_\mathrm{c}) & \mathrm{if}\, \gamma_\mathrm{m}>\gamma_\mathrm{c}\\
       A\, \frac{\gamma_\mathrm{c}^\mathrm{syn}}{\gamma_\mathrm{m}}\, \frac{I_2}{I_0}\, 
   h(x_\mathrm{c}) & \mathrm{if}\, \gamma_\mathrm{m}<\gamma_\mathrm{c} \,\mathrm{and}\, Y(\gamma_\mathrm{c})<1\\
       \left(A\, \frac{\gamma_\mathrm{c}^\mathrm{syn}}{\gamma_\mathrm{m}}\, \frac{I_2}{I_0}\, 
   h(x_\mathrm{c})\right)^{1/3} & \mathrm{if}\, \gamma_\mathrm{m}<\gamma_\mathrm{c} \,\mathrm{and}\, Y(\gamma_\mathrm{c})>1\\
\end{array}\right. ,
\label{eq:Yc_all_cases}
\end{equation}
where $A=\frac{3}{4}\frac{K_\mathrm{IC}}{K_P}\, \frac{p-2}{p-1}\, \frac{\epsilon_\mathrm{e}}{\epsilon_\mathrm{B}}$ is a constant as the microphysics parameters $\epsilon_\mathrm{B}$, $\epsilon_\mathrm{e}$, and $p$ are assumed to be constant during the forward-shock propagation. We note that $x_\mathrm{c}=1$ in the slow-cooling regime $\gamma_\mathrm{m}<\gamma_\mathrm{c}$. A second useful relation is obtained from Eq.~(\ref{eq:defh}) and the definition of $\gamma_0$, 
\begin{equation}
    Y(\gamma_\mathrm{c})=\frac{Y(\gamma_\mathrm{c})}{Y(\gamma_0)}=\frac{h(x_\mathrm{c})}{h(x_0)}\, .
    \label{eq:Yc_xc_x0}
\end{equation}
Then, $\gamma_\mathrm{c}$ is computed from the following implicit equation, which is solved iteratively,
\begin{equation}
    \gamma_\mathrm{c} =
    \left\lbrace\begin{array}{rl}
    \gamma_\mathrm{c}^\mathrm{syn} & \mathrm{if}\, Y(\gamma_\mathrm{c}) < 1\\
    \gamma_\mathrm{c}^\mathrm{syn} / Y(\gamma_\mathrm{c}) & \mathrm{if}\, Y(\gamma_\mathrm{c}) > 1\\
    \end{array}\right.\, .
    \label{eq:gc_from_Yc}
\end{equation}
We start by assuming $Y(\gamma_\mathrm{c})=Y^\mathrm{no\, KN}$ as defined in Sect.~\ref{sec:SSCnoKN} and then iterate as follows:
\begin{enumerate}
    \item Compute $\gamma_\mathrm{c}$ from Eq.~(\ref{eq:gc_from_Yc}).
    \item Knowing $\gamma_\mathrm{m}$, $\gamma_\mathrm{c}$, and $\gamma_\mathrm{self}$, scan the different possible cases for the ordering of 
     $\gamma_\mathrm{m}$, $\gamma_\mathrm{c}$, $\gamma_0$, $\widehat{\gamma}_\mathrm{m}$, and $\widehat{\gamma_\mathrm{c}}$ (and additional characteristic Lorentz factors for some cases) as listed in Appendix~\ref{sec:sol_fgh} and compute $\gamma_0$ from Eq.~(\ref{eq:Yc_xc_x0}), which can be analytically inverted as provided for each case in Appendix~\ref{sec:sol_fgh}. We stop the scan of the possible cases when the ordering of all characteristic Lorentz factors including the obtained value of $\gamma_0$ is the correct one.
    \item Having identified the correct case and therefore knowing the expression of $f(x)$, compute the integrals $I_0$ and $I_2$.
    \item Compute an updated value of $Y(\gamma_\mathrm{c})$ from Eq.~(\ref{eq:Yc_all_cases}).
    \item Start a new iteration at step 1 until convergence. 
\end{enumerate}
We stop the procedure when the relative variation of $\gamma_\mathrm{c}$ in an iteration falls below $\epsilon_\mathrm{tol}$. We use  $\epsilon_\mathrm{tol}=10^{-4}$, which is reached in most cases in fewer than $20$ iterations. 
In the context of a full afterglow light-curve calculation, as we expect $Y$ to vary smoothly during the jet propagation, we use the value of $Y(\gamma_\mathrm{c})$ obtained at the previous step of the dynamics instead of $Y^\mathrm{no\, KN}$ to start the iterative procedure. Then, we usually reach convergence in only a few iterations.

\subsubsection{Final calculation of the emission in the comoving frame}

The self-consistent spectrum in the comoving frame, that is, $p_\nu^\mathrm{syn}$ and $p_\nu^\mathrm{SSC}$, the synchrotron and SSC powers per electron and per unit frequency averaged over the timescale $t_\mathrm{dyn}$, can be computed by the following procedure based on the previous paragraphs:
\begin{enumerate}
\item The input parameters are $n_{\mathrm{e}}^{\mathrm{acc}}$, $t_\mathrm{dyn}$, $B$, $u_\mathrm{e}/u_\mathrm{B}=\epsilon_\mathrm{e}/\epsilon_\mathrm{B}$, $\gamma_\mathrm{m}$, and $p$.
\item The following quantities can immediately be deduced: $\gamma_\mathrm{c}^\mathrm{syn}$ from Eq.~(\ref{eq:gamma_c_y_c}), $\gamma_\mathrm{self}$ from Eq.~(\ref{eq:gamma_self}), $\tau_\mathrm{T}$ from Eq.~(\ref{eq:tau_Thomson}), and $Y^\mathrm{no\, KN}$ from Eq.~(\ref{eq:YSSCnoKN}). 
\item Then, the iterative procedure described above allows us to identify the spectral regime among all the possibilities listed in Appendix~\ref{sec:sol_fgh}, and we can compute $\gamma_\mathrm{c}$. At the end of this step, all breaks and slopes in the functions $f(x)$ and $h(x)$ are known. The corresponding integrals $I_0$ and $I_2$ are computed by a numerical integration of Eqs.~(\ref{eq:I_0}) and~(\ref{eq:I_2}).
\item The function $g(y)$ can immediately be deduced from $f(x)$, using Eq.~(\ref{eq:defg}). The corresponding integral $J_0$ is computed by a numerical integration of Eq.~(\ref{eq:J_0}), and the synchrotron spectrum per electron, $p_\nu^\mathrm{syn}$, is deduced from Eq.~(\ref{eq:pnusynpere}).  
\item Finally, the function $G(y)$ is computed by an numerical integration of Eq.~(\ref{eq:G_y_SSC}) and the SSC spectrum per electron, $p_\nu^\mathrm{SSC}$, is deduced from Eq.~(\ref{eq:pnuSSCpere}).
\end{enumerate}

\subsubsection{Pair production}\label{sec:pair_production}

At VHE, pair production $\gamma\gamma \xrightarrow{} e^+e^-$ can start to contribute to the SSC flux depletion. We account for this mechanism by a simplified treatment, assuming an isotropic distribution of the low-energy seed photons, as for the SSC calculation, and approximating the cross-section for pair production by a Dirac function at twice the threshold of the interaction. Then, VHE photons of frequency $\nu$ can produce pairs by interacting with low-energy photons at a frequency $\nu_\mathrm{seed}=2\left( m_\mathrm{e} c^2\right)^2/\left(h^2 \nu\right)$, and the characteristic timescale of this interaction is given by
\begin{equation}
    t_{\gamma\gamma}(\nu) = \frac{h}{\sigma_{\rm T}c} \left[ u_{\nu_\mathrm{seed}=2\frac{\left( m_\mathrm{e} c^2\right)^2}{h^2 \nu}} \right]^{-1}\, ,
\end{equation}
with $u_{\nu} = u^\mathrm{syn}_\nu+u^\mathrm{SSC}_\nu = n_\mathrm{e}^\mathrm{acc} \left(p_\nu^\mathrm{syn}+p_\nu^\mathrm{SSC}\right) t_\mathrm{dyn}$, leading to
\begin{equation}
    t_{\gamma\gamma}(\nu) = \frac{h}{\tau_{\rm T}} \left[ \left(p_{\nu_\mathrm{seed}}^\mathrm{syn}+p_{\nu_\mathrm{seed}}^\mathrm{SSC}\right)_{\nu_\mathrm{seed}=2\frac{\left( m_\mathrm{e} c^2\right)^2}{h^2 \nu}} \right]^{-1}\, .
\end{equation}
For very high frequencies $\nu$, the seed photons for pair production have a low frequency and $p_{\nu_\mathrm{seed}}^\mathrm{syn}+p_{\nu_\mathrm{seed}}^\mathrm{SSC}\simeq p_{\nu_\mathrm{seed}}^\mathrm{syn}$. The total emitted power per electron $p_\nu = p_\nu^\mathrm{syn}+p_\nu^\mathrm{SSC}$ is then corrected for by a factor $\frac{t_{\gamma\gamma}(\nu)}{t_\mathrm{dyn}}\left(1-e^{-t_\mathrm{dyn}/t_{\gamma\gamma}(\nu)}\right) $, which in practice only attenuates the SSC component at high frequency, where $t_{\gamma\gamma}(\nu)\ll t_\mathrm{dyn}$. 

The implementation of the model allows us to select the level of approximation for the SSC emission (synchrotron only, Thomson, or full calculation) and to include or exclude the attenuation due to the pair production.

\subsection{Observed flux}
\label{sec:eatsurfaces}

\begin{figure}
    \centering
    \resizebox{\hsize}{!}{\includegraphics{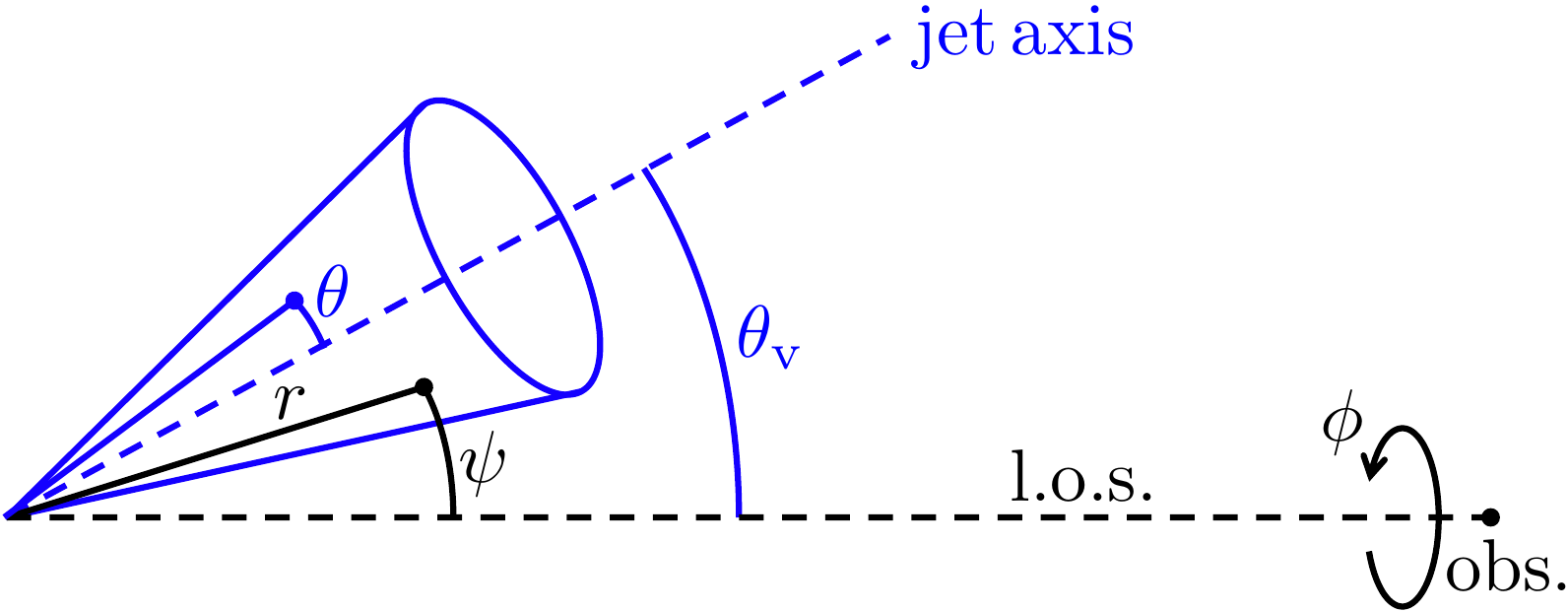}}
    \caption{Coordinates and notations used in the expression of the observed flux.}
    \label{fig:integration_angle_schema}
\end{figure}

When the emissivity in the comoving frame is known (Sect.~\ref{sec:emicom}), the flux measured by a distant observer with a viewing angle $\theta_\mathrm{v}$ is computed by an integration over equal-arrival time surfaces, taking into account relativistic Doppler boosting and relativistic beaming, as well as the effect of cosmological redshift. This leads to the following expression of the observer frame flux density ($\mathrm{erg}\cdot\mathrm{s}^{-1}\cdot\mathrm{cm}^{-2}\cdot\mathrm{Hz}^{-1}$) measured at time $t_{\rm obs}^z$ and frequency $\nu_{\rm obs}^z$\citep{1999ApJ...523..187W}:
\begin{eqnarray}\label{eq:f_nuobs_1}
F_{\nu_{\rm obs}^z}\left(t_{\rm obs}^z\right) & = & \frac{1 + z}{4\pi D_{\rm L}^2} \int_0^{\infty}\mathrm{d}{r}\, \int_0^{\pi}\mathrm{d}\psi\,\int_0^{2\pi}\mathrm{d}{\phi}\, r^2 \sin{\psi}\\\nonumber
& & \!\!\!\!\!\!\times
\left[\mathcal{D}^2(r,\psi,t)\, 4\pi j'_{\nu'}\left(r,\psi, \phi,t\right)\right]_{t=\frac{t_{\rm obs}^z}{1+z}+\frac{r}{c}\cos{\psi}}\, ,
\end{eqnarray}
where we use spherical coordinates $\left(r;\psi;\phi\right)$ with the polar axis equal to the line of sight, $\psi$ and $\phi$ the colatitude and longitude, and the jet axis in the direction $\left(\psi=\theta_\mathrm{v},\phi=0\right)$ (see Fig.~\ref{fig:integration_angle_schema}), and where $z$ and $D_{\rm L}$ are the redshift and the luminosity distance of the source. The angle $\psi$ differs from $\theta$ used in Sect.~\ref{sec:dynamics}, which is measured from the jet axis. The time $t$  (source frame) and frequency $\nu'$ (comoving frame) are given by
\begin{eqnarray}
 t & = & \frac{t_{\rm obs}^z}{1+z} +\frac{r}{c}\cos{\psi}\\
 \nu' & = & (1+z)\frac{\nu_{\rm obs}^z}{\mathcal{D}\left(r,\psi,t\right)}\, .
\end{eqnarray}
The Doppler factor is expressed as
\begin{equation}
    \mathcal{D}(r,\psi,t) = \frac{1}{\Gamma(r,\psi,t)\left(1-\beta(r,\psi,t)\cos{\psi}\right)},
\end{equation}
and  $j'_{\nu'}$ is the emissivity ($\mathrm{erg \cdot cm^{-3} \cdot s^{-1} \cdot Hz^{-1} \cdot sr^{-1}}$) in the comoving frame. 

In practice, following the discretisation of the jet structure described in Sect.~\ref{sec:geometry_structure}, the observer-frame flux density $F_{\nu_{\rm obs}^z}\left(t_{\rm obs}^z\right) $ is computed as $F_{\nu_{\rm obs}^z}\left(t_{\rm obs}^z\right) =\sum_{i=0}^N F_{\nu_{\rm obs}^z}^{(i)}\left(t_{\rm obs}^z\right)$, where $F_{\nu_{\rm obs}^z}^{(i)}\left(t_{\rm obs}^z\right)$ are the contributions of the core jet ($i=0$) and lateral rings ($i=1\to N$). To compute each contribution, the thin-shell approximation allows us to reduce the double integral over $r$ and $\psi$ in Eq.~(\ref{eq:f_nuobs_1})  to a simple integral over $r$, or equivalently, over $\psi$ or $t$. In addition, to improve the computation time, the remaining integral on $\phi$ is computed analytically. This leads to 
\begin{eqnarray}
F_{\nu_{\rm obs}^z}^{(i)}\left(t_{\rm obs}^z\right) \!\!\! & = & \!\!\!\frac{1+z}{4\pi D_{\rm L}^2} \int\limits_{t_{\mathrm{min}}^{(i)}(t_{\rm obs}^z)}^{t_{\mathrm{max}}^{(i)}(t_{\rm obs}^z)} \!\!\!\! \frac{c\, \mathrm{d}t}{2\Gamma_i(t)R_i(t)}\, \frac{\Delta\phi_i\left(\theta_\mathrm{v};\psi_i(t^z_\mathrm{obs};t)\right)}{2\pi}\nonumber\\
& & \times
\mathcal{D}_i^2\left(t^z_\mathrm{obs};t\right)\,
\left[ 4\pi N_\mathrm{e}\left(R_i(t)\right)\right]\, p'^{(i)}_{\nu'}(t)\, , 
\end{eqnarray}
where 
\begin{eqnarray}
\cos{\left(\psi_i(t^z_\mathrm{obs};t)\right)} & = & c\frac{t-\frac{t^z_\mathrm{obs}}{1+z}}{R_i(t)}\, ,\\
\mathcal{D}_i(t^z_\mathrm{obs};t) & = & \frac{1}{\Gamma_i(t)\left(1-\beta_i(t)\cos{\psi_i(t^z_\mathrm{obs};t)}\right)}\, ,
\end{eqnarray} 
$N_\mathrm{e}(r)=\zeta M_\mathrm{ext}(r)/m_\mathrm{p}$ is the number of shock-accelerated electrons per unit solid angle, and $p'^{(i)}_{\nu'}(t)$ is the power per unit frequency and per electron in the comoving frame computed in Sect.~\ref{sec:emicom}  and evaluated  at time $t$ and comoving frequency $\nu'=(1+z)\nu^z_\mathrm{obs}/\mathcal{D}_i\left(t^z_\mathrm{obs};t\right)$. 

The limits of the integral $t_{\mathrm{min}}^{(i)}(t_{\rm obs}^z)$ and $t_{\mathrm{max}}^{(i)}(t_{\rm obs}^z)$ are defined by the condition $\psi_{\mathrm{min}, i} \le \psi_i\left(t^z_\mathrm{obs};t\right) \le \psi_{\mathrm{max}, i}$, where
\begin{equation}
    \psi_{\mathrm{min}, i} = \left\lbrace\begin{array}{rll}
\theta_{\mathrm{min}, i} - \theta_\mathrm{v} & \mathrm{if} & \theta_\mathrm{v} \le \theta_{\mathrm{min}, i}\\
0 & \mathrm{if} & \theta_{\mathrm{min}, i} \le \theta_\mathrm{v} \le \theta_{\mathrm{max}, i}\\
\theta_\mathrm{v} - \theta_{\mathrm{max}, i} & \mathrm{if} & \theta_{\mathrm{max}, i} \le \theta_\mathrm{v} \le \pi/2
\end{array}\right.\, ,
\end{equation}
and $\psi_{\mathrm{max}, i} = \theta_\mathrm{v} + \theta_{\mathrm{max}, i}$, that is,
\begin{equation}
t - \frac{R_i(t)}{c}\cos{\left(\psi_{\mathrm{max},i}\right)} \le \frac{t_{\rm obs}^z}{1+z} \le 
t - \frac{R_i(t)}{c}\cos{\left(\psi_{\mathrm{min},i}\right)}
\, .
\end{equation}
Finally, we use the exact analytical calculation of the geometrical term $\Delta \phi_i$ defined by $\Delta \phi_i(\theta_\mathrm{v};\psi)=\int_0^{2\pi}\mathrm{d}\phi f_i(\theta_\mathrm{v};\psi;\phi)$, where $f_i(\theta_\mathrm{v};\psi;\phi)=1$ when the direction $\left(\psi;\phi\right)$ is contained in the component $i$ (core jet or ring) and $0$ otherwise.

\section{Comparison with other afterglow models} \label{sec:comparison_other_models} 

\subsection{Dynamics and synchrotron emission}
To test our calculation of the dynamics of the forward external shock and the associated synchrotron emission, we compared our results with two other models of the afterglow of a structured jet: the model presented in \citet{2018MNRAS.478.4128G}, and \texttt{afterglowpy}, a public Python module for calculating GRB afterglow light curves and spectra, based on \citet{2020ApJ...896..166R}.

The first comparison is straightforward: With our default values of the normalisation coefficients $K_{\nu}$ and $K_{\mathrm{P}}$, the assumptions (dynamics, microphysics, radiation, calculation of observed quantities) of our model in the purely synchrotron case are exactly the same as in \citet{2018MNRAS.478.4128G}. We verified that we exactly reproduced the different cases in their Fig.~4.

\begin{figure}[t]
    \centering
    \resizebox{\hsize}{!}{\includegraphics{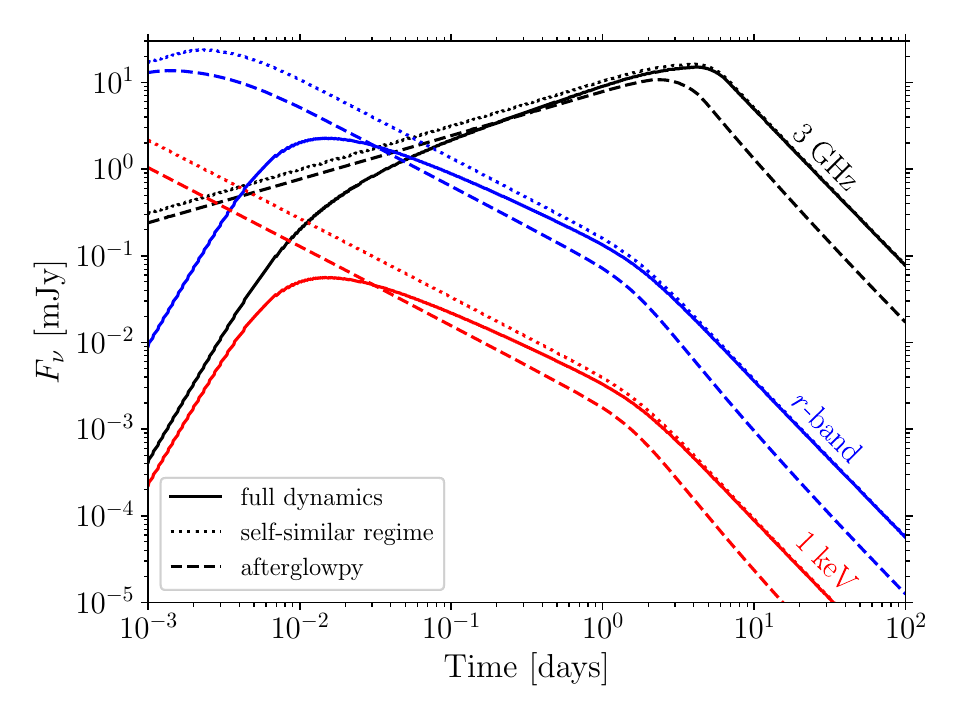}}
    \caption{Synthetic afterglow light curves of a top-hat jet viewed on-axis in the synchrotron-only radiation regime. The solid lines show the results obtained using our model with full dynamics treatment (using Eq.~(\ref{eq:gamma_shock})). The dotted lines follow the self-similar solution at all times. The dashed lines are obtained with \texttt{afterglowpy} without lateral expansion. The fluxes are computed in radio at $3~\mathrm{GHz}$ (black), in optical in \textit{r} band at $5.06\times10^{14}~\mathrm{Hz}$ (blue), and in X-rays at $1~\mathrm{keV}$ (red). The parameters used for this figure are $\mathrm{E_{\mathrm{0, iso}}} = 10^{52}~\mathrm{erg}$, $\theta_{\mathrm{c}} = 4~\mathrm{deg}$, $n_{\mathrm{ext}} = 10^{-3}~\mathrm{cm}^{-3}$, $\epsilon_{\mathrm{e}} = 10^{-1}$, $\epsilon_{\mathrm{B}} = 10^{-1}$, $p = 2.2$, and $D_{\mathrm{L}} = 100~\mathrm{Mpc}$.}
    \label{fig:boulodrome_vs_afterglowpy}
\end{figure}

The case of \texttt{afterglowpy} requires a more detailed comparison as there are differences in some treatments of the afterglow physics.\\
\noindent (1) Early dynamics: In \texttt{afterglowpy}, even the early-time dynamics is computed assuming the  self-similar  Blandford \& McKee regime, that is, $\Gamma(t;\theta) \propto E(\theta)^{1/2} n_0^{1/2} t^{-3/2}$ (see Sect.~2.1 in \citealt{2020ApJ...896..166R}), whereas we included the coasting phase with $\Gamma\sim\mathrm{cst}$ before a smooth transition towards the self-similar regime at the deceleration radius, as described in Sect.~\ref{sec:dynamics}. Fig.~\ref{fig:boulodrome_vs_afterglowpy} compares for a typical set of parameters the afterglow light curves from a top-hat jet viewed on-axis in radio, optical, and X-rays obtained with  our model (solid line), \texttt{afterglowpy} (dashed line), and our model where the dynamics was forced to be in the self-similar regime at all times (dotted line). This allows us to confirm that the two models with the same self-similar dynamics agree well, except for the flux normalisation, as discussed below, and that the model with the full dynamics progressively converges towards the same solution: Following the peak at the deceleration radius (at $\sim 10^{-2}\, \mathrm{days}$ in this example), the light curve obtained with the full dynamics smoothly converges towards the self-similar solution, and the two light curves are identical at late times (typically after the jet break at $\sim 2~\mathrm{days}$ in this example). On the other hand, the self-similar approximation strongly over-estimates the fluxes until the deceleration radius, and the full dynamics should always be included when considering early observations (see also \citealt{2024arXiv240219359W}). In the case of GW~170817 discussed in the next section, the earliest detection by \textit{Chandra} was obtained $\sim 9$~days after the merger, by which time, the core jet dynamics had reached the self-similar regime.\\
\noindent (2) Late dynamics: In contrast to \texttt{afterglowpy}, the current version of our model does not include lateral expansion of the ejecta. When we compared it with \texttt{afterglowpy}, we therefore deactivated this option in the latter, which led to the excellent late-time agreement shown in Fig.~\ref{fig:boulodrome_vs_afterglowpy}. The impact of the lateral spreading is expected to be very limited as long as the core jet is relativistic \citep[see e.g.][]{1999ApJ...523..187W,2012MNRAS.421..570G,2012ApJ...747L..30V,2018ApJ...865...94D}, so that we stopped the jet propagation when the core Lorentz factor reached $\Gamma = 2$ in the simulations of the afterglow of GW~170817 discussed in the next section. In addition, we excluded observations after 400 days from the data set used for the afterglow fitting, which corresponds to $\Gamma \gtrsim 3-4$ for the core jet in our best-fit models.\\
\noindent (3) Flux normalisation: \texttt{afterglowpy} relies on a scaling to \texttt{boxfit} \citep{2012ApJ...749...44V}, while in our model, this normalisation is derived from analytical approximations detailed in Sect.~\ref{sec:vhe_model}. This leads to the difference in the normalisation shown in Fig.~\ref{fig:boulodrome_vs_afterglowpy} when \texttt{afterglowpy} is compared with our model, where the same self-similar dynamics is forced. The flux ratio varies between $1$ and $5$ at all wavelengths. When exploring a broad parameter space, varying the angle, energy injection, and microphysical parameters, we verified that the flux ratio never exceeded $5$.

To conclude this comparison, we note that \texttt{afterglowpy} and our model (in purely synchrotron mode) have been used by \citet{2023ApJ...948L..12K} for two independent Bayesian inferences of the parameters of the afterglow of GRB~221009A using the same set of early observational data. The two codes converged towards very similar solutions.

\begin{figure}[t]

    \centering
    \resizebox{\hsize}{!}{\includegraphics{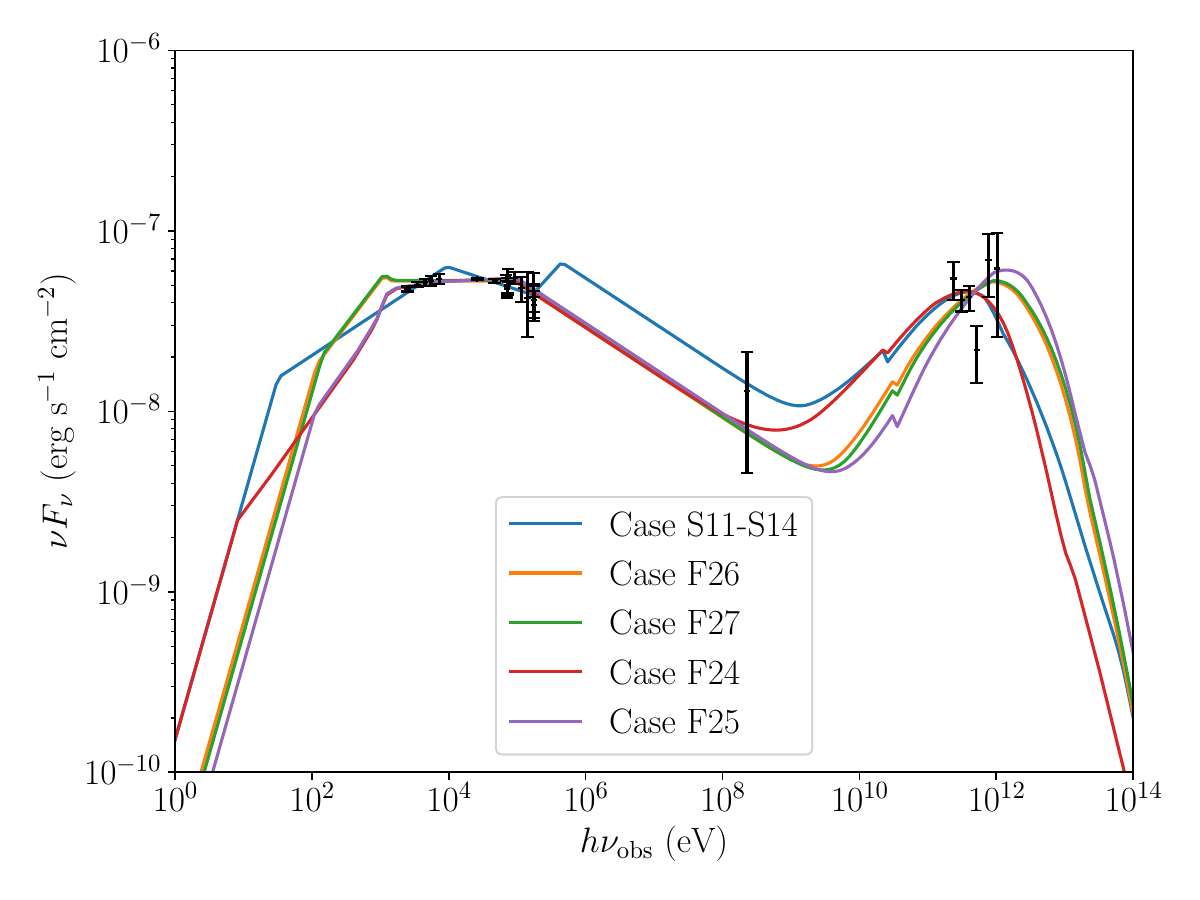}}
    \caption{Spectral fits to the afterglow observations of GRB~190114C at $90~\mathrm{s}$ (black data points taken from  \citealt{2019Natur.575..459M}). Each colour represents the best-fit spectrum for each of the possible radiative regimes (see text and Fig.~\ref{fig:yamasaki_posteriors}). The spectral cases are detailed in Appendix~\ref{sec:sol_fgh}. This figure can be compared to Fig.~6 in \citet{2022MNRAS.512.2142Y}, which shows very similar fits.}
    \label{fig:yamasaki_best_fit_models}
\end{figure}

\begin{figure*}[t]
    \centering
    \includegraphics[width=\textwidth]{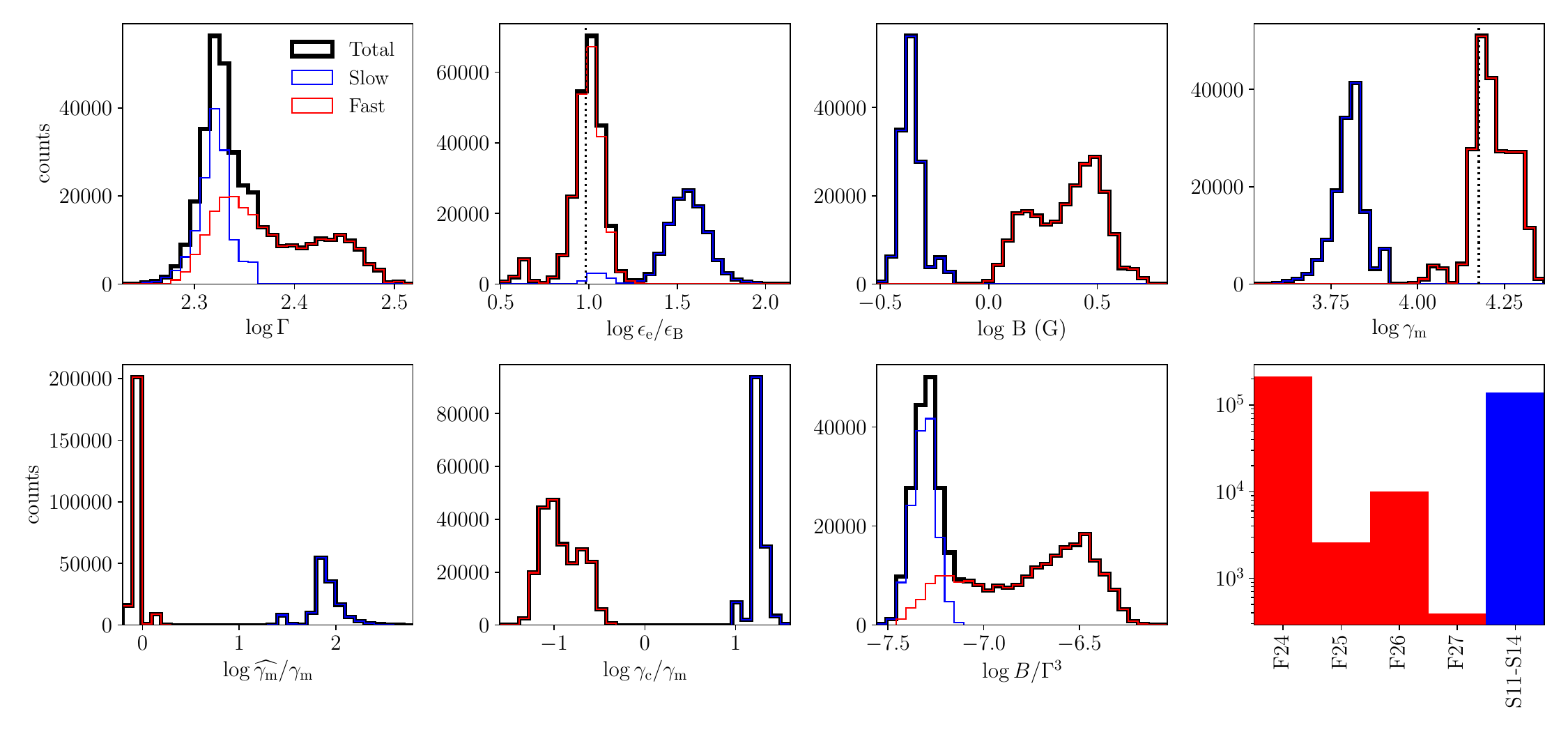}
    \caption{Marginalised posterior distributions for the spectral fit of the afterglow observations of GRB~190114C at $t_\mathrm{obs}= 90~\mathrm{s}$. The total (black) is subdivided between the slow (blue) and fast (red) cooling cases. The first row shows the four free parameters explored in this MCMC analysis ($\Gamma$, $\epsilon_\mathrm{e} / \epsilon_\mathrm{B}$, $B$ and $\gamma_\mathrm{m}$). In the second row, we show some derived quantities: $\widehat{\gamma}_\mathrm{m} / \gamma_\mathrm{m}$, $\gamma_\mathrm{c} / \gamma_\mathrm{m}$ and $B / \Gamma^3$. The bottom right panel shows the distribution of the radiative regimes found in the posterior sample (see Appendix~\ref{sec:sol_fgh} for their description). The best-fit model for each radiative regime is shown in Fig.~\ref{fig:yamasaki_best_fit_models}.}
    \label{fig:yamasaki_posteriors}
\end{figure*}

\subsection{Synchrotron self-Compton emission}

We now test the predictions of our model in the SSC regime. Several GRBs have been observed at VHE. Two events were particularly well followed up at all wavelengths, GRB~190114C \citep{2019Natur.575..455M, 2019Natur.575..459M} and GRB~221009A \citep{LHAASO}. In both cases, the VHE detection occurred at very early times, but the analysis favoured a dominant external forward-shock origin for the emission. However, the case of GRB~221009A, which is an extreme GRB (the brightest ever detected by far) is quite complex. 
According to several studies, the early VHE emission and the later observations at lower frequencies probe different regions of a structured jet seen on-axis and may also include a significant contribution of the reverse shock at some frequencies \citep[see e.g.][]{2023ApJ...946L..23L, 2023SciA....9I1405O, 2023MNRAS.524L..78G, 2023MNRAS.522L..56S, 2024ApJ...966..141Z, 2024JHEAp..41...42Z,2024MNRAS.530..347D}.
We therefore decided to rather focus on GRB~190114C to test our model. This long GRB has rich observational follow-up from radio to VHE at simultaneous times.
Following \citet{2019Natur.575..455M}, we assumed that the spectrum shows two distinct components.

To test the radiative component of our model independently of the dynamics, we decided to analyse the observed spectrum at one fixed time, $t_\mathrm{obs} = 90~\mathrm{s}$, following the approach proposed by \citet{2022MNRAS.512.2142Y}. Their spectral model was also based on \citet{2009ApJ...703..675N}, which allows an easy comparison. For this purpose, we used their values $K_\mathrm{IC} = 1$ and $w_\mathrm{KN} = 0.2$ in this subsection. We also considered a one-zone model of the emitting region, characterised by a radius $R$, a Lorentz factor $\Gamma$, and $N_\mathrm{e}$ accelerated electrons radiating in a magnetic field $B'$. The electrons were injected with an initial distribution with a minimum Lorentz factor $\gamma_\mathrm{m}$ and a slope $p$. Following \citet{2022MNRAS.512.2142Y}, we fixed $p=2.5$ and let four parameters free for the spectral fitting: $\Gamma$, $B'$, $\gamma_\mathrm{m}$, and $\epsilon_\mathrm{e} / \epsilon_\mathrm{B}$. Then, $R$ and $t'_\mathrm{dyn}$ were deduced from $\Gamma$ and $t_\mathrm{obs}$, $n_\mathrm{e}^\mathrm{acc}$ from $B'$, $\epsilon_\mathrm{e} / \epsilon_\mathrm{B}$, $\gamma_\mathrm{m}$, and $p$, and finally, $N_\mathrm{e}$ from $n_\mathrm{e}^\mathrm{acc}$, $R$, and $\Gamma$ \citep[see Sect.~2 in][]{2022MNRAS.512.2142Y}.

We used the same observational data as \citet{2022MNRAS.512.2142Y}, shown in black in Fig.~\ref{fig:yamasaki_best_fit_models}. We then fit the data using an MCMC procedure with log-uniform priors, such that $1.3 < \log \Gamma < 3.5$; $0 < \log \epsilon_\mathrm{e} / \epsilon_\mathrm{B} < 4$; $-2 < \log B < 3$; $1 < \log \gamma_\mathrm{m} < 7$.

The marginalised posterior distributions for these four parameters are provided on the first row of Fig.~\ref{fig:yamasaki_posteriors}. As also discussed in \citet{2022MNRAS.512.2142Y}, several spectral solutions are possible. We show for the further discussion the posterior distributions of $\widehat{\gamma}_\mathrm{m} / \gamma_\mathrm{m}$, $\gamma_\mathrm{c} / \gamma_\mathrm{m}$, and $B / \Gamma^3$, as well as the distribution of spectral cases in the bottom row of Fig.~\ref{fig:yamasaki_posteriors}. The best-fit model for each of the spectral cases is shown in Fig.~\ref{fig:yamasaki_best_fit_models}. We find a high overall agreement with the results of \citet{2022MNRAS.512.2142Y}.
\begin{itemize}
\item The identified radiative regimes in the best-fit models are the same: Cases F24 and F25 correspond to their cases IIc and III (top row in their Fig.~6), F26 and F27 to their case I (middle row in their Fig.~6), and S11-S14 to one subcase of their slow case I (bottom right panel in their Fig.~6).
\item Most best models are in fast cooling (see the distribution of $\gamma_\mathrm{c}/\gamma_\mathrm{m}$ in Fig.~\ref{fig:yamasaki_posteriors}), with electrons at $\gamma_\mathrm{m}$ close to the transition between the Thomson and the KN regimes (see the distribution of $\widehat{\gamma}_\mathrm{m}/\gamma_\mathrm{m}$ in Fig.~\ref{fig:yamasaki_posteriors}).   
\item The peak of the posterior distributions of $\epsilon_\mathrm{e}/\epsilon_\mathrm{B}$ and $\gamma_\mathrm{m}$ in Fig.~\ref{fig:yamasaki_posteriors} is very close to the assumed values in their Fig.~6 (as indicated by the vertical dotted black lines).
\item Based on the analytical estimates from \citet{2021ApJ...923..135D}, \citet{2022MNRAS.512.2142Y} assumed a strong correlation between the Lorentz factor and the magnetic field with $B/\Gamma^3\sim 10^{-7}$. Our MCMC exploration confirms this result, but gives an estimate of the dispersion (see the distribution of $B/\Gamma^3$ in Fig.~\ref{fig:yamasaki_posteriors}).
\end{itemize}
A detailed comparison shows that our results differ slightly for the exact parameter values, owing to some differing normalisation constants. It should therefore be kept in mind that the absolute parameter values from spectral fits from different GRB afterglow studies should only be compared when the models that were used are the same. 

The good agreement of our results with \citet{2022MNRAS.512.2142Y} allows us to validate our radiative model, as both studies are independent implementations of the synchrotron+SSC model using the same level of approximation. In addition, we confirm that this model is a convincing candidate to reproduce the broadband emission of GRB~190114C.

\section{Very high energy afterglow of GW 170817}
\label{sec:170817}

Having validated the calculation of the dynamics and synchrotron + SSC emission in our model, we now present the results obtained by fitting the observations of the GRB afterglow of GW~170817 using a Bayesian analysis.

\subsection{Observational data}\label{sec:170817_data}

In its numerical implementation, our model allows for simultaneous multi-wavelength modelling. We used $94$ data points, from radio to X-rays between $9.2$ and $380$ days\footnote{Data points are provided by K. Mooley on \href{https://github.com/kmooley/GW170817/}{github.com/kmooley/GW170817/}}, as compiled and reprocessed homogeneously by \citet{2021ApJ...922..154M}. We also included the five constraining early-time upper limits also reported in \citet{2021ApJ...922..154M}, as shown in Table~\ref{table:170817_early_ul}. Early-time constraints are useful to restrict the posterior models to those with a single peak in flux density, as discussed in Sect.~\ref{sec:dynamics}. All these flux measurements and upper limits were fitted simultaneously.

The likelihood that we used for our study is based on a modified $\chi^2$ calculation to account for the upper limits,
\begin{equation}
   \ln{\mathcal{L}} = - \frac{1}{2}\, 
    \left[ 
    \sum_{i} \frac{\left(m_i - d_i\right)^2}{\sigma_i^2} 
    \!+\! \sum_{j}
    \frac{\left(\max{\{ u_j; m_j \}} - u_j \right)^2}{\sigma_j^2} 
    \right]\, ,
\end{equation}
where the first sum concerns observational data (subscripts~$i$), while the second sum concerns upper limits (subscripts~$j$). The quantities $m_i$ and $m_j$ are the model flux densities at each observing time and frequency of the observational data set, $d_i$ is the flux density value of a detection, $\sigma_i$ is its associated uncertainty, $u_j$ is the upper flux limit presented in Table~\ref{table:170817_early_ul}, and $\sigma_j$ is an equivalent uncertainty that we arbitrarily defined as $\sigma_j = 0.2 u_j$, which roughly corresponds to the typical uncertainties for the detections. Adding this element accounts for some models whose predictions are slightly above the reported upper limit. The quantity $\max\{ u_j; m_j \}$ in the second term implies that it behaves as a penalty that is only added for predicted fluxes $m_j$ above the observed upper limits. Another option would be to force the likelihood to diverge as soon as one of the modelled fluxes is above the corresponding upper limit.

We chose to exclude late-time observations after $400~\mathrm{days}$ in the fits. We expect the lateral spreading of the jet to play a role at late times, when the shock front reaches Lorentz factors $\Gamma \lesssim 3$, by which point our model no longer accurately describes the jet dynamics. In addition, other emission sites (e.g. the kilonova afterglow; \citealt{2011Natur.478...82N}) could also contribute to the observed flux at late times. It is also difficult to include the latest observations \citep{2022ApJ...927L..17H}, which were not included in the compilation by \citet{2021ApJ...922..154M}: Late-time data in X-rays consist of only a few photons, and a correct modelling in this case should be expressed in photon counts and should account for the instrumental response and low-statistics effects. The result of such an analysis is currently unclear, with an uncertain late rebrightening \citep{2022ApJ...927L..17H} (see the discussion in  \citealt{2022MNRAS.510.1902T}). An interpretation of the late evolution of the afterglow of GW~170817 appears to require a dedicated study, and we therefore excluded this phase from the analysis presented here. Finally, we did not include the H.E.S.S. upper limit at the peak \citep{2020ApJ...894L..16A}, or the \textit{Fermi} large area telescope (LAT) and H.E.S.S. early upper limits \citep{2018ApJ...861...85A, 2017ApJ...850L..22A}, as they do not constrain the fit (see Sect.~\ref{sec:discussion_170817_vhe}).

\begin{table}[t]
    \caption{Early-time upper limits used in the afterglow fitting. }
\centering
\begin{tabular}{c c c c}
\hline\hline
Time & Frequency & Instrument & Upper limit \\
\hline 
    $0.57$ & $9.7\times 10^{9}$ & VLA & $144$ \\
    $0.70$ & $1.2\times 10^{18}$ & NuSTAR & $7.3\times 10^{-4}$\\
    $1.44$ & $1.0\times 10^{10}$ & VLA & $13.8$\\
    $2.40$ & $2.41\times 10^{17}$ & Chandra & $2.3\times 10^{-4}$\\
    $3.35$ & $3\times 10^{9}$ & VLA & $19$\\
\hline
\end{tabular}
\tablefoot{The times are given in days, the frequencies in Hz, and the upper limits in $\mu\mathrm{Jy}$. These upper flux limits are selected as the most constraining of all the early observations listed in the compilation by \citet{2021ApJ...922..154M}.}
\label{table:170817_early_ul}
\end{table}

\begin{figure*}[t]
    \centering
    \includegraphics[width=\textwidth]{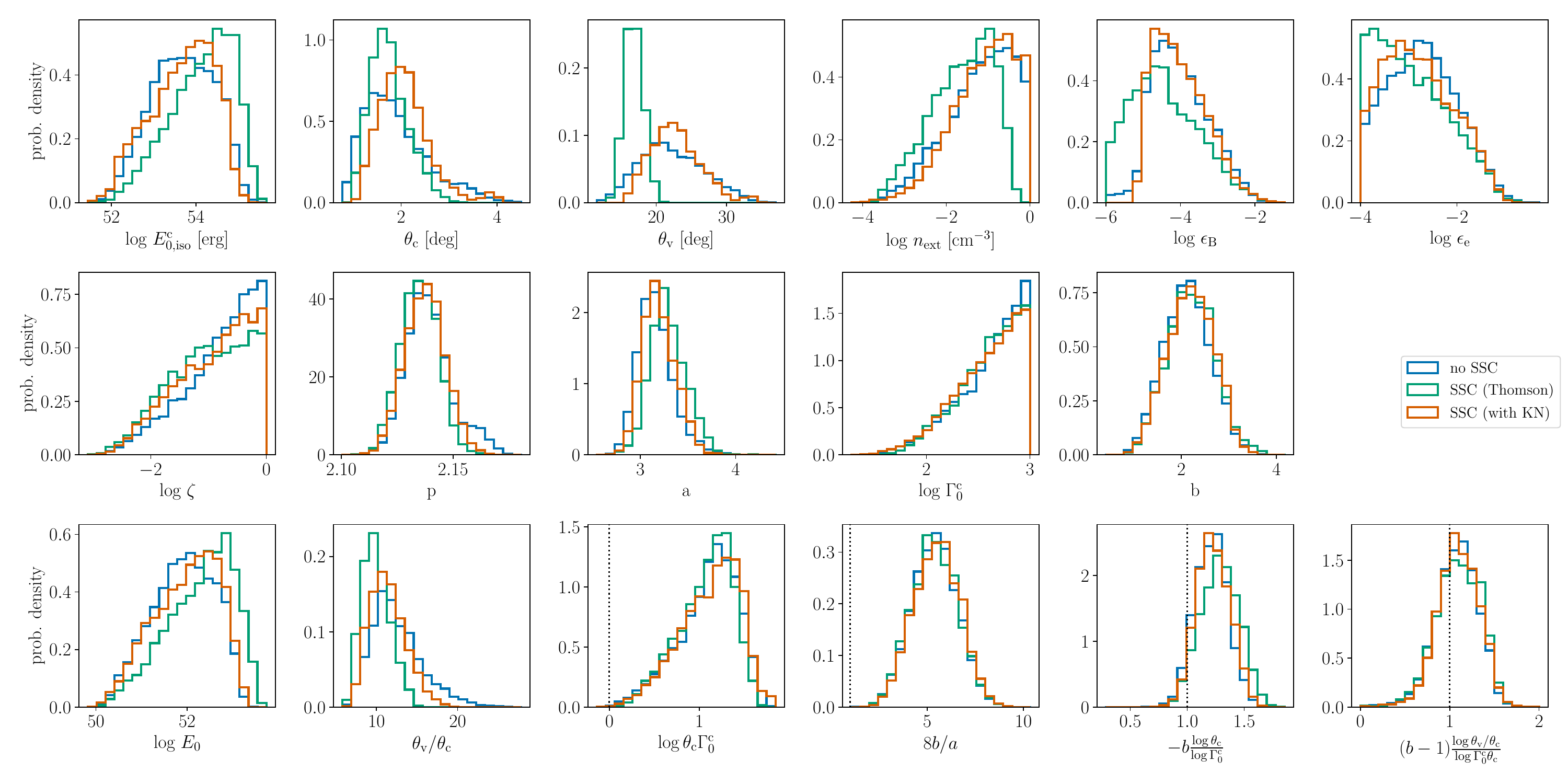}
    \caption{Comparison of the marginalised posterior distributions for the three fits of the afterglow of GW~170817 presented in Sect.~\ref{sec:170817}, which differ only by the treatment of the radiative processes: "no SSC" in blue; "SSC (Thomson)" in green; "SSC (with KN)" in orange. The first two rows show the inferred distributions of the free model parameters. The last row shows the distributions of several quantities derived from these parameters: The true energy of the jet $E_0$ (see Eq.~(\ref{eq:energy_0})), the ratio $\theta_{\mathrm{v}} / \theta_{\mathrm{c}}$, and the four quantities used in the conditions for single-peaked light curves from \citet{2020MNRAS.493.3521B} and described in Sect.~\ref{sec:dynamics}, $\log{\left(\theta_{\mathrm{c}} \Gamma_0^\mathrm{c}\right)}$, $8b/a$, $-b \frac{\log \theta_\mathrm{c}}{\log \Gamma_0^\mathrm{c}}$, and $(b-1)\frac{\log{\left(\theta_\mathrm{v} / \theta_\mathrm{c}\right)}}{\log{\left(\Gamma_0^\mathrm{c}\theta_\mathrm{c}\right)}}$.}
    \label{fig:posterior_model_comparison}
\end{figure*}

\subsection{Results from the afterglow fitting}\label{sec:170817_fits}

\begin{table}[b]
\caption{Free parameters of the MCMC and their prior bounds and shape.}
\centering
\begin{tabular}{c c c}
\hline\hline
Parameter & Bounds & Type \\
\hline 
    $E_{\rm 0, iso}^\mathrm{c}$ (erg) & $10^{50}-10^{56}$ & log-uniform \\
    $\theta_{\rm c}$ (deg) & $0.5 - 10$ & uniform \\
    $\theta_{\rm v}$ (deg) & $0 - 50$ & uniform \\
    $n_{\rm ext}$ (cm$^{-3}$) & $10^{-6} - 10^{0}$ & log-uniform \\
    $\epsilon_{\rm B}$ & $10^{-6} - 1$ & log-uniform \\
    $\epsilon_{\rm e}$ & $10^{-4} - 1$ & log-uniform \\
    $\zeta$ & $10^{-4} - 1$ & log-uniform \\
    $p$ & $2.1 - 2.4$ & uniform \\
    $a$ & $0.1 - 7$ & uniform \\
    $\Gamma_0^\mathrm{c}$ & $10^1 - 10^3$ & log-uniform \\
    $b$ & $0 - 6$ & uniform \\
\hline
\end{tabular}
\label{table:mcmc_priors}
\end{table}

We performed a Bayesian analysis of the afterglow data of GW~170817 using the Markov chain Monte Carlo (MCMC) algorithm of the Python suite \texttt{emcee} \citep{2013PASP..125..306F, 2019JOSS....4.1864F}. We ran three separate fits that differed only in the assumptions for the radiative processes in the shocked region.
\begin{enumerate}
    \item "no SSC": Emission is only produced by synchrotron radiation, as described in \citet{1998ApJ...497L..17S} and Sect.~\ref{sec:pureSYN}.
    \item "SSC (Thomson)": SSC is taken into account assuming that all scatterings occur in the Thomson regime, as described in \citet{2001ApJ...548..787S} and Sect.~\ref{sec:SSCnoKN}.
    \item "SSC (with KN)": SSC is taken into account including the KN regime, as described in \citet{2009ApJ...703..675N} and Sect.~\ref{sec:fgh}.
\end{enumerate}

The priors used for these three fits are identical and presented in Table~\ref{table:mcmc_priors}. We used (log-)uniform priors over large intervals in order to remain as agnostic as possible. Their bounds are shown in Table~\ref{table:mcmc_priors}. Additional constraints on the viewing angle can be obtained by combining the GW signal and the accurate distance and localisation of the host galaxy \citep{2018ApJ...860L...2F}, or by combining the afterglow photometry and VLBI imagery \citep{2023MNRAS.524..403G}, and the density of the external medium can be constrained from direct observation of the host galaxy \citep{2017Sci...358.1579H,2019ApJ...886L..17H}. We did not include them to focus on the constraints obtained from the afterglow modelling alone.

For each of the three fits, we initialised $50$ chains and ran $10\,000$ iterations per chain. After studying the convergence speed, we removed the first $2\,000$ iterations for each chain. We show the posterior distributions for the remaining $400\,000$ parameter samples.

We present the marginalised posterior distribution of the free parameters for the three fits in Fig.~\ref{fig:posterior_model_comparison}, complemented by Fig.~\ref{fig:MCMC_3_models}, which shows the corresponding joint and marginalised posterior distributions at $3\sigma$ credibility (corner plot). We also show in Fig.~\ref{fig:posterior_model_comparison} the distribution of some derived quantities, the true initial kinetic energy of the ejecta $E_0$ as deduced from Eq.~(\ref{eq:energy_0}), the ratio of the viewing angle over the opening angle of the core jet $\theta_\mathrm{v} / \theta_\mathrm{c}$, and the four conditions for single-peaked light curves described by \citet{2020MNRAS.493.3521B} and in Sect.~\ref{sec:dynamics}. The three fits lead to similar distributions across the parameter space, with a few exceptions that are discussed in Sect.~\ref{sec:discussion_170817}. The median parameter values as well as their 90\% credible intervals are reported in Table~\ref{table:best_fit_values}. We finally show in Table~\ref{table:derived_best_fit_values} the median values of the derived quantities shown in Fig.~\ref{fig:posterior_model_comparison}.

\begin{figure*}[t]
    \centering
    \includegraphics[width=\textwidth]{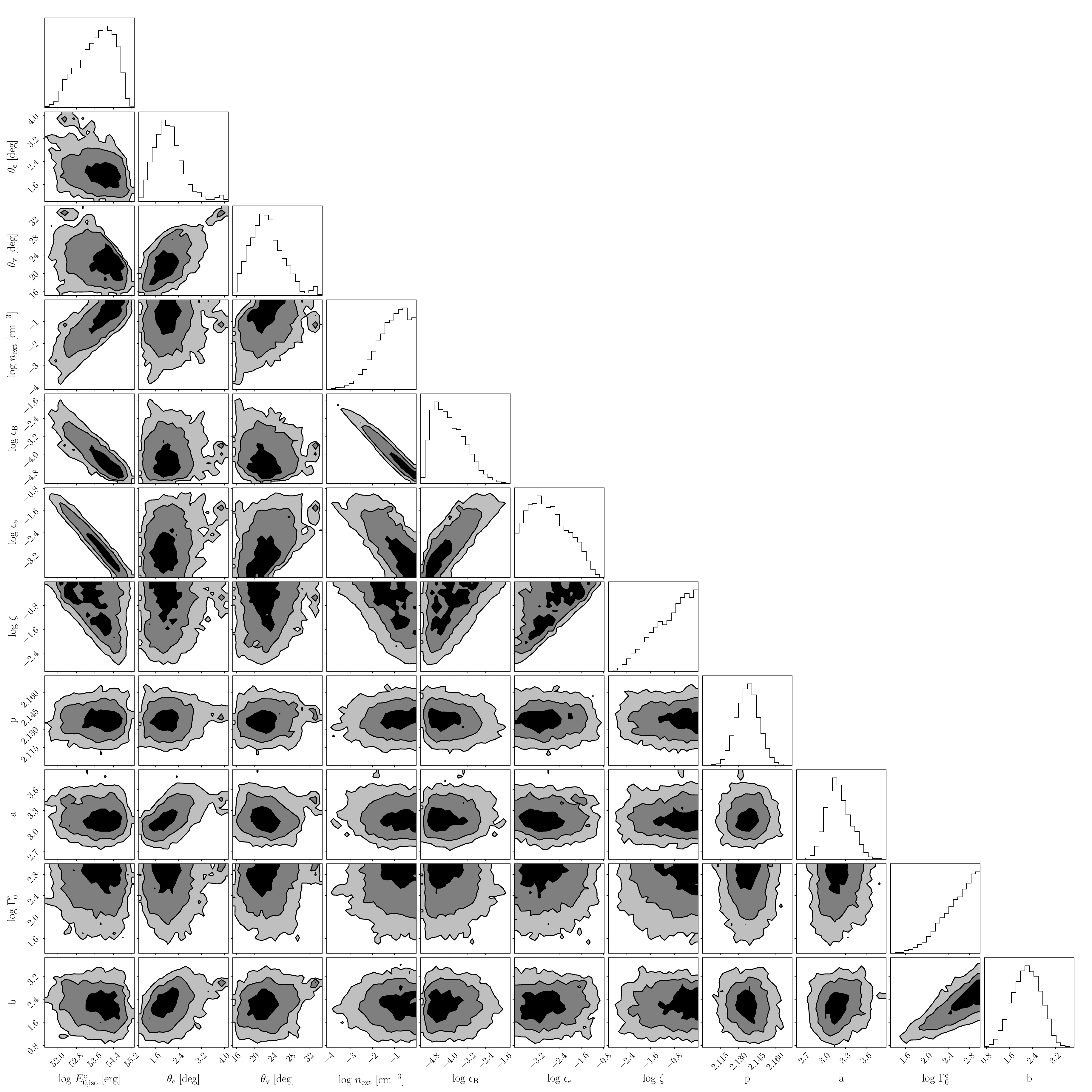}
    \caption{Posterior joint and marginalised distributions of the free model parameters, $E_\mathrm{0,iso}^\mathrm{c}$, $\theta_{\mathrm{c}}$, $\theta_{\mathrm{v}}$, $n_{\mathrm{ext}}$, $\epsilon_{\mathrm{B}}$, $\epsilon_{\mathrm{e}}$, $\zeta$, $p$ $a$, $\Gamma_0^\mathrm{c}$, $b$, for the "SSC (with KN)" fit of the afterglow of GW~170817 presented in Sect.~\ref{sec:170817}, which includes SSC scatterings in the Thomson and KN regimes. The priors are all uniform or log-uniform and are shown in Table~\ref{table:mcmc_priors}. The fitted data include all points until $t_{\rm max} = 400~\mathrm{days}$, as well as five early-time upper limits (see Sect.~\ref{sec:170817_data}). The coloured contours correspond to the $1\sigma$, $2\sigma$, and $3\sigma$ confidence intervals for each parameter.}
    \label{fig:MCMC_SSC}
\end{figure*}

\begin{table}[b]
\caption{Posterior values for the free parameters of the three fits of the afterglow of GW~170817.}
\centering
\resizebox{\linewidth}{!}{
\begin{minipage}{9.75cm}
\begin{tabular}{c | c c c}
\hline\hline
Free & \multicolumn{3}{c}{Model} \\
Parameter & no SSC & SSC (Thomson) & SSC (KN) \\
\hline 
    $\log E_{\rm 0, iso}^\mathrm{c}$ [$\mathrm{erg}$] & $53.70^{+1.21}_{-1.15}$ & $54.11^{+1.14}_{-1.30}$ & $53.70^{+1.12}_{-1.32}$\\
    $\theta_{\rm c}$  [$\degr$] & $1.77^{+1.12}_{-0.89}$ & $1.85^{+1.28}_{-0.82}$ & $2.07^{+0.83}_{-0.82}$\\
    $\theta_{\rm v}$ [$\degr$] & $21.81^{+7.69}_{-7.02}$ & $17.16^{+2.50}_{-2.35}$ & $22.67^{+5.94}_{-5.58}$\\
    $\log n_{\rm ext}$ [$\mathrm{cm^{-3}}$] & $-1.09^{+1.09}_{-1.28}$ & $-1.61^{+1.08}_{-1.23}$ & $-1.03^{+1.03}_{-1.14}$\\
    $\log \epsilon_{\rm B}$ & $-4.09^{+1.36}_{-1.01}$ & $-4.43^{+1.39}_{-1.54}$ & $-4.08^{+1.19}_{-0.97}$\\
    $\log \epsilon_{\rm e}$ & $-2.73^{+0.99}_{-1.27}$ & $-2.97^{+1.13}_{-1.03}$ & $-2.81^{+1.15}_{-1.13}$\\
    $\log \zeta$ & $-0.68^{+0.68}_{-1.05}$ & $-0.96^{+0.96}_{-0.96}$ & $-0.81^{+0.81}_{-1.07}$\\
    $p$ & $2.14^{+0.02}_{-0.02}$ & $2.14^{+0.01}_{-0.01}$ & $2.14^{+0.01}_{-0.02}$\\
    $a$ & $3.14^{+0.30}_{-0.29}$ & $3.34^{+0.66}_{-0.50}$ & $3.19^{+0.30}_{-0.27}$\\
    $\log \Gamma_0^\mathrm{c}$ & $2.68^{+0.32}_{-0.53}$ & $2.62^{+0.38}_{-0.49}$ & $2.64^{+0.36}_{-0.51}$\\
    $b$ & $2.13^{+0.84}_{-0.78}$ & $2.20^{+1.07}_{-0.97}$ & $2.24^{+0.79}_{-0.80}$\\
\hline
\end{tabular}
\end{minipage}}
\tablefoot{
The three fits are presented in Sect.~\ref{sec:170817_fits}. The table reports the median value of the posterior distribution of each parameter. The reported uncertainties are the 90\% credible intervals.
}
\label{table:best_fit_values}
\end{table}

\begin{figure}[t]
    \centering
    \resizebox{\hsize}{!}{\includegraphics{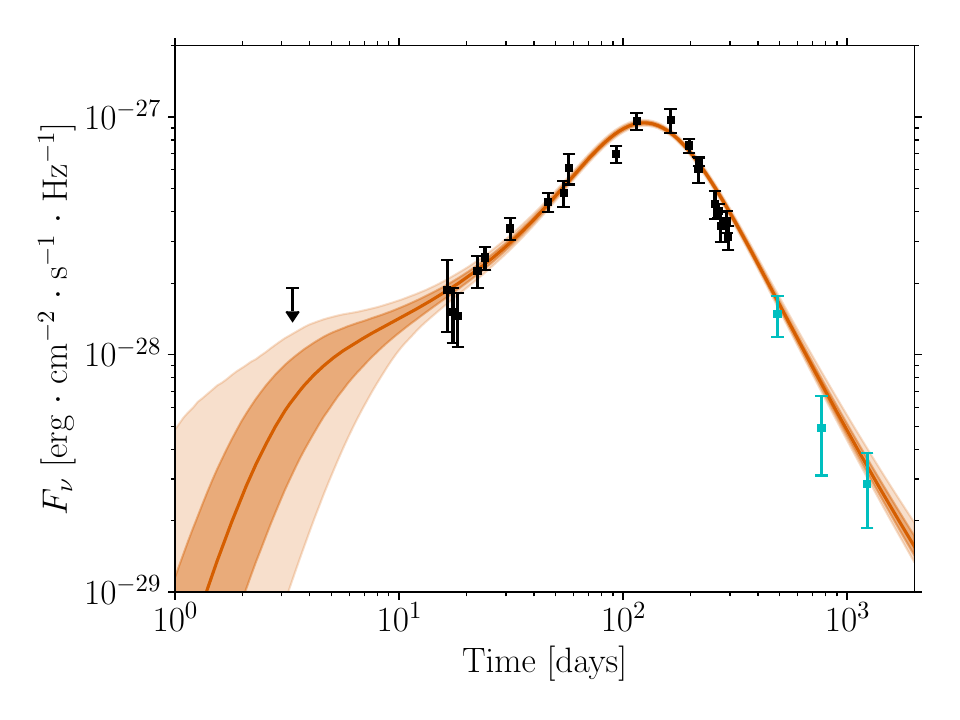}}
    \caption{Posterior distribution of the radio light curves at $\nu_{\rm obs} = 3~\mathrm{GHz}$ for the "SSC (with KN)" fit of the afterglow of GW~170817. The solid lines represent the median value at each observing time, the dark contours are the $68\%$ confidence interval, and the light contours show the $97.5\%$ confidence interval. Late-time observations that were not used in the fit are shown in blue.}
    \label{fig:lc_posterior_radio}
\end{figure}

\begin{figure}[t]
    \centering
    \resizebox{\hsize}{!}{\includegraphics{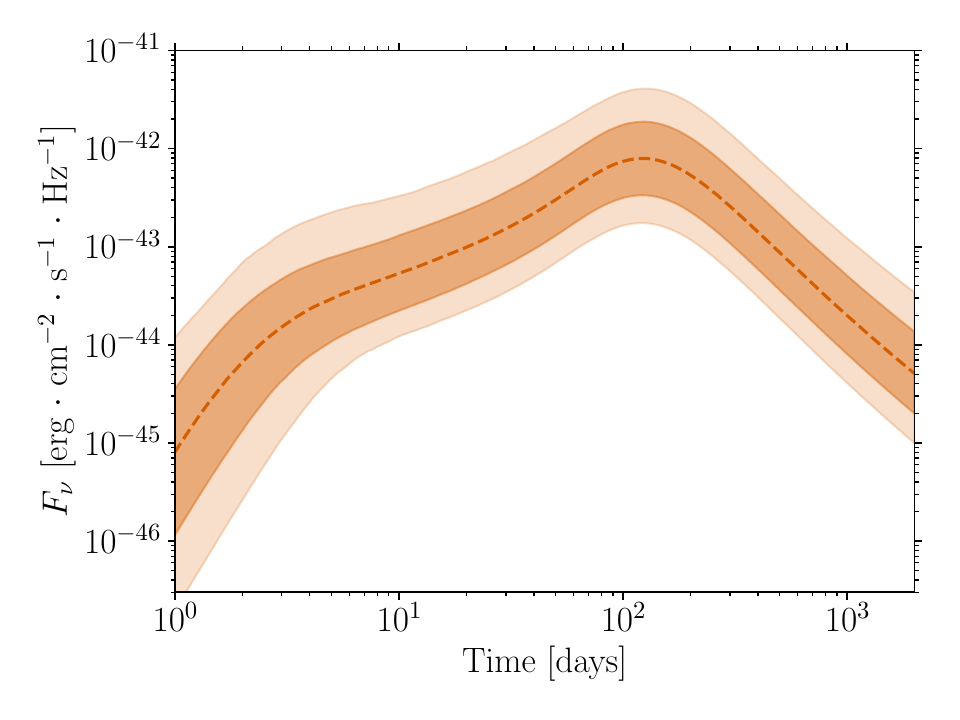}}
    \caption{Posterior distribution of the VHE light curves at $h\nu_{\rm obs} = 1~\mathrm{TeV}$ for the "SSC (with KN)" fit of the afterglow of GW~170817. Same as in Fig.~\ref{fig:lc_posterior_radio}. This is a prediction of the model. No observational constraint is included in the fit.}
    \label{fig:lc_posterior_vhe}
\end{figure}

The detailed joint posterior distributions for the "SSC (with KN)" model where emission is produced by synchrotron radiation and SSC scatterings in the Thomson and KN regimes are shown in Fig.~\ref{fig:MCMC_SSC}. In this figure, which is representative of the results obtained with the three fits, we observe typical correlations between some of the afterglow parameters, such as $\theta_{\mathrm{c}}$ and $\theta_{\mathrm{v}}$, $E_{\mathrm{0, iso}}^\mathrm{c}$ and $n_{\mathrm{ext}}$, or $\epsilon_{\mathrm{e}}$ and $\epsilon_{\mathrm{B}}$. Some other parameters are anti-correlated, such as $E_{\mathrm{0, iso}}^\mathrm{c}$ and $\epsilon_{\mathrm{B}}$, or $n_{\mathrm{ext}}$ and $\epsilon_{\mathrm{B}}$. These degeneracies in the model parameters are expected when only the synchrotron component is observed \citep[see e.g.][]{2022MNRAS.511.2848A}. In the case of GW~170817, the afterglow light curves can to first approximation be described by five quantities: its peak time, peak flux, the temporal slopes of the rising and decreasing phase, and the spectral slope. This does not allow us to constrain all the free parameters listed in Table~\ref{table:mcmc_priors}.

As a post-processing step, we resampled $20\,000$ light curves at several observing frequencies ($3$~GHz, $1$~keV, and $1$~TeV) and spectra at several observing times ($20$~days, $110$~days, and $400$~days), where the set of parameters was directly drawn from the posterior sample for each of the three fits. At each observing time (for the light curves) or at each frequency (for the spectra), we then determined the $68\%$ and $97.5\%$ credible intervals for the distribution of predicted flux densities, which we used to draw confidence contours around the median value of the light curves and spectra. Because the posterior samples are similar for the three models, we only show the results for the most realistic case ("SSC (with KN)" model) in Figs.~\ref{fig:lc_posterior_radio} and~\ref{fig:lc_posterior_vhe} (radio and VHE light curve), and in Fig.~\ref{fig:sp_posterior} (spectrum at 110 days, close to the peak).

\subsection{Discussion: Inferred parameters}\label{sec:discussion_170817}

\begin{figure}[t]
\vspace*{-0.3cm}
    \centering
    \resizebox{\hsize}{!}{\includegraphics{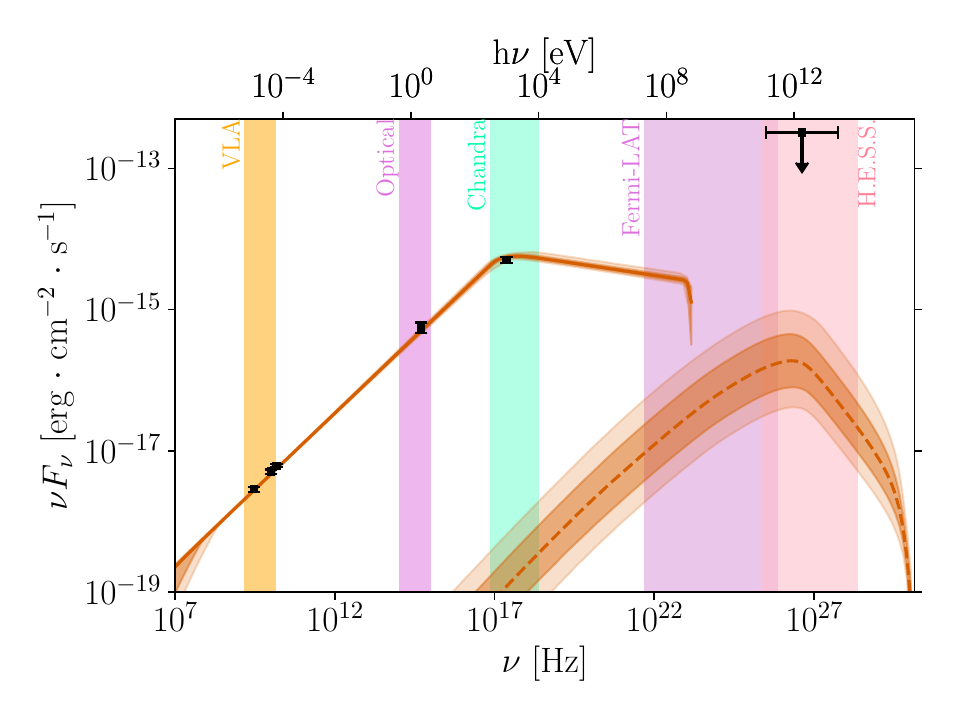}}
    \caption{Posterior distribution of the afterglow spectrum around its peak ($t_\mathrm{obs} = 110~\mathrm{days}$) for the "SSC (with KN)" fit of the afterglow of GW~170817. The data points show the multi-wavelength observations at $t_\mathrm{obs} \pm 4~\mathrm{days}$. The upper limit from H.E.S.S. is also indicated \citep{2020ApJ...894L..16A}. The low-energy component (solid line) is produced by synchrotron radiation, while the high-energy emission (dashed line) is powered by SSC scattering. The thick lines represent the median value at each observing frequency, the dark contours show the $68\%$ confidence interval, and the light contours present the $97.5\%$ confidence interval. Some instrument observing spectral ranges are shown in colours.}
    \label{fig:sp_posterior}
\end{figure}

As shown in Fig.~\ref{fig:lc_posterior_radio}, the afterglow of GW~170817 is very well fitted by the model. The model accurately predicts the flux at all times, with a very small dispersion. The dispersion of the predicted flux is larger at very early or very late times, when no data point is included in the fit. The quality of the model prediction is the same for the spectrum in Fig.~\ref{fig:sp_posterior}. The frequency $\nu_\mathrm{m,obs}$ is below the data points at the lowest frequency, with some dispersion in absence of observational constraints. The critical frequency $\nu_\mathrm{c,obs}$ is found to be in X-rays, just above the highest-frequency data points.

The inferred values of most model parameters (Fig.~\ref{fig:posterior_model_comparison} and Table~\ref{table:best_fit_values}) are very similar to the results obtained in previous studies that modelled the same afterglow with the synchrotron radiation from a decelerating structured jet \citep[][]{2019MNRAS.489.1919T,2019ApJ...870L..15L,2019Sci...363..968G,2020ApJ...896..166R}, with some exceptions that are due to different priors. A first difference concerns the viewing and core jet opening angles $\theta_\mathrm{v}$ and $\theta_\mathrm{c}$. Even though the values inferred by the "SSC (with KN)" fit are comparable with values found in previous studies, the corresponding ratio  $\theta_\mathrm{v}/\theta_\mathrm{c}\sim 11$ (Table~\ref{table:derived_best_fit_values}) is at the higher end of the other predictions. \citet{2021ApJ...909..114N} reported a compilation of these predictions and showed that afterglow light curves can only constrain this ratio and not the two parameters independently, which is confirmed by the correlation observed in Fig.~\ref{fig:MCMC_SSC}. This difference is probably due to the flat and broad priors we used for these two parameters, whereas most previous studies used much stricter priors based on an external constraint, either on the viewing angle derived from GW data using the value of the Hubble constant from Planck \citep[as in][]{2019MNRAS.489.1919T,2020ApJ...896..166R}, or a constraint on the viewing angle and the core jet using VLBI imagery \citep[as in][]{2019Sci...363..968G,2019ApJ...870L..15L}. The marginalised distributions of $\theta_\mathrm{v}$ and $\theta_\mathrm{c}$ show that a restricted region of the parameter space has been explored by the MCMC chains for these parameters, as is also visible in Fig.~\ref{fig:MCMC_3_models}.

When compared with the results of \citet{2020ApJ...896..166R}, whose assumptions are closest to those of our "no SSC" fit and who computed the light curves using \texttt{afterglowpy}, with which we compared our model in Sect.~\ref{sec:comparison_other_models}, we also find that our predicted values for $E_{0\mathrm{, iso}}^\mathrm{c}$ and $n_{\mathrm{ext}}$ are higher by about one order of magnitude. Fig.~\ref{fig:MCMC_SSC} shows that it is also an effect of our different priors for the angles. However, we observe that the expected strong correlation between these two parameters and the inferred ratio $E^\mathrm{c}_\mathrm{0,iso}/n_\mathrm{ext} \sim 10^{55}~\mathrm{erg\cdot cm^{-3}}$ (Table~\ref{table:derived_best_fit_values}) is close to the value obtained by \citet{2020ApJ...896..166R}. Using similar priors for the angles, or including a constraint on the external density from the observation of the host galaxy \citep{2017Sci...358.1579H,2019ApJ...886L..17H}, would then reduce the inferred values for these two parameters. We discuss below the fact that detections in the VHE range would help us to break this degeneracy and allow for a more precise determination of the energy and external density, independently of these external constraints.

\begin{table}[b]
\caption{Posterior values for several derived quantities from the three fits of the afterglow of GW~170817.}

\centering
\resizebox{\linewidth}{!}{
\begin{minipage}{11.5cm}
\begin{tabular}{c | c c c}
\hline\hline
Derived quantity & \multicolumn{3}{c}{Model} \\
 & no SSC & SSC (Thomson) & SSC (KN) \\
\hline 
    $\log E_0$ [$\mathrm{erg}$] & $51.95^{+1.05}_{-1.13}$ & $52.41^{+1.04}_{-1.22}$ & $52.07^{+1.01}_{-1.19}$\\
    $\theta_{\rm v} / \theta_{\rm c}$ & $11.95^{+5.29}_{-4.54}$ & $9.84^{+2.80}_{-2.59}$ & $11.19^{+3.58}_{-3.45}$\\
    $\log{\left({E_{0,\mathrm{iso}}^\mathrm{c}}/{n_\mathrm{ext}}\right)}$ [$\mathrm{erg \cdot cm^3}$] & $54.90^{+1.08}_{-1.18}$ & $55.77^{+0.42}_{-0.43}$ & $54.79^{+0.86}_{-0.85}$\\
    \hline
    $\log{\left(\theta_\mathrm{c} \Gamma_0^\mathrm{c}\right)}$ & $1.15^{+0.48}_{-0.58}$ & $1.12^{+0.43}_{-0.54}$ & $1.17^{+0.44}_{-0.61}$\\
    $8b / a$ & $5.40^{+1.81}_{-2.05}$ & $5.36^{+1.95}_{-1.94}$ & $5.54^{+1.91}_{-1.96}$\\
    $-b \log{\theta_\mathrm{c}}/\log{\Gamma_0^\mathrm{c}}$ & $1.21^{+0.23}_{-0.25}$ & $1.29^{+0.27}_{-0.28}$ & $1.22^{+0.25}_{-0.22}$\\
    $(b-1)\log {\left(\theta_\mathrm{v} / \theta_\mathrm{c}\right)}/\log{\left( \Gamma_0^\mathrm{c}\theta_\mathrm{c}\right)}$ & $1.09^{+0.38}_{-0.38}$ & $1.10^{+0.39}_{-0.39}$ & $1.10^{+0.40}_{-0.34}$\\
\hline
\end{tabular}
\end{minipage}}
\tablefoot{The three fits are presented in Sect.~\ref{sec:170817_fits} and the posterior values of the free parameters are reported in Table~\ref{table:best_fit_values}. The table reports the median value of posterior distribution of each quantity. The uncertainties reported are the 90\% credible intervals.
 The four last rows correspond to the quantities that appear in the conditions for single-peak light curves (Sect.~\ref{sec:dynamics}).}
\label{table:derived_best_fit_values}
\end{table}

The inferred values for the microphysics parameters $\epsilon_\mathrm{B}$ and $p$ are very close to those obtained in previous studies. The value of $\epsilon_\mathrm{e}$ is slightly lower, but we also find that the median value for the fraction of accelerated electrons is $\sim 0.15$, whereas this parameter is  not included in past studies. Both parameters are strongly correlated, as shown in Fig.~\ref{fig:MCMC_SSC}. Overall, the values of these microphysics parameters, including $\zeta$, agree well with the current understanding of the plasma physics at work in relativistic collisionless shocks \citep[see e.g.][]{2015SSRv..191..519S}, and they are comparable to the values obtained for cosmological short GRBs by \citet{2015ApJ...815..102F}. The inferred value $\zeta < 1$ leads to the question of a possible contribution to the radiation of the remaining thermal electrons \citep[see e.g.][]{2022ApJ...924...40W}. Assuming $\zeta=1$ as in previous studies is relevant in the case of GW~170817 as the observed emission is dominated by the synchrotron radiation, and the cooling break $\nu_\mathrm{c,obs}$ is not detected. However, these microphysics parameters impact the synchrotron and SSC components in different manners, as discussed below, and they must therefore be considered to predict the VHE emission.

As shown in the second row of Fig.~\ref{fig:posterior_model_comparison} and in Table~\ref{table:derived_best_fit_values}, the inferred values of the parameters describing the initial lateral structure of the ejecta agree well with the conditions established in \citet{2020MNRAS.493.3521B} and listed in Sect.~\ref{sec:dynamics} to obtain a single-peak light curve: In practice, double-peak light curves are not excluded a priori in our MCMC, but are disfavoured due to the inclusion of early upper limits (Table~\ref{table:170817_early_ul}). Finally, we note that the initial distribution of the Lorentz factor ($\Gamma^\mathrm{c}_0$ and $b$) is poorly constrained, which was expected as most of the observed emission is produced in the self-similar stage of the deceleration. A correlation is observed between $\Gamma_0^\mathrm{c}$ and $b$. This can be explained by the fact that the lateral structure is required to reproduce the early-time slow rise of the light curve \citep{2018MNRAS.478..407N}. Specifically, some material needs to have lower initial Lorentz factors to overcome effects of Doppler boosting out of the line of sight. With a higher core $\Gamma_0^\mathrm{c}$, a steeper decrease in $\Gamma_0(\theta)$ is required for material at a given latitude $\theta$ to have a sufficiently low initial Lorentz factor that the observed slow rise can be reproduced, leading to higher values of $b$.

\begin{figure}[t]
    \centering
    \resizebox{\hsize}{!}{\includegraphics{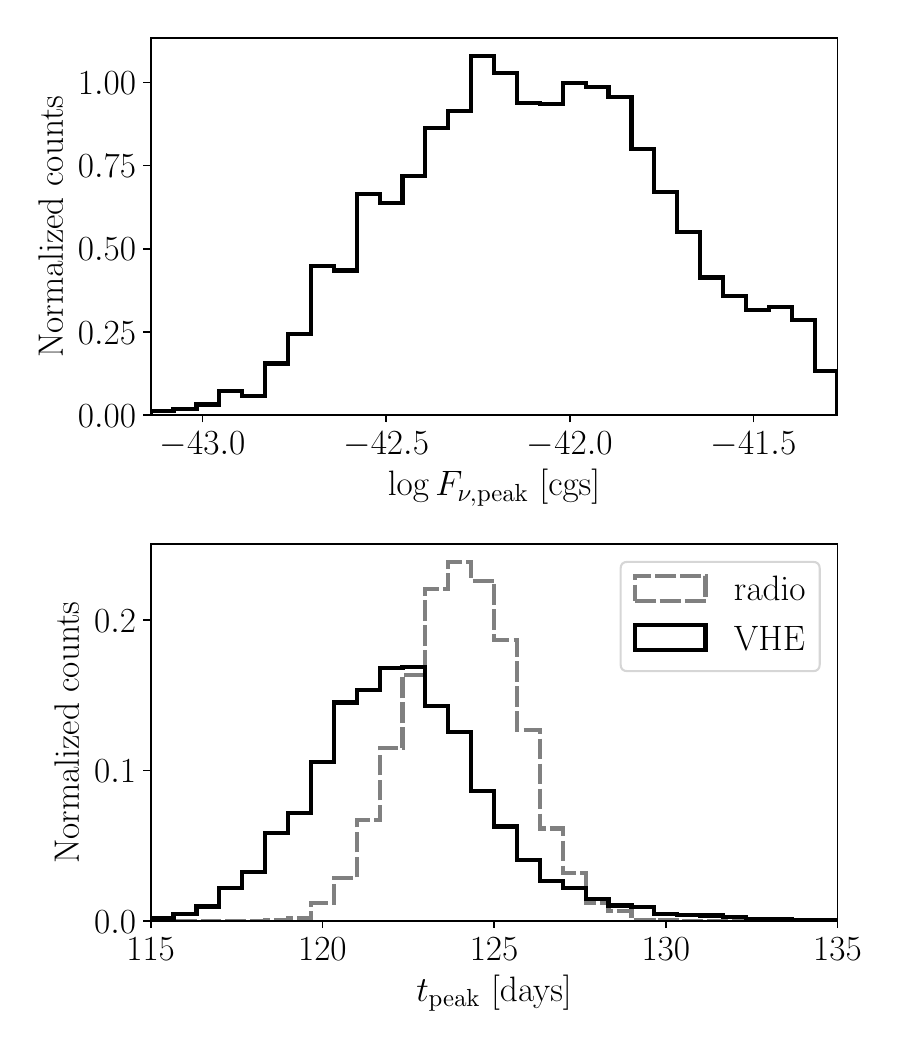}}
    \caption{Posterior distributions of the VHE peak flux density (top) and the radio (dashed grey line) and VHE (solid black line) peak times (bottom) for the "SSC (with KN)" fit of the afterglow of GW~170817.}
    \label{fig:vhe_peak}
\end{figure}

\subsection{Discussion: Predicted very high energy emission of GW~170817}\label{sec:discussion_170817_vhe}

We now discuss the predicted VHE emission of the afterglow of GW~170817. The predicted light curve at 1 TeV is plotted in Fig.~\ref{fig:lc_posterior_vhe}, and the VHE component of the spectrum at the peak of the light curve is shown in Fig.~\ref{fig:sp_posterior}, where we also indicate the upper limit obtained by H.E.S.S. \citep{2020ApJ...894L..16A}. It appears that (i) the synchrotron emission extends at most up to the GeV range\footnote{The emission in the $0.1-1\, \mathrm{GeV}$ range is then dominated by the synchrotron radiation. For comparison with Fig.~\ref{fig:sp_posterior}, the sensitivity of \textit{Fermi}/LAT above 100 MeV is of the order of $10^{-11}\, \mathrm{erg}\cdot \mathrm{s}^{-1} \cdot \mathrm{cm}^{-2}$ for a $\sim 1\, \mathrm{day}$ observation \citep{2020ApJS..247...33A, 2022ApJS..260...53A}. The afterglow of GW~170817 was therefore undetectable by \textit{Fermi}/LAT at any time.} (due to the synchrotron burnoff limit; see Appendix~\ref{ap:gamma_max}) and the VHE emission is then entirely due to the SSC process; (ii) even when the dispersion is taken into account, the predicted flux at 1 TeV is at least two orders of magnitude below the upper limit of $3.2 \times 10^{-13}~\mathrm{erg}\cdot \mathrm{s}^{-1} \cdot \mathrm{cm}^{-2}$ obtained around the peak by H.E.S.S. \citep{2020ApJ...894L..16A}, which is therefore not constraining. It is also clear when comparing Figs.~\ref{fig:lc_posterior_radio} and~\ref{fig:lc_posterior_vhe} that the dispersion of the predicted VHE flux is much larger than for the synchrotron component, which is well constrained by the observations from radio to X-rays. This is also shown in the distribution of the predicted peak flux and peak time in radio and at 1 TeV in Fig.~\ref{fig:vhe_peak}. The VHE light curve peaks a few days before the light curves from radio to X-rays because the synchrotron and SSC components evolve differently. This result is due to the fact that the dependence of the SSC component on the model parameters is complex and very different from the dependences of the synchrotron component identified by \citet{1998ApJ...497L..17S}. This also highlights the interest of detecting the VHE emission for the inference of model parameters, as some of the degeneracies visible in Fig.~\ref{fig:MCMC_SSC} would be broken.

\begin{table}[t]
\caption{Parameters of the two reference cases.}
\centering
\begin{tabular}{ccc}
\hline\hline
Parameter & Moderate  & Optimistic\\
\hline 
    $E_{\rm 0, iso}^\mathrm{c}$ [erg] & $4.12 \times 10^{52}$ & $5.07 \times 10^{52}$ \\
    $\theta_{\rm c}$ [$\degr$] & $1.75$ & $1.31$ \\
    $\theta_{\rm v}$ [$\degr$] & $21.25$ & $24.40$ \\
    $n_{\rm ext}$ [cm$^{-3}$] & $4.25 \times 10^{-3}$ &  $1.52 \times 10^{-2}$ \\
    $\epsilon_{\rm B}$ & $1.61 \times 10^{-3}$ & $4.79 \times 10^{-4}$ \\
    $\epsilon_{\rm e}$ & $1.87 \times 10^{-2}$ & $3.54 \times 10^{-2}$  \\
    $\zeta$ & $1$ & $3.57 \times 10^{-1}$\\
    $p$ & $2.139$ &  $2.149$ \\
    $a$ & $2.98$ & $2.95$  \\
    $\Gamma_0^\mathrm{c}$ & $184$ & $164$ \\
    $b$ & $2.01$ & $1.34$ \\
\hline
\end{tabular}
\tablefoot{
The two parameter sets are in the 97.5\% confidence interval of the distributions inferred by the "SSC (with KN)" fit of the afterglow of GW~170817. They differ by the predicted VHE flux, which is either close to the median (moderate case) or at the higher end of the confidence interval (optimistic case).}
\label{table:ref_params}
\end{table}

\begin{figure}
\vspace*{-0.3cm}
    \centering
    \resizebox{\hsize}{!}{\includegraphics{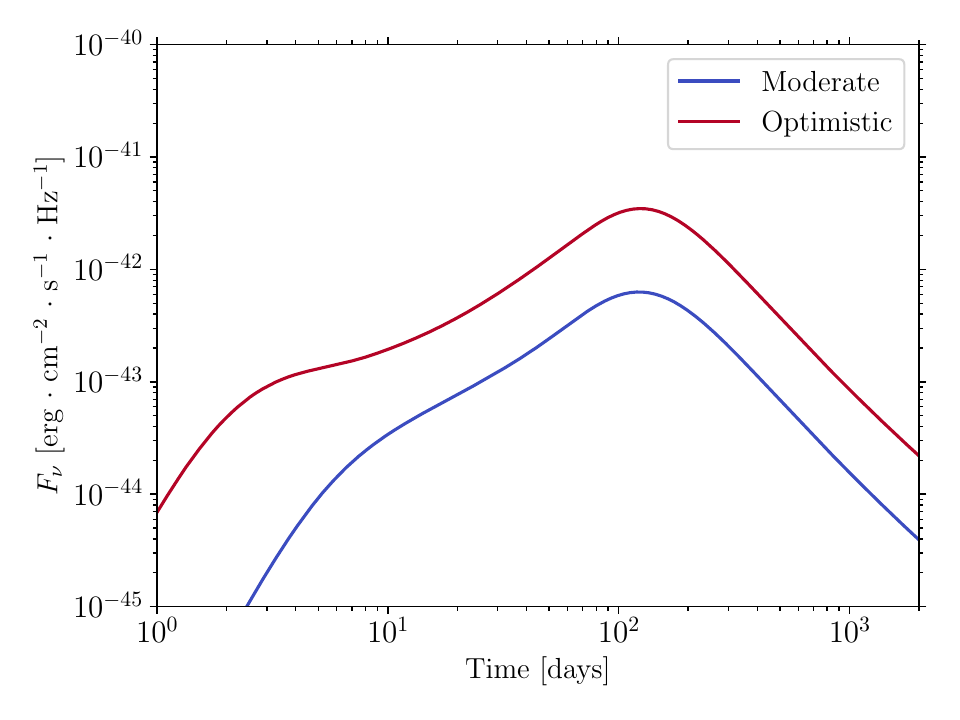}}
    \resizebox{\hsize}{!}{\includegraphics{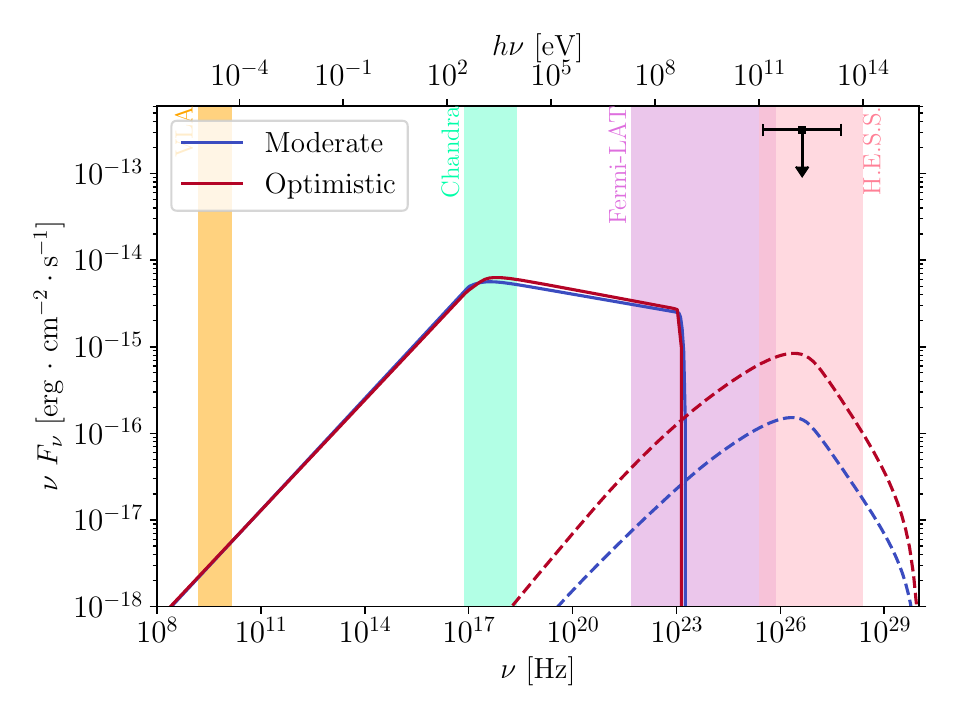}}
    \caption{VHE light curve at $h\nu_\mathrm{obs}=1~\mathrm{TeV}$ (top) and spectrum around the peak at $t_\mathrm{obs}=110~\mathrm{days}$ (bottom) for the moderate (in blue) and optimistic (in red) reference cases (see Table~\ref{table:ref_params}). Both predict similar fluxes in the synchrotron range due to the observational constraints, but they differ in the SSC regime. The upper limit from H.E.S.S. is indicated  for comparison \citep{2020ApJ...894L..16A}.
    }
    \label{fig:ref_lc_vhe}
\end{figure}

In the afterglow of GW~170817, the synchrotron spectrum is in the slow-cooling regime, with a spectral peak determined by the critical Lorentz factor $\gamma_\mathrm{c}$. If IC scattering occurred in Thomson regime, the peak of the SSC component would be due to the scattering of photons at $\nu_\mathrm{c}$ by the electrons at $\gamma_\mathrm{c}$ \citep{2001ApJ...548..787S}, and this peak should be intense as $\epsilon_\mathrm{e}\gg \epsilon_\mathrm{B}$ (see Eq.~(\ref{eq:YSSCnoKN})). However, the predicted VHE emission is very weak because the KN attenuation is strong. To illustrate this effect, we select in Table~\ref{table:ref_params} two reference sets of parameters that are taken from the posterior sample. They are in the 68\% and 97.5\% confidence intervals for the "SSC (with KN)" fit and fit the observed light curves perfectly well. The first set of parameters (moderate) corresponds to a predicted VHE flux that is close to the median, whereas the second set of parameters (optimistic) corresponds to a predicted VHE flux at the highest end of the confidence interval (see the two predicted VHE light curves and peak spectra in Fig.~\ref{fig:ref_lc_vhe}). In both cases, we selected parameter sets with a low density and therefore, with a low energy (see the discussion in Sect.~\ref{sec:discussion_170817}).

\begin{figure}[t]
    \centering
    \resizebox{\hsize}{!}{\includegraphics{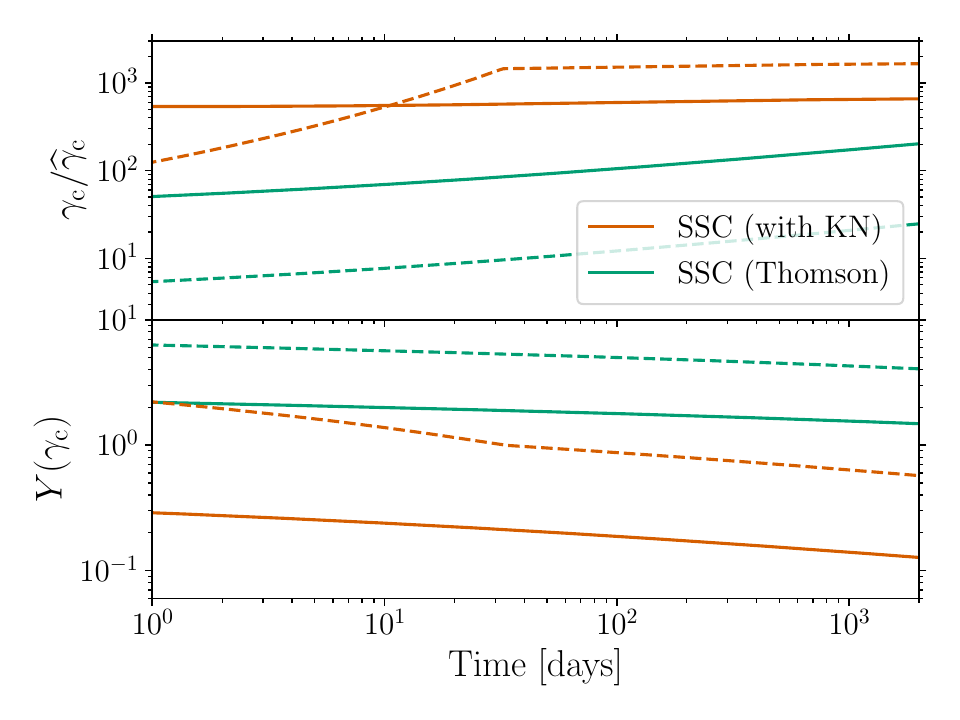}}
    \caption{Evolution of $\gamma_\mathrm{c} / \widehat{\gamma}_\mathrm{c}$ (top) and of the Compton parameter at $\gamma_\mathrm{c}$ , $Y(\gamma_\mathrm{c})$ (bottom) in the core jet. These quantities are plotted as a function of the observer time for the moderate (solid line) and optimistic (dashed line) reference cases (see Table~\ref{table:ref_params}) in the "SSC (Thomson)" case (green) and in the "SSC (with KN)" case (orange).}
    \label{fig:Ygc_wc}
\end{figure}

Fig.~\ref{fig:Ygc_wc} shows for these two reference cases the evolution of $Y(\gamma_\mathrm{c})$ and $\gamma_\mathrm{c}/\widehat{\gamma}_\mathrm{c}$ for the core jet. The SSC emission is either computed by assuming that all scatterings are in Thomson regime, as in the "SSC (Thomson)" fit, or by including the KN attenuation, as in the "SSC (with KN)" fit. The figure shows that in the full calculation, $\gamma_\mathrm{c}/\widehat{\gamma}_\mathrm{c}\gg 1$ during the whole evolution, so that the scattering of photons at $\nu_\mathrm{c}$ by electrons at $\gamma_\mathrm{c}$ is strongly reduced in the KN regime. The Compton parameter $Y(\gamma_\mathrm{c})$ then remains very low.  This confirms the origin of the weak VHE emission in the afterglow of GW~170817. In this regime of weak SSC emission, the synchrotron component is not affected by the IC cooling, $\gamma_\mathrm{c}\simeq \gamma_\mathrm{c}^\mathrm{syn}$, and this explains why the results of the "no SSC" and "SSC (with KN)" fits are so close (Table~\ref{table:best_fit_values}). On the other hand, Fig.~\ref{fig:Ygc_wc} also shows that $Y^\mathrm{no\, KN}$   overestimates $Y(\gamma_\mathrm{c})$ in this case, leading to $Y(\gamma_\mathrm{c})> 1$. The critical Lorentz factor is therefore given by $\gamma_\mathrm{c}\simeq \gamma_\mathrm{c}^\mathrm{syn}/Y(\gamma_\mathrm{c})$, which leads to an overestimation of $\gamma_\mathrm{c}^\mathrm{syn}$ in the MCMC to maintain the cooling break above the X-ray measurements (increased $E_{0\mathrm{, iso}}$ and lower $\epsilon_{\mathrm{B}}$). This explains the observed difference between the values of the model parameters inferred in the "SSC (Thomson)" and "SSC (with KN)" fits in Table~\ref{table:best_fit_values}. The value of $\gamma_\mathrm{c}/\widehat{\gamma}_\mathrm{c}\gg 1$ in this case (green curves in the top panel of Fig.~\ref{fig:Ygc_wc}) proves that the Thomson regime is not justified.

As the upper limit of H.E.S.S. was obtained close to the predicted peak with an already long exposure time of 53.9~hours \citep{2020ApJ...894L..16A}, our prediction shows that it was unfortunately impossible to detect the VHE afterglow of GW~170817 with currently available instruments. We discuss the conditions that would make a post-merger afterglow detectable at VHE in the future in the next section, especially in the context of the increased sensitivity expected with the CTA.

\section{Detectability of post-merger afterglows at very high energy, and prospects for the CTA}\label{sec:conditions_tev}

Even if the VHE afterglow of GW~170817 was not detectable, deeper observations of similar events can be expected in the future, in particular, with the CTA. The off-axis afterglow phase is of particular interest because after a few days, full-night observations can be conducted without a significant intrinsic source flux variability over the observation time. We investigate in this section the conditions that would make such an event detectable by current instruments and the upcoming CTA. For this study, we assumed a CTA detection sensitivity at $1~\mathrm{TeV}$ about six times better than that of H.E.S.S., as expected from \citet{2019scta.book.....C}. For an exposure time of $\sim 50\, \mathrm{hours}$ as for the H.E.S.S. deep observations of GW~170817 \citep{2020ApJ...894L..16A}, this leads to a detection sensitivity at $1~\mathrm{TeV}$ of $5 \times 10^{-14}~\mathrm{erg}\cdot \mathrm{s}^{-1} \cdot \mathrm{cm}^{-2}$.

\begin{figure}[t]
    \centering
    \resizebox{\hsize}{!}{\includegraphics{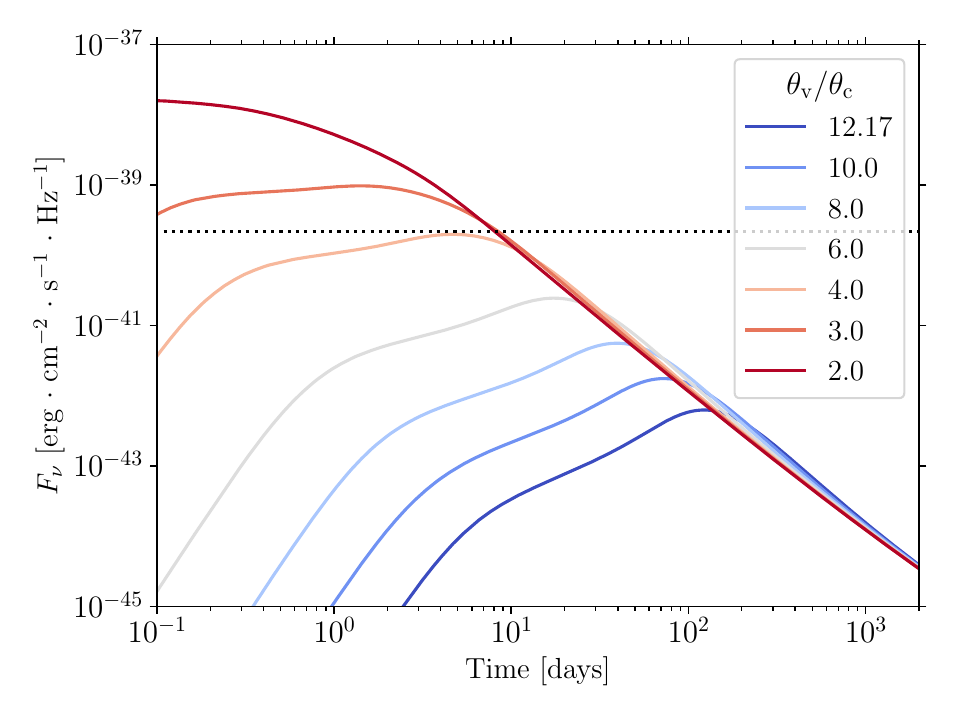}}   \resizebox{\hsize}{!}{\includegraphics{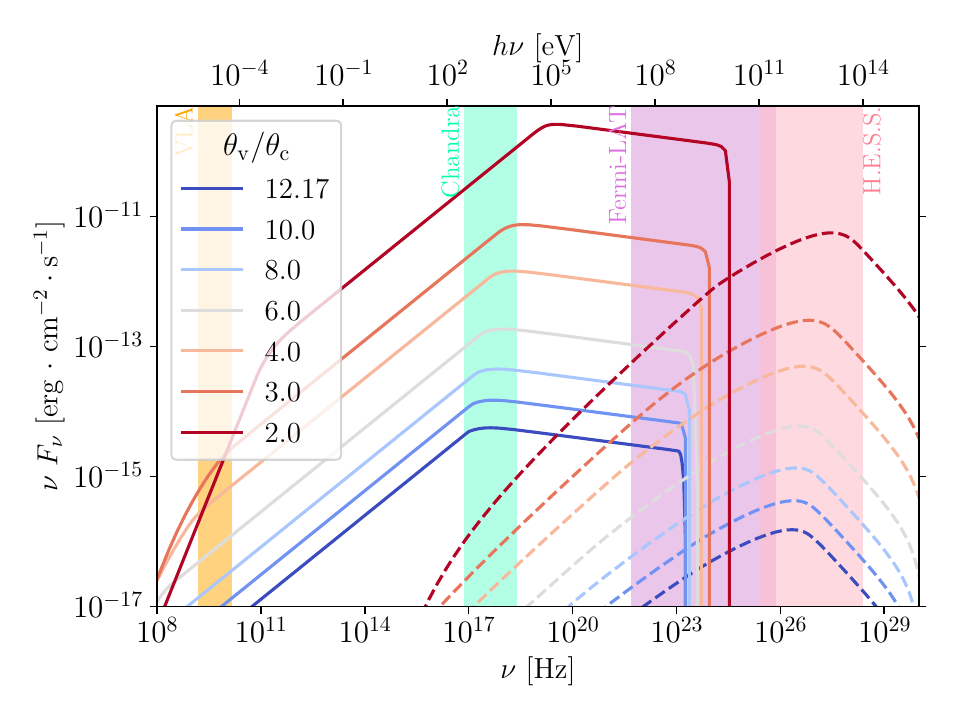}}
    \caption{Light curves at $1~\mathrm{TeV}$ (top panel) and spectra at light-curve peak (bottom panel) of the moderate reference case (see Table~\ref{table:ref_params}) for varying viewing angles. The case fitting the afterglow of GW~170817 corresponds to $\theta_\mathrm{v} / \theta_\mathrm{c} = 12.17$. 
    The peak times at which spectra are calculated are (in the ordering of the legend) $123$, $73$, $39$, $17$, $4.7$, $1.4$, and $0.06$ days. The dotted line in the top panel corresponds to the assumed CTA sensitivity in Sect.~\ref{sec:conditions_tev}.
    }
    \label{fig:vary_theta_obs}
\end{figure}

A geometrical effect such as a lower viewing angle naturally increases the peak of the observed flux, and it has a similar effect on the synchrotron and SSC components. In Fig.~\ref{fig:vary_theta_obs}, we show the light curves (top) and spectra at the time of the $1~\mathrm{TeV}$ emission peak (bottom) for the moderate reference case, and for gradually decreasing viewing angles. The $1~\mathrm{TeV}$ light curves peak at earlier times when the viewing angle decreases, and the peak flux reaches higher values. As expected, the relative flux increase at the time of peak is similar for the synchrotron and the SSC components (Fig.~\ref{fig:vary_theta_obs}, bottom), and all light curves follow a similar long-term evolution when the entire jet emission becomes visible to the off-axis observer. For the moderate (optimistic) reference case (Table~\ref{table:ref_params}), we find that a ratio $\theta_\mathrm{v} / \theta_\mathrm{c} \lesssim 3$ ($\theta_\mathrm{v} / \theta_\mathrm{c} \lesssim 6$) would have made the afterglow of GW~170817 detectable by H.E.S.S., provided the conditions had allowed for a very early deep observation. With the CTA, the VHE emission could have been detected up to $\theta_\mathrm{v} / \theta_\mathrm{c} \simeq 4$ ($\theta_\mathrm{v} / \theta_\mathrm{c} \simeq 8$). In this scenario, the emission peaks at $\sim 5~\mathrm{days}$ ($13~\mathrm{days}$).
This opens up interesting perspectives even when the expected detection rate of nearby events at small viewing angles is low \citep[see e.g.][]{2019A&A...631A..39D,2021A&A...651A..83M}.

\begin{figure}
    \centering
    \resizebox{\hsize}{!}{\includegraphics{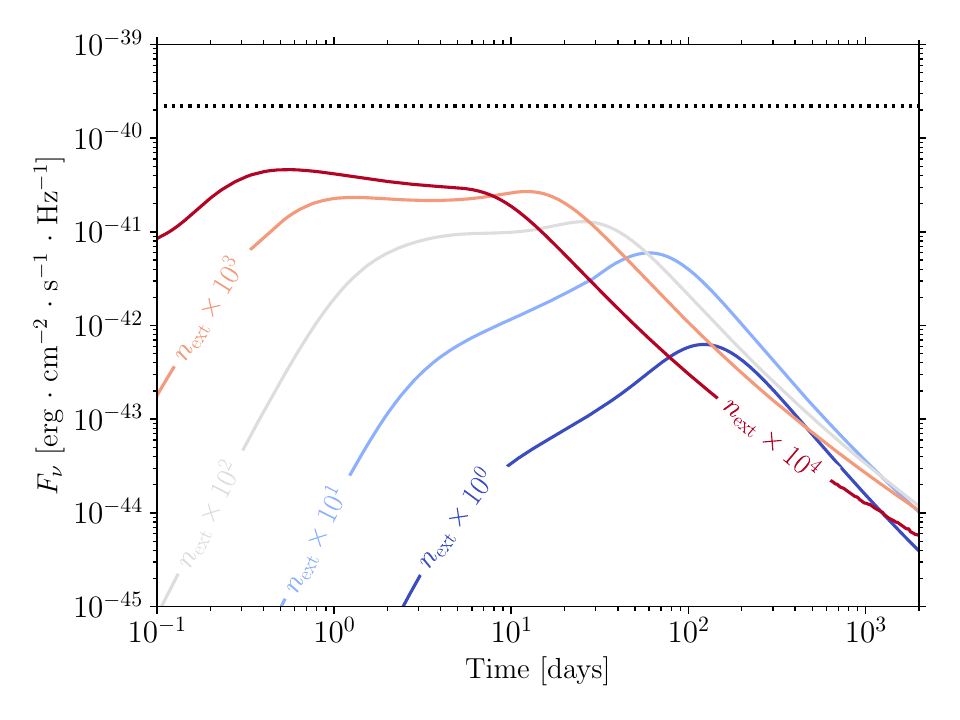}}
    \resizebox{\hsize}{!}{\includegraphics{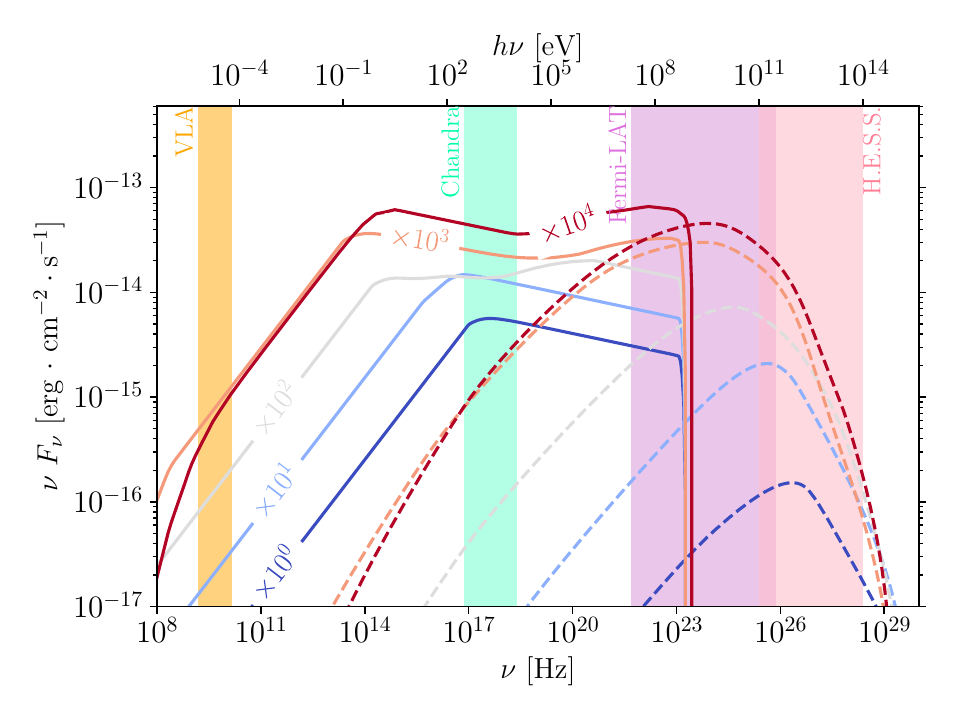}}
    \caption{Light curves at $1~\mathrm{TeV}$ (top panel) and spectra at light-curve peak (bottom panel) of the moderate reference case (see Table~\ref{table:ref_params}) for varying external medium densities $n_\mathrm{ext}$. The case fitting the afterglow of GW~170817 corresponds to $n_\mathrm{ext} = 4.25 \times 10^{-3}~\mathrm{cm}^{-3}$. Each line is labelled with the multiplication factor applied to the reference $n_\mathrm{ext}$. The peak times at which spectra are calculated are $123$, $60$, $26$, $12$, and $0.55$ days. The dotted line in the top panel corresponds to the assumed CTA sensitivity in Sect.~\ref{sec:conditions_tev}.}
    \label{fig:vary_n_ext}
\end{figure}

Under some conditions, the predicted VHE emission could also be intrinsically brighter. As discussed in Sect.~\ref{sec:discussion_170817_vhe}, the VHE afterglow of GW~170817 is weak because of the strong KN attenuation ($\gamma_\mathrm{c}/\widehat{\gamma}_\mathrm{c}\gg 1$). The regime for the scattering of photons at $\nu_\mathrm{c}$ by electrons at $\gamma_\mathrm{c}$ is affected by several parameters \citep{2007PhR...442..166N}. When the scattering occurs in the Thomson regime, the predicted scaling in the core jet is
\begin{equation}
\frac{\gamma_\mathrm{c}}{\widehat{\gamma}_\mathrm{c}} \propto 
\frac{   1}{(1+Y_\mathrm{c})^3}\frac{1}{\epsilon_\mathrm{B}^{5/2} n_\mathrm{ext}^{3/2}\epsilon^\mathrm{c}_0 }\, .
\end{equation}
Therefore, the Thomson regime is favoured by a high magnetic field, a high density, or a high kinetic energy. As $\epsilon_\mathrm{B}$ is determined by plasma instabilities at the ultra-relativistic shock \citep{2015SSRv..191..519S}, it should not vary much from an afterglow to the next. On the other hand, the external density and the kinetic energy may differ between mergers. We focus here on the effect of the external density, which is slightly stronger. Fig.~\ref{fig:vary_n_ext} shows the evolution of the afterglow emission for the moderate reference case when $n_\mathrm{ext}$  increases up to $\sim 10\, \mathrm{cm^{-3}}$, all other parameters being kept constant. A higher external density leads to an earlier peak time at higher luminosities, but affects the synchrotron and the SSC flux differently: As expected, the SSC process becomes more efficient and the relative SSC-to-synchrotron ratio increases. The transition from the weak SSC emission due to a strong KN attenuation described in Sect.~\ref{sec:discussion_170817_vhe} to a more efficient case in Thomson regime occurs in this reference case above $n_\mathrm{ext}\sim 1~\mathrm{cm}^{-3}$. This corresponds to the radiative regime S15 in Appendix~\ref{sec:sol_fgh}, with spectral breaks at $\nu_\mathrm{m} < \nu_\mathrm{c} < \widehat{\nu}_\mathrm{c} < \nu_0$. This effect can clearly be observed in the bottom panel of Fig.~\ref{fig:vary_n_ext}, showing the observed spectrum at the peak emission. At the highest external densities, the increase in the VHE flux stalls for two reasons. Firstly, the strong SSC emission starts to affect the synchrotron spectrum significantly, which features two peaks (visible on the bottom panel of Fig.~\ref{fig:vary_n_ext}). The cooling frequency $\nu_\mathrm{c}$ at the peak also decreases, shifting the SSC peak to lower frequencies so that the flux at 1~TeV does not increase as significantly as the SSC peak flux. A second limitation, dominant at the highest densities, comes from the pair production.

Even when these limitations are taken into account, a higher external density clearly favours a brighter VHE emission, even when both the moderate and optimistic reference cases remain undetectable by H.E.S.S. or the CTA at the reference viewing angle. We conclude that post-merger VHE afterglows should become detectable in the future only under several favourable conditions, such as a higher external density and a slightly lower viewing angle. For instance, the moderate (optimistic) reference case with an external density $n_\mathrm{ext}= 1~\mathrm{cm^{-3}}$ and a viewing angle $\theta_\mathrm{v}/\theta_\mathrm{c}=6$ should be detectable by the CTA\footnote{On the other hand, even with $n_\mathrm{ext}=1\, \mathrm{cm^{-3}}$ and $\theta_\mathrm{v}/\theta_\mathrm{c}=6$, the emission in the \textit{Fermi}/LAT range at the peak is still dominated by the synchrotron radiation and remains undetectable at $40\, \mathrm{Mpc}$ for a $\sim 1\, \mathrm{day}$ observation.} up to a distance of $\sim 90~\mathrm{Mpc}$ ($\sim 400~\mathrm{Mpc}$). However, two additional effects should be included in the optimistic case and may reduce this maximum distance: (i) At 400 Mpc, the attenuation by the extragalactic background light is no longer negligible at 1 TeV (see e.g. \citealt{2011MNRAS.410.2556D}), and (ii) in this case, the VHE afterglow peaks very early ($\sim 1$ day). This would require an early follow-up, which will necessarily be less deep due to a limited exposure time. Detections with the CTA above 100 Mpc remain possible in the optimistic case. As some parameters that are naturally degenerate in an afterglow fit including only synchrotron radiation, such as $n_{\mathrm{ext}}$ and $\epsilon_{\mathrm{B}}$, have different effects on the SSC emission, these VHE detections would therefore better constrain the parameter space and allow for a better understanding of the physical conditions in the post-merger relativistic ejecta.

Another direct consequence of the impact of $n_\mathrm{ext}$ on SSC emission is that short GRB afterglows detected at VHE are more likely to originate from mergers that occur in a denser environment. These VHE detections could therefore be unique probes of high-density media at the location of BNS mergers, which are sought clues for binaries with short merger times. Fast mergers are indeed expected to occur close to their formation sites, where the external medium density is expected to be higher \citep[see e.g. the discussion in][]{2020A&A...639A..15D}. Population synthesis studies have shown that the as yet poorly constrained common-envelope phase during the binary star evolution can lead to the production of BNS with very short initial separations that merge rapidly, as discussed for instance by \citet{2012ApJ...759...52D}. Favourable kick intensities and orientations during the supernovae could also tend to shrink the initial orbital separation and increase the orbital eccentricity for part of the BNS population \citep{1996ApJ...471..352K}. Observationally, the projected offset distribution of BNS mergers in their host galaxy seems to feature a tail below one half-light radius for $\sim 20\%$ of the systems \citep{2013ApJ...776...18F, 2014ARA&A..52...43B}. These observations have several observational biases, such as afterglow detection, host galaxy association, and projection effects, which probably leads to sample incompleteness, but they give evidence for a population of fast mergers. If BNS mergers are the dominant site for \textit{r}-process nucleosynthesis \citep[see e.g.][]{1974ApJ...192L.145L,2011ApJ...738L..32G,2015MNRAS.448..541J,2017ApJ...836..230C}, another more indirect evidence is provided by the observed large scatter of [Eu/Fe] abundances in extremely metal-poor stars \citep[see the compilation by][]{2008PASJ...60.1159S}, which requires fast mergers that allow the production of \textit{r}-process elements when the environment is still extremely metal poor \citep[see e.g.][]{2016MNRAS.455...17V,2019ApJ...875..106C,2021MNRAS.506.4374D}. It has already been suggested by \citet{2020A&A...639A..15D} that multi-messenger observations of BNS, including the detection of the afterglow, will be biased in favour of a high-density environment, thus allowing us to probe this potential population of fast mergers. We showed here that future VHE afterglow detections either following the GW detection of a BNS merger or the detection of a short GRB can also probe this population directly.

\section{Conclusion}\label{sec:conclusion}

We developed a detailed model of the afterglow of a laterally structured jet from radio to VHE. Emission is produced in the shocked external medium behind the forward external shock, and the dynamics does not include the late lateral expansion of the ejecta. These effects, as well as the contribution of the reverse shock, will be implemented in the future. The main challenge addressed in this paper is the self-consistent calculation of the synchrotron and SSC emission while accounting for the two different IC regimes, the Thomson and KN regime. We based our approach on the method proposed by \citet{2009ApJ...703..675N}, which we extended to include additional effects: The maximum electron Lorentz factor $\gamma_\mathrm{max}$ was also computed self-consistently, leading to a realistic estimate of the high-energy cutoff of the synchrotron component, and the attenuation at high-energy due to pair production was included as well. The values of $\gamma_\mathrm{c}$ and $Y(\gamma_\mathrm{c})$ and of the spectral regimes were determined numerically. Synchrotron self-absorption, which is relevant at low radio frequencies, is not yet included. The implementation of the model is computationally efficient and allows a fitting of afterglow data.

We used this model to fit the multi-wavelength afterglow of GW~170817, including radio to X-ray observations up to 400~days and five early-time upper limits. We inferred the model parameters using three different assumptions for the emission: (i) Purely synchrotron, (ii) synchrotron and SSC, assuming that scatterings occur in Thomson regime, and (iii) self-consistent synchrotron and SSC calculation taking the KN regime into account. We obtained excellent fits in the three cases, with similar constraints on the parameters for the purely synchrotron case and the full calculation with SSC in KN regimes, whereas the case in which scattering was assumed to be in the Thomson regime deviated significantly. The SSC emission in the afterglow of GW~170817 is indeed weak due to the strong KN attenuation, which confirms the need to include the self-consistent calculation of the SSC emission in the KN regime in afterglow models. The predicted VHE flux at the peak has a large dispersion, but remains about two orders of magnitude below the upper limit obtained around the peak by H.E.S.S. This shows that the VHE afterglow of GW~170817 was undetectable by current instruments, but that future detections of similar events would break some degeneracies among the model parameters and would provide better constraints on the physics of the relativistic ejecta.

We then studied how these post-merger VHE afterglows may become detectable in the future in the context of the improved sensitivity of the CTA. Beyond the evident effect of a less off-axis viewing angle, we showed that the efficiency of the SSC emission becomes higher in external media with higher densities than GW~170817. For instance, a similar ultra-relativistic jet in a medium with a density of $\sim 1\, \mathrm{cm^{-3}}$ and seen slightly less off-axis than GW~170817 may become detectable by the CTA at 100-400~Mpc.

This bias of VHE afterglow detections in favour of high-density environments offers a new probe of a possible population of fast mergers. With a short merger time, a merger should indeed occur in more central, and denser, regions of its host galaxy. Thus, future detections of these VHE afterglows would not only advance our physical understanding of relativistic jets that are produced after a BNS merger, but would also constrain the physics of the evolution of massive binary systems.

\begin{acknowledgements}

Clément Pellouin acknowledges funding support from the Initiative Physique des Infinis (IPI), a research training program of the Idex SUPER at Sorbonne University. The authors acknowledge the Centre National d’Études Spatiales (CNES) for financial support in this research project. This work made use of the Infinity computing cluster at IAP. The authors thank K. Mooley for providing GW~170817 afterglow observation data at \href{https://github.com/kmooley/GW170817/}{github.com/kmooley/GW170817/}. This work made use of Python 3.8 and of the Python packages \texttt{arviz} \citep{2019JOSS....4.1143K}, \texttt{astropy} \citep{2022ApJ...935..167A}, \texttt{corner} \citep{2016JOSS....1...24F}, \texttt{emcee} \citep{2019JOSS....4.1864F, 2013PASP..125..306F}, \texttt{matplotlib} \citep{2007CSE.....9...90H}, \texttt{multiprocessing}, \texttt{numpy} \citep{2020Natur.585..357H}, \texttt{pandas} \citep{mckinneyprocscipy2010}, \texttt{scipy} \citep{2020NatMe..17..261V}.
\end{acknowledgements}

\bibliographystyle{aa}
\bibliography{biblio}

\appendix

\section{Maximum electron Lorentz factor and synchrotron burnoff limit}\label{ap:gamma_max}

The maximum Lorentz factor $\gamma_\mathrm{max}$ of accelerated electrons is defined in Sect.~\ref{sec:acc_el_b_field} and is reached when the acceleration timescale becomes longer than the cooling timescale, 
\begin{equation}\label{eq:gamma_max_how_to}
    t_{\mathrm{acc}}(\gamma_{\mathrm{max}}) = \mathrm{min} (t_{\mathrm{rad}}(\gamma_{\mathrm{max}});~t_{\mathrm{dyn}})\, ,
\end{equation}
where the dynamical timescale $t_{\mathrm{dyn}}$  and the acceleration timescale $ t_{\mathrm{acc}}$ are defined by Eqs.~(\ref{eq:tdyn}) and~(\ref{eq:tacc_gamma}). From Eqs.~(\ref{eq:trad_gamma}) and (\ref{eq:gamma_c_y_c}), the radiative timescale $t_{\mathrm{rad}}$ is given by
\begin{equation}
    t_{\mathrm{rad}}(\gamma) = \frac{t_{\rm dyn}}{1 + Y(\gamma)} \frac{\gamma_{\rm c}^{\rm syn}}{\gamma}\, .
\end{equation}
Eq.~(\ref{eq:gamma_max_how_to}) can then be rewritten
\begin{equation}\label{eq:to_solve_for_gamma_max}
    \mathrm{min}\left[ 1; \frac{\gamma_{\rm c}^{\rm syn}}{\gamma_{\rm max} (1+Y(\gamma_{\rm max}))} \right] = \widetilde{K} \gamma_{\rm max}\, ,
\end{equation}
with 
\begin{equation}\label{eq:k_tilde}
    \widetilde{K} = K_{\rm acc} \frac{m_{\rm e} c}{eBt_{\rm dyn}}\, .
\end{equation}
Then $\gamma_{\rm max}$ is the solution of 
\begin{equation}
\gamma_{\rm max}^2 [1+Y(\gamma_{\rm max})] = \frac{\gamma_{\rm c}^{\rm syn}}{\widetilde{K}}
\label{eq:gmax_to_solve}
\end{equation}
except in unlikely cases where electrons at $\gamma_\mathrm{max}$ are slow cooling ($\gamma_\mathrm{max}<\gamma_\mathrm{c}$), that lead to $\gamma_{\rm max} = 1/\widetilde{K}$. 

In the current version of the model, we assume \mbox{$\gamma_\mathrm{max}\gg\gamma_\mathrm{m}$} so that the calculation of $\gamma_\mathrm{c}$ and the identification of the radiative regime and associated normalised electron distribution $f(x)$ and Compton parameter $h(x)$ are independent of $\gamma_\mathrm{max}$ (see Sect.~\ref{sec:fgh}). Therefore the maximum Lorentz factor is determined once this radiative regime is identified. We follow an  iterative procedure where the segments of $h(x)$ are explored successively starting from the last segment $x>\widehat{x}_1$ (highest Lorentz factors):
\begin{enumerate}
\item We check if there is a solution of Eq.~(\ref{eq:gmax_to_solve}) in this segment, with the usual approximation for $1+Y$, 
\begin{equation}
\frac{\gamma_{\rm c}^{\rm syn}}{\widetilde{K}} = \left\lbrace\begin{array}{cl}
\gamma_{\rm max}^2 & \mathrm{if}\, Y(\gamma_\mathrm{max}) < 1 \\
\gamma_{\rm max}^2 Y(\gamma_{\rm max}) & \mathrm{if}\, Y(\gamma_\mathrm{max}) > 1 \\
\end{array}\right.\, ,
\label{eq:gmax_to_solve_num}
\end{equation}
where $Y(\gamma_\mathrm{max})$ is computed from Eq.~(\ref{eq:defh}), which leads to
\begin{equation}
Y(\gamma_\mathrm{max}) = 
\left\lbrace\begin{array}{cl}
    A\, \frac{\gamma_\mathrm{m}}{\gamma_\mathrm{c}^\mathrm{syn}}\, \frac{I_2}{I_0}\, 
   h(x_\mathrm{max}) & \mathrm{if}\, \gamma_\mathrm{m}>\gamma_\mathrm{c}\\
       A\, \frac{\gamma_\mathrm{c}^\mathrm{syn}}{\gamma_\mathrm{m}}\, \frac{I_2}{I_0}\, 
   h(x_\mathrm{max}) & \mathrm{if}\, \gamma_\mathrm{m}<\gamma_\mathrm{c}\, \\
   & \mathrm{and}\, Y(\gamma_\mathrm{max})<1\\
       \left(A\, \frac{\gamma_\mathrm{c}^\mathrm{syn}}{\gamma_\mathrm{m}}\, \frac{I_2}{I_0}\, 
   h(x_\mathrm{max})\right)^{1/3} & \mathrm{if}\, \gamma_\mathrm{m}<\gamma_\mathrm{c} \, \\
   & \mathrm{and}\, Y(\gamma_\mathrm{max})>1\\
\end{array}\right. \, .
\label{eq:Ymax_all_cases}
\end{equation}
The pre-factor $A$ is the same as in Eq.~(\ref{eq:Yc_all_cases}). 
\item If there is a solution in the considered segment, we stop the iterative procedure and keep the following value for the maximum electron Lorentz factor:
\begin{itemize}
\item the solution $\gamma_\mathrm{max}$ found in the segment if $\gamma_\mathrm{max} > \gamma_\mathrm{c}$ (fast cooling);
\item the solution $\gamma_\mathrm{max}=1/\widetilde{K}$ otherwise (slow cooling).
\end{itemize}
\item If there is no solution in the considered segment:
\begin{itemize}
\item If the next segment is still at least partially in fast-cooling regime (i.e. the upper bound is above $\gamma_\mathrm{c}$), we start a new iteration at step 1. using this new segment;
\item otherwise we stop the iterative procedure and keep for $\gamma_\mathrm{max}$ the slow-cooling solution $\gamma_\mathrm{max}=1/\widetilde{K}$. 
\end{itemize}
\end{enumerate}

In most cases, the maximum electron Lorentz factor is large enough to be in fast-cooling regime ($\gamma_\mathrm{max}>\gamma_\mathrm{c}$ so that $t_\mathrm{rad}(\gamma_\mathrm{max})<t_\mathrm{dyn}$) with a negligible IC cooling ($Y(\gamma_\mathrm{max})<1$). In this case, the maximum Lorentz factor $\gamma_\mathrm{max}=\sqrt{{\gamma_{\rm c}^{\rm syn}}/{\widetilde{K}}}$ leads to the usual synchrotron burnoff limit \citep[see e.g.][]{2010ApJ...718L..63P} for the high-frequency cutoff of the synchrotron spectrum in the comoving frame $ h\nu_{\mathrm{max}} =h\nu_\mathrm{syn}(\gamma_\mathrm{max})=K_2 B {\gamma_{\rm c}^{\rm syn}}/{\widetilde{K}}$:
\begin{equation}
    h\nu_{\mathrm{max}} 
    =   \frac{1}{K_{P_\mathrm{max}}\frac{ 3 h e^2 }{ \sigma_\mathrm{T} m_\mathrm{e} c} \, K_\mathrm{acc} } 
    =  \frac{160\, \mathrm{MeV}}{K_{P_\mathrm{max}}\, K_\mathrm{acc} }\, .
\end{equation}
For the best-fit parameters of the afterglow of GW~170817 presented in Sect.~\ref{sec:170817}, the Lorentz factor of the core jet at $t_\mathrm{peak}$ is $\Gamma \simeq 10$. The limit is thus expected at $\sim 1~\mathrm{GeV}$ in the observer frame, as seen in Fig.~\ref{fig:sp_posterior}.

\section{Radiative regimes and corresponding electron distributions}
\label{sec:sol_fgh}

\begin{figure*}
\vspace*{-4ex}

\begin{center}
\begin{tabular}{cc}
 \includegraphics[width=0.49\textwidth]{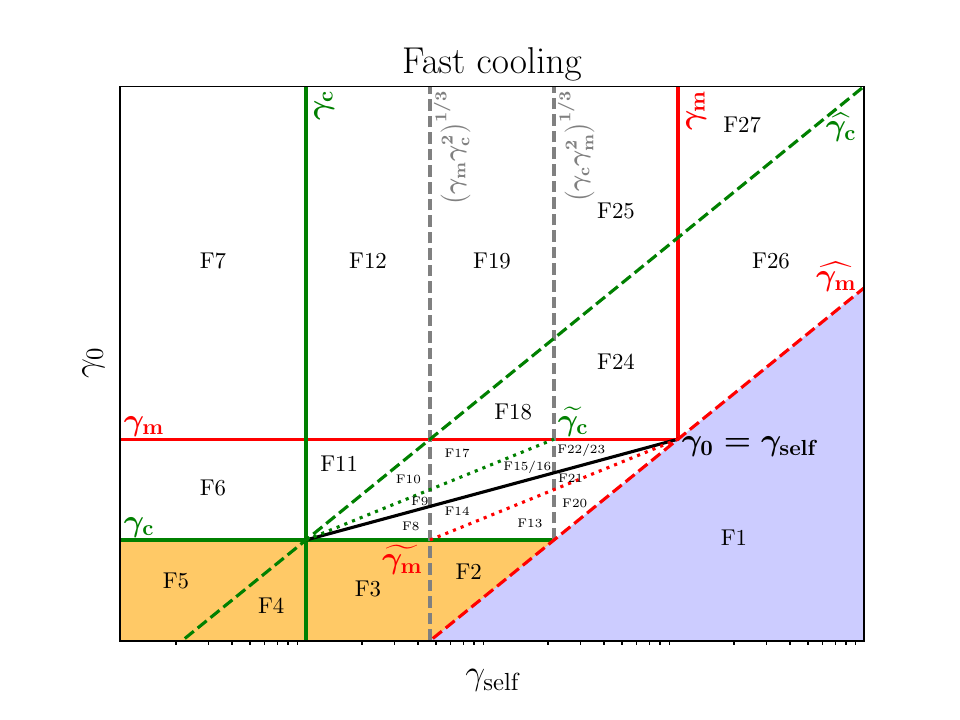}
 &
 \includegraphics[width=0.49\textwidth]{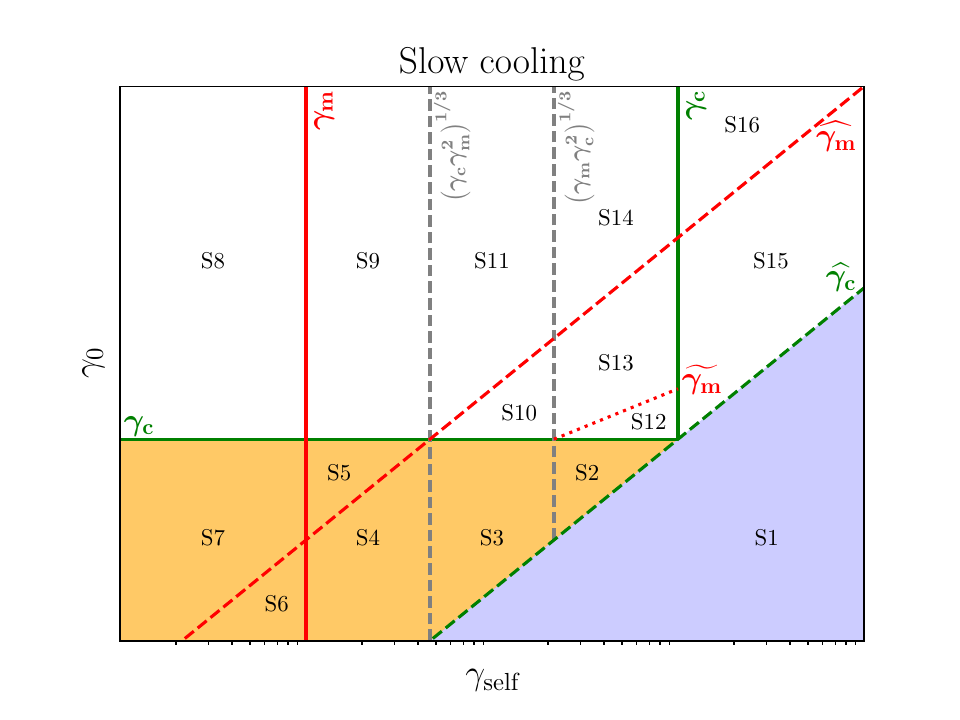}
 \\
\end{tabular}
\end{center}
\vspace*{-4ex}

\caption{Definition of all possible cases for the radiative regime and the distribution of electrons in the plane $\gamma_0$ vs $\gamma_\mathrm{self}$. The Lorentz factors $\gamma_\mathrm{m}$  and $\gamma_\mathrm{c}$ are indicated by red and green solid lines either in the fast-cooling regime with $\gamma_\mathrm{m}>\gamma_\mathrm{c}$ (left) or in the slow-cooling regime with $\gamma_\mathrm{m}< \gamma_\mathrm{c}$ (right). The black solid line in the left panel corresponds to  $\gamma_0=\gamma_\mathrm{self}$. The two  Lorentz factors $\widehat{\gamma}_\mathrm{m}$ (red) and $\widehat{\gamma}_\mathrm{c}$ (green) are plotted in dashed lines. In some regions, additional Lorentz factors relevant to identify the radiative regime are also plotted: $\widetilde{\gamma}_\mathrm{m}$ (red dotted line) and $\widetilde{\gamma}_\mathrm{c}$ (green dotted line). The regions where there is no impact of the SSC cooling on the distribution of electrons are shaded in orange ($\widehat{\gamma}_\mathrm{m,c}<\gamma_0 < \gamma_\mathrm{c}$) and blue ($\gamma_0<\widehat{\gamma}_\mathrm{m,c}$). This diagram allows to identify all possible orderings of $\gamma_0$, $\gamma_\mathrm{m}$, $\gamma_\mathrm{c}$, $\widehat{\gamma}_\mathrm{m}$, $\widehat{\gamma}_\mathrm{m}$ defining all possible radiative regimes.
The corresponding limits for $\gamma_\mathrm{self}$ are indicated by vertical black dashed lines. All the possible cases are listed in Table~\ref{tab:cases_F} (fast cooling) and~\ref{tab:cases_S} (slow cooling). In practice, several cases have the same breaks and slopes for $f(x)$ and $h(x)$ and can therefore be treated simultaneously in the numerical implementation of the model. This allows to group
$\left(F1,F2,F3,F4,F5\right)$,
$\left(F8,F13,F14\right)$,
$\left(F9,F15,F16\right)$,
$\left(F10,F17\right)$,
$\left(F20,F21\right)$,
$\left(F22,F23\right)$,
$\left(S1,S2,S3,S4,S5,S6,S7\right)$,
$\left(S8,S9\right)$,
$\left(S10,S12,S13\right)$
and $\left(S11,S14\right)$.
}
\label{fig:all_cases}
\end{figure*}

The IC cooling impacts the distribution of electrons if $\max{\left(\gamma_\mathrm{c};\tilde{\gamma}_\mathrm{p}\right)}<\gamma<\gamma_0$ (Eq.~(\ref{eq:impactSSC})).
As
\begin{equation}\label{eq:gamma_hat_gamma_self}
    \widehat{\gamma}=\frac{\gamma_\mathrm{self}^3}{\gamma^2}\, ,
\end{equation}
all possibles cases regarding the radiative regime and the distribution of electrons can be identified in the $\left(\gamma_\mathrm{self}; \gamma_0\right)$ plane shown in Fig.~\ref{fig:all_cases}. As an electron of Lorentz factor $\gamma$ can upscatter its own synchrotron photon only if $\gamma\le\widehat{\gamma}$, that is, if $\gamma\le\gamma_\mathrm{self}$, the x-axis quantifies the importance of the KN suppression: the lower $\gamma_\mathrm{self}$ (left side of the diagram), the more electrons are affected by KN effects. The y-axis quantifies the importance of IC cooling: the IC power of all electrons with $\gamma>\gamma_0$ is negligible compared to the synchrotron power. Therefore, the lower $\gamma_0$ (bottom of the diagram), the more electrons have a negligible IC cooling. We list in Tables~\ref{tab:cases_F} (fast cooling) and~\ref{tab:cases_S} (slow cooling) all these cases and provide the corresponding self-consistent solution (breaks and slopes) for the normalised distribution of electrons $f(x)$ (Eq.~(\ref{eq:f_x})) and the normalised Compton parameter $h(x)$ (Eq.~(\ref{eq:defh})), as obtained following the detailed method described in \cite{2009ApJ...703..675N}. The break corresponding to the peak of the synchrotron spectrum is highlighted in red. We also provide in each case the expression of $\gamma_0$ obtained from the relation $Y(\gamma_\mathrm{c})=h(x_\mathrm{c})/h(x_0)$ (Eq.~(\ref{eq:Yc_xc_x0})), as needed in the iterative procedure to determine $Y(\gamma_\mathrm{c})$: see Sect.~\ref{sec:fgh}.

The distribution of electrons is unaffected by the SSC cooling when $\gamma_0<\gamma_\mathrm{c}$ or $\gamma_\mathrm{c}<\gamma_0 < \widehat{\gamma}_\mathrm{m,c}$. These two regions are shaded in orange and blue in Fig.~\ref{fig:all_cases}. When $\gamma_0 <\widehat{\gamma}_\mathrm{m,c}$ (cases F1 and S1), the IC cooling is negligible for all electrons and $\gamma_0$ is undefined. The solution is provided by the purely synchrotron case (Sect.~\ref{sec:pureSYN}). When $\widehat{\gamma}_\mathrm{m,c}<\gamma_0 < \gamma_\mathrm{c}$ (cases F2 to F5 and S2 to S6), the normalised electron distribution $f(x)$ is also the same as in the purely synchrotron case but $\gamma_0$ is defined and its expression depends on the considered regime.

In all other cases, the SSC cooling affects the distribution of electrons and the synchrotron spectrum, with additional breaks ($N\ge 3$), as listed in Tables~\ref{tab:cases_F} and~\ref{tab:cases_S}. There are more subcases in fast cooling. The most relevant ones for GRB afterglows are discussed in \citet{2009ApJ...703..675N}: their case I with its two subcases correspond to F26 and F27; their case IIa to F20-F21, their case IIb to F22-F23; their cases IIc and III to some subcases of F24 and F25.

Cases F24 and F25 are indeed more complex and require a specific treatment (see Sect.~3.2 \citealt{2009ApJ...703..675N}), as a large number of breaks can appear in the distribution, that must be found iteratively. These two cases correspond to $\gamma_\mathrm{c} < \widehat{\gamma}_\mathrm{m} < \gamma_\mathrm{self} < \gamma_\mathrm{m} < \widehat{\gamma}_\mathrm{c}$. New breaks will appear in the spectral shape $f(x)$ every time that $^{(2n)}\widehat{\gamma}_\mathrm{m} = \widehat{\gamma}_\mathrm{c}$, where using Eq.~(\ref{eq:gamma_hat_gamma_self}), we define the successive $^{(n)}\widehat{\gamma}$ as
\begin{eqnarray}
    ^{(n)}\widehat{\gamma} & = & \left\lbrace\begin{array}{cc}
    \frac{\gamma_\mathrm{self}^{2^{n}+1}}{\gamma^{2^n}} & \mathrm{if~} n \mathrm{~is~odd}\\
    \frac{\gamma^{2^n}}{\gamma_\mathrm{self}^{2^{n}-1}} & \mathrm{if~} n \mathrm{~is~even}\\
    \end{array}\right.\label{eq:gamma_hat_nth}
    \, .
\end{eqnarray}
This condition is equivalent to $\gamma_\mathrm{self} = \gamma_\mathrm{self, cr, 2n}$, where
\begin{equation}
    \gamma_{\mathrm{self,cr,2n}} = \left( \gamma_\mathrm{m}^{2^{2n-1}} \gamma_\mathrm{c} \right)^{\frac{1}{2^{2n-1}+1}}\, ,
\end{equation}
with $n \geq 1$.
For $n=1$, the first value $\gamma_\mathrm{self, cr, 2}=\left( \gamma_\mathrm{m}^{2} \gamma_\mathrm{c} \right)^{\frac{1}{3}}$ is plotted in Fig.~\ref{fig:all_cases} and is the frontier at low $\gamma_\mathrm{self}$ of the cases F24 and F25. On the other hand $\gamma_\mathrm{self, cr, 2n}$ tends to $ \gamma_\mathrm{m}$ when $n$ tends to infinity, that is, the frontier at high $\gamma_\mathrm{self}$. Then, the values of $\gamma_{\mathrm{self,cr,2n}}$ for $n>1$ divide F24 and F25 in many sub-regions. We define the column $n$ by the condition $\gamma_\mathrm{self,cr, 2n} < \gamma_\mathrm{self} < \gamma_\mathrm{self,cr,2n+2}$. In practice, $\gamma_\mathrm{self,cr, 2n}$ rapidly converges, and the spectral shapes in these narrower columns rapidly become closer and closer to each other. This allows numerically to not distinguish columns above a certain $n_\mathrm{max}$. We use $n_\mathrm{max}=7$.

In each column $n$, cases F24 and F25 are then divided in several subcases depending on the relative position of $\gamma_0$ and 
$\widetilde{\gamma}_\mathrm{c}$,
$\widehat{\gamma}_\mathrm{c}$
and  
$^{(2k)}\widehat{\gamma}_\mathrm{m}$ with $k\ge 1$. In each subcase, the list of breaks and slopes can be determined for $f(x)$ and $h(x)$, as well as 
the position of the peak of the synchrotron spectrum. 
We show for illustration  the results in the two first columns in Table~\ref{tab:cases_F2425_12}.

\begin{table*}
    \caption{Fast cooling: list of all possible radiative regimes and corresponding electron distribution.}
    \vspace*{-5.5ex}
    
\begin{center}
\resizebox{\linewidth}{!}{\begin{minipage}{26.5cm}
\begin{tabular}{|l|}
\hline
\\
\textbf{Case F1}\\
\\
\hline
\end{tabular}
\begin{tabular}{lc|ccccccccccc}
\hline
& breaks & & $\gamma_{\mathrm{c}}$ & & \red{$\gamma_{\mathrm{m}}$} & &  $\widehat{\gamma_{\mathrm{m}}}$ & & $\widehat{\gamma_{\mathrm{c}}}$ &  \\
\hline
f & slopes & & | & $2$ & \red{|} & \multicolumn{5}{c}{$p+1$}  \\
\hline
h & slopes & & \multicolumn{4}{c}{$0$} & | & $1/2$ & | & $4/3$ \\
\hline
\end{tabular}
\begin{tabular}{|l|}
\hline
\\
$\gamma_0$ undefined\\
\\
\hline
\end{tabular}

\begin{tabular}{|l|}
\hline
\\
\textbf{Case F2}\\
\\
\hline
\end{tabular}
\begin{tabular}{lc|ccccccccccc}
\hline
 & breaks & & $\widehat{\gamma_{\mathrm{m}}}$ & & \blue{$\gamma_0$} & & $\gamma_{\mathrm{c}}$ & & \red{$\gamma_{\mathrm{m}}$} & & $\widehat{\gamma_{\mathrm{c}}}$ &  \\
\hline
f & slopes & & & & & & | & $2$ & \red{|} & \multicolumn{3}{c}{$p+1$} \\
\hline
h & slopes & $0$ & | & \multicolumn{7}{c}{$1/2$} & | & $4/3$ \\
\hline
\end{tabular}
\begin{tabular}{|l|}
\hline
\\
$\gamma_0 = \gamma_{\mathrm{c}} Y_{\mathrm{c}}^{2}$\\
\\
\hline
\end{tabular}

\begin{tabular}{|l|}
\hline
\\
\textbf{Case F3}\\
\\
\hline
\end{tabular}
\begin{tabular}{lc|ccccccccccc}
\hline
& breaks & & $\widehat{\gamma_{\mathrm{m}}}$ & & \blue{$\gamma_0$} & & $\gamma_{\mathrm{c}}$ & & $\widehat{\gamma_{\mathrm{c}}}$ & & \red{$\gamma_{\mathrm{m}}$} &  \\
\hline
f & slopes & & & & & & | & \multicolumn{3}{c}{$2$} & \red{|} & $p+1$ \\
\hline
h & slopes & $0$ & | & \multicolumn{5}{c}{$1/2$} & | & \multicolumn{3}{c}{$4/3$} \\
\hline
\end{tabular}
\begin{tabular}{|l|}
\hline
\\
$\gamma_0 = \gamma_{\mathrm{c}} Y_{\mathrm{c}}^{2}$\\
\\
\hline
\end{tabular}

\begin{tabular}{|l|}
\hline
\\
\textbf{Case F4}\\
\\
\hline
\end{tabular}
\begin{tabular}{lc|ccccccccccc}
\hline
& breaks & & $\widehat{\gamma_{\mathrm{m}}}$ & & \blue{$\gamma_0$} & & $\widehat{\gamma_{\mathrm{c}}}$ & & $\gamma_{\mathrm{c}}$ & & \red{$\gamma_{\mathrm{m}}$} &  \\
\hline
f & slopes & & & & & & & & | & $2$ & \red{|} & $p+1$ \\
\hline
h & slopes & $0$ & | & \multicolumn{3}{c}{$1/2$} & | & \multicolumn{5}{c}{$4/3$} \\
\hline
\end{tabular}
\begin{tabular}{|l|}
\hline
\\
$\gamma_0 = \gamma_{\mathrm{c}}^{8/3} \widehat{\gamma_{\mathrm{c}}}^{-5/3} Y_{\mathrm{c}}^{2}$\\
\\
\hline
\end{tabular}

\begin{tabular}{|l|}
\hline
\\
\textbf{Case F5}\\
\\
\hline
\end{tabular}
\begin{tabular}{lc|ccccccccccc}
\hline
& breaks & & $\widehat{\gamma_{\mathrm{m}}}$ & & $\widehat{\gamma_{\mathrm{c}}}$ & & \blue{$\gamma_0$} & & $\gamma_{\mathrm{c}}$ & & \red{$\gamma_{\mathrm{m}}$} &  \\
\hline
f & slopes & & & & & & & & | & $2$ & \red{|} & $p+1$ \\
\hline
h & slopes & $0$ & | & $1/2$ & | & \multicolumn{7}{c}{$4/3$} \\
\hline
\end{tabular}
\begin{tabular}{|l|}
\hline
\\
$\gamma_0 = \gamma_{\mathrm{c}} Y_{\mathrm{c}}^{3/4}$\\
\\
\hline
\end{tabular}

\begin{tabular}{|l|}
\hline
\\
\textbf{Case F6}\\
\\
\hline
\end{tabular}
\begin{tabular}{lc|ccccccccccccc}
\hline
& breaks & & $\widehat{\gamma_{\mathrm{m}}}$ & & $\widehat{\gamma_0}$ & & $\widehat{\gamma_{\mathrm{c}}}$ & & $\gamma_{\mathrm{c}}$ & & $\gamma_0$ & & \red{$\gamma_{\mathrm{m}}$} & \\
\hline
f & slopes & & & & & & & & | & $2/3$ & | & $2$ & \red{|} & $p+1$ \\
\hline
h & slopes & $0$ & | & $1/2$ & | & $7/6$ & | &\multicolumn{7}{c}{$4/3$} \\
\hline
\end{tabular}
\begin{tabular}{|l|}
\hline
\\
$\gamma_0 = \gamma_{\mathrm{c}} Y_{\mathrm{c}}^{3/4}$\\
\\
\hline
\end{tabular}

\begin{tabular}{|l|}
\hline
\\
\textbf{Case F7}\\
\\
\hline
\end{tabular}
\begin{tabular}{lc|ccccccccccccc}
\hline
& breaks & & $\widehat{\gamma_0}$ & & $\widehat{\gamma_{\mathrm{m}}}$ & & $\widehat{\gamma_{\mathrm{c}}}$ & & $\gamma_{\mathrm{c}}$ & & $\gamma_{\mathrm{m}}$ & & \red{$\gamma_0$} & \\
\hline
f & slopes & & & & & & & & | & $2/3$ & | & $ p-1/3$ & \red{|} & $p+1$ \\
\hline
h & slopes & $0$ & | & $(10-3p)/6$ & | & $7/6$ & | & \multicolumn{7}{c}{$4/3$} \\
\hline
\end{tabular}
\begin{tabular}{|l|}
\hline
\\
$\gamma_0 = \gamma_{\mathrm{c}} Y_{\mathrm{c}}^{3/4}$\\
\\
\hline
\end{tabular}

\begin{tabular}{|l|}
\hline
\\
\textbf{Case F8}\\
\\
\hline
\end{tabular}
\begin{tabular}{lc|ccccccccccccc}
\hline
& breaks & & $\widehat{\gamma_{\mathrm{m}}}$ & & $\gamma_{\mathrm{c}}$ & & $\gamma_0$ & & $\widehat{\gamma_0}$ & & $\widehat{\gamma_{\mathrm{c}}}$ & & \red{$\gamma_{\mathrm{m}}$} & \\
\hline
f & slopes & & & & | & $3/2$ & | & \multicolumn{5}{c}{$2$} & \red{|} & $p+1$ \\
\hline
h & slopes & $0$ & | & \multicolumn{5}{c}{$1/2$} & | & $3/4$ & | &\multicolumn{3}{c}{$4/3$} \\
\hline
\end{tabular}
\begin{tabular}{|l|}
\hline
\\
$\gamma_0 = \gamma_{\mathrm{c}} Y_{\mathrm{c}}^2$\\
\\
\hline
\end{tabular}

\begin{tabular}{|l|}
\hline
\\
\textbf{Case F9}\\
\\
\hline
\end{tabular}
\begin{tabular}{lc|ccccccccccccccc}
\hline
& breaks & & $\widehat{\gamma_{\mathrm{m}}}$ & & $\gamma_{\mathrm{c}}$ & & $\widehat{\gamma_0}$ & & $\gamma_0$ & & $\widehat{\widehat{\gamma_0}}$ & & $\widehat{\gamma_{\mathrm{c}}}$ & & \red{$\gamma_{\mathrm{m}}$} &  \\
\hline
f & slopes & & & & | & $3/2$ & | & $1$ & | & \multicolumn{5}{c}{$2$} & \red{|} & $p+1$ \\
\hline
h & slopes & $0$ & | & \multicolumn{3}{c}{$1/2$} & | & \multicolumn{3}{c}{$1$} & | & $3/4$ & | & \multicolumn{3}{c}{$4/3$} \\
\hline
\end{tabular}
\begin{tabular}{|l|}
\hline
\\
$\gamma_0 = \gamma_{\mathrm{c}}^{1/2} \widehat{\gamma_0}^{1/2} Y_{\mathrm{c}} = \gamma_{\mathrm{c}}^{1/4} \gamma_{\mathrm{self}}^{3/4} Y_{\mathrm{c}}^{1/2}$\\
\\
\hline
\end{tabular}

\begin{tabular}{|l|}
\hline
\\
\textbf{Case F10}\\
\\
\hline
\end{tabular}
\begin{tabular}{lc|ccccccccccccc}
\hline
& breaks & & $\widehat{\gamma_{\mathrm{m}}}$ & & $\widehat{\gamma_0}$ & & $\gamma_{\mathrm{c}}$ & & $\gamma_0$ & & $\widehat{\gamma_{\mathrm{c}}}$ & & \red{$\gamma_{\mathrm{m}}$} & \\
\hline
f & slopes & & & & & & | & $1$ & | & \multicolumn{3}{c}{$2$} & \red{|} & $p+1$ \\
\hline
h & slopes &  $0$ & | & $1/2$ & | & \multicolumn{5}{c}{$1$} & | &\multicolumn{3}{c}{$4/3$} \\
\hline
\end{tabular}
\begin{tabular}{|l|}
\hline
\\
$\gamma_0 = \gamma_{\mathrm{c}} Y_{\mathrm{c}}$\\
\\
\hline
\end{tabular}

\begin{tabular}{|l|}
\hline
\\
\textbf{Case F11}\\
\\
\hline
\end{tabular}
\begin{tabular}{lc|ccccccccccccccc}
\hline
& breaks & & $\widehat{\gamma_{\mathrm{m}}}$ & & $\widehat{\gamma_0}$ & & $\widehat{\widehat{\gamma_{\mathrm{c}}}}$ & & $\gamma_{\mathrm{c}}$ & & $\widehat{\gamma_{\mathrm{c}}}$ & & $\gamma_0$ & & \red{$\gamma_{\mathrm{m}}$} &  \\
\hline
f & slopes & & & & & & & & | & $1$ & | & $2/3$ & | & $2$ & \red{|} & $p+1$ \\
\hline
h & slopes & $0$ & | & $1/2$ & | & $7/6$ & | & \multicolumn{3}{c}{$1$} & | &\multicolumn{5}{c}{$4/3$} \\
\hline
\end{tabular}
\begin{tabular}{|l|}
\hline
\\
$\gamma_0 = \gamma_{\mathrm{c}}^{3/4} \widehat{\gamma_{\mathrm{c}}}^{1/4} Y_{\mathrm{c}}^{3/4}$\\
\\
\hline
\end{tabular}

\begin{tabular}{|l|}
\hline
\\
\textbf{Case F12}\\
\\
\hline
\end{tabular}
\begin{tabular}{lc|ccccccccccccccc}
\hline
& breaks & & $\widehat{\gamma_0}$ & & $\widehat{\gamma_{\mathrm{m}}}$ & & $\widehat{\widehat{\gamma_{\mathrm{c}}}}$ & & $\gamma_{\mathrm{c}}$ & & $\widehat{\gamma_{\mathrm{c}}}$ & & $\gamma_{\mathrm{m}}$ & & \red{$\gamma_0$} &  \\
\hline
f &  slopes & & & & & & & & | & $1$ & | & $2/3$ & | & $p-1/3$ & \red{|} & $p+1$ \\
\hline
h & slopes & $0$ & | & $(10-3p)/6$ & | & $7/6$ & | & \multicolumn{3}{c}{$1$} & | &\multicolumn{5}{c}{$4/3$} \\
\hline
\end{tabular}
\begin{tabular}{|l|}
\hline
\\
$\gamma_0 = \gamma_{\mathrm{c}}^{3/4} \widehat{\gamma_{\mathrm{c}}}^{1/4} Y_{\mathrm{c}}^{3/4}$\\
\\
\hline
\end{tabular}

\begin{tabular}{|l|}
\hline
\\
\textbf{Case F13}\\
\\
\hline
\end{tabular}
\begin{tabular}{lc|ccccccccccccc}
\hline
& breaks & & $\widehat{\gamma_{\mathrm{m}}}$ & & $\gamma_{\mathrm{c}}$ & & $\gamma_0$ & & $\widehat{\gamma_0}$ & & \red{$\gamma_{\mathrm{m}}$} & & $\widehat{\gamma_{\mathrm{c}}}$ & \\
\hline
f & slopes & & & & | & $3/2$ & | & \multicolumn{3}{c}{$2$} & \red{|} & \multicolumn{3}{c}{$p+1$} \\
\hline
h & slopes & $0$ & | & \multicolumn{5}{c}{$1/2$} & | & \multicolumn{3}{c}{$3/4$} & | & $4/3$ \\
\hline
\end{tabular}
\begin{tabular}{|l|}
\hline
\\
$\gamma_0 = \gamma_{\mathrm{c}} Y_{\mathrm{c}}^2$\\
\\
\hline
\end{tabular}

\begin{tabular}{|l|}
\hline
\\
\textbf{Case F14}\\
\\
\hline
\end{tabular}
\begin{tabular}{lc|ccccccccccccc}
\hline
& breaks & & $\widehat{\gamma_{\mathrm{m}}}$ & & $\gamma_{\mathrm{c}}$ & & $\gamma_0$ & & \red{$\gamma_{\mathrm{m}}$} & & $\widehat{\gamma_0}$ & & $\widehat{\gamma_{\mathrm{c}}}$ & \\
\hline
f & slopes & & & & | & $3/2$ & | & $2$ & \red{|} & \multicolumn{5}{c}{$p+1$} \\
\hline
h &slopes &  $0$ & | & \multicolumn{7}{c}{$1/2$} & | & $3/4$ & | & $4/3$ \\
\hline
\end{tabular}
\begin{tabular}{|l|}
\hline
\\
$\gamma_0 = \gamma_{\mathrm{c}} Y_{\mathrm{c}}^2$\\
\\
\hline
\end{tabular}

\begin{tabular}{|l|}
\hline
\\
\textbf{Case F15}\\
\\
\hline
\end{tabular}
\begin{tabular}{lc|ccccccccccccccc}
\hline
& breaks & & $\widehat{\gamma_{\mathrm{m}}}$ & & $\gamma_{\mathrm{c}}$ & & $\widehat{\gamma_0}$ & & $\gamma_0$ & & $\widehat{\widehat{\gamma_0}}$ & & \red{$\gamma_{\mathrm{m}}$} & & $\widehat{\gamma_{\mathrm{c}}}$ &  \\
\hline
f & slopes & & & & | & $3/2$ & | & $1$ & | & \multicolumn{3}{c}{$2$} & \red{|} & \multicolumn{3}{c}{$p+1$} \\
\hline
h & slopes & $0$ & | & \multicolumn{3}{c}{$1/2$} & | & \multicolumn{3}{c}{$1$} & | & \multicolumn{3}{c}{$3/4$} & | & $4/3$ \\
\hline
\end{tabular}
\begin{tabular}{|l|}
\hline
\\
$\gamma_0 = \gamma_{\mathrm{c}}^{1/2} \widehat{\gamma_0}^{1/2} Y_{\mathrm{c}} = \gamma_{\mathrm{c}}^{1/4} \gamma_{\mathrm{self}}^{3/4} Y_{\mathrm{c}}^{1/2}$\\
\\
\hline
\end{tabular}

\begin{tabular}{|l|}
\hline
\\
\textbf{Case F16}\\
\\
\hline
\end{tabular}
\begin{tabular}{lc|ccccccccccccccc}
\hline
& breaks & & $\widehat{\gamma_{\mathrm{m}}}$ & & $\gamma_{\mathrm{c}}$ & & $\widehat{\gamma_0}$ & & $\gamma_0$ & & \red{$\gamma_{\mathrm{m}}$} & & $\widehat{\widehat{\gamma_0}}$ & & $\widehat{\gamma_{\mathrm{c}}}$ &  \\
\hline
f & slopes & & & & | & $3/2$ & | & $1$ & | & $2$ & \red{|} & \multicolumn{5}{c}{$p+1$} \\
\hline
h & slopes & $0$ & | & \multicolumn{3}{c}{$1/2$} & | & \multicolumn{5}{c}{$1$} & | & $3/4$ & | & $4/3$ \\
\hline
\end{tabular}
\begin{tabular}{|l|}
\hline
\\
$\gamma_0 = \gamma_{\mathrm{c}}^{1/2} \widehat{\gamma_0}^{1/2} Y_{\mathrm{c}} = \gamma_{\mathrm{c}}^{1/4} \gamma_{\mathrm{self}}^{3/4} Y_{\mathrm{c}}^{1/2}$\\
\\
\hline
\end{tabular}

\begin{tabular}{|l|}
\hline
\\
\textbf{Case F17}\\
\\
\hline
\end{tabular}
\begin{tabular}{lc|ccccccccccccc}
\hline
& breaks & & $\widehat{\gamma_{\mathrm{m}}}$ & & $\widehat{\gamma_0}$ & & $\gamma_{\mathrm{c}}$ & & $\gamma_0$ & & \red{$\gamma_{\mathrm{m}}$} & & $\widehat{\gamma_{\mathrm{c}}}$ & \\
\hline
f & slopes & & & & & & | & $1$ & | & $2$ & \red{|} & \multicolumn{3}{c}{$p+1$} \\
\hline
h & slopes & $0$ & | & $1/2$ & | & \multicolumn{7}{c}{$1$} & | & $4/3$ \\
\hline
\end{tabular}
\begin{tabular}{|l|}
\hline
\\
$\gamma_0 = \gamma_{\mathrm{c}} Y_{\mathrm{c}}$\\
\\
\hline
\end{tabular}

\begin{tabular}{|l|}
\hline
\\
\textbf{Case F18}\\
\\
\hline
\end{tabular}
\begin{tabular}{lc|ccccccccccccc}
\hline
& breaks & & $\widehat{\gamma_0}$ & & $\widehat{\gamma_{\mathrm{m}}}$ & & $\gamma_{\mathrm{c}}$ & & $\gamma_{\mathrm{m}}$ & & \red{$\gamma_0$} & & $\widehat{\gamma_{\mathrm{c}}}$ & \\
\hline
f & slopes & & & & & & | & $1$ & | & $p$ & \red{|} & \multicolumn{3}{c}{$p+1$} \\
\hline
h & slopes & $0$ & | & $(3-p)/2$ & | & \multicolumn{7}{c}{$1$} & | & $4/3$ \\
\hline
\end{tabular}
\begin{tabular}{|l|}
\hline
\\
$\gamma_0 = \gamma_{\mathrm{c}} Y_{\mathrm{c}}$\\
\\
\hline
\end{tabular}

\begin{tabular}{|l|}
\hline
\\
\textbf{Case F19}\\
\\
\hline
\end{tabular}
\begin{tabular}{lc|ccccccccccccccc}
\hline
& breaks & & $\widehat{\gamma_0}$ & & $\widehat{\widehat{\gamma_{\mathrm{c}}}}$ & & $\widehat{\gamma_{\mathrm{m}}}$ & & $\gamma_{\mathrm{c}}$ & & $\gamma_{\mathrm{m}}$ & & $\widehat{\gamma_{\mathrm{c}}}$ & & \red{$\gamma_0$} &  \\
\hline
f & slopes & & & & & & & & | & $1$ & | & $p$ & | & $p-1/3$ & \red{|} & $p+1$ \\
\hline
h & slopes & $0$ & | & $(10-3p)/6$ & | & $(3-p)/2$ & | & \multicolumn{5}{c}{$1$} & | &\multicolumn{3}{c}{$4/3$} \\
\hline
\end{tabular}
\begin{tabular}{|l|}
\hline
\\
$\gamma_0 = \gamma_{\mathrm{c}}^{3/4} \widehat{\gamma_{\mathrm{c}}}^{1/4} Y_{\mathrm{c}}^{3/4}$\\
\\
\hline
\end{tabular}

\begin{tabular}{|l|}
\hline
\\
\textbf{Case F20}\\
\\
\hline
\end{tabular}
\begin{tabular}{lc|ccccccccccccccc}
\hline
& breaks & & $\gamma_{\mathrm{c}}$ & & $\widehat{\gamma_{\mathrm{m}}}$ & & $\gamma_0$ & & $\widehat{\gamma_0}$ & & \red{$\gamma_{\mathrm{m}}$} & & $\widehat{\widehat{\gamma_{\mathrm{m}}}}$ & & $\widehat{\gamma_{\mathrm{c}}}$ &  \\
\hline
f & slopes & & | & $2$ & | & $3/2$ & | & \multicolumn{3}{c}{$2$} & \red{|} & \multicolumn{5}{c}{$p+1$} \\
\hline
h & slopes & \multicolumn{3}{c}{$0$} & | & \multicolumn{3}{c}{$1/2$} & | & \multicolumn{3}{c}{$3/4$} & | & $1/2$ & | & $4/3$ \\
\hline
\end{tabular}
\begin{tabular}{|l|}
\hline
\\
$\gamma_0 = \widehat{\gamma_{\mathrm{m}}} Y_{\mathrm{c}}^2$\\
\\
\hline
\end{tabular}

\begin{tabular}{|l|}
\hline
\\
\textbf{Case F21}\\
\\
\hline
\end{tabular}
\begin{tabular}{lc|ccccccccccccccc}
\hline
& breaks & & $\gamma_{\mathrm{c}}$ & & $\widehat{\gamma_{\mathrm{m}}}$ & & $\gamma_0$ & & \red{$\gamma_{\mathrm{m}}$} & & $\widehat{\gamma_0}$ & & $\widehat{\widehat{\gamma_{\mathrm{m}}}}$ & & $\widehat{\gamma_{\mathrm{c}}}$ &  \\
\hline
f & slopes & & | & $2$ & | & $3/2$ & | & $2$ & \red{|} & \multicolumn{7}{c}{$p+1$} \\
\hline
h & slopes & \multicolumn{3}{c}{$0$} & | & \multicolumn{5}{c}{$1/2$} & | & $3/4$ & | & $1/2$ & | & $4/3$ \\
\hline
\end{tabular}
\begin{tabular}{|l|}
\hline
\\
$\gamma_0 = \widehat{\gamma_{\mathrm{m}}} Y_{\mathrm{c}}^2$\\
\\
\hline
\end{tabular}

\begin{tabular}{|l|}
\hline
\\
\textbf{Case F22}\\
\\
\hline
\end{tabular}
\begin{tabular}{lc|ccccccccccccccccc}
\hline
& breaks & & $\gamma_{\mathrm{c}}$ & & $\widehat{\gamma_{\mathrm{m}}}$ & & $\widehat{\gamma_0}$ & & $\gamma_0$ & & $\widehat{\widehat{\gamma_0}}$ & & \red{$\gamma_{\mathrm{m}}$} & & $\widehat{\widehat{\gamma_{\mathrm{m}}}}$ & & $\widehat{\gamma_{\mathrm{c}}}$ &  \\
\hline
f & slopes & & | & $2$ & | & $3/2$ & | & $1$ & | & \multicolumn{3}{c}{$2$} & \red{|} & \multicolumn{5}{c}{$p+1$} \\
\hline
h & slopes & \multicolumn{3}{c}{$0$} & | & $1/2$ & | & \multicolumn{3}{c}{$1$} & | & \multicolumn{3}{c}{$3/4$} & | & $1/2$ & | & $4/3$ \\
\hline
\end{tabular}
\begin{tabular}{|l|}
\hline
\\
$\gamma_0 = \widehat{\gamma_{\mathrm{m}}}^{1/2} \widehat{\gamma_0}^{1/2} Y_{\mathrm{c}} = \widehat{\gamma_{\mathrm{m}}}^{1/4} \gamma_{\mathrm{self}}^{3/4} Y_{\mathrm{c}}^{1/2}$\\
\\
\hline
\end{tabular}

\begin{tabular}{|l|}
\hline
\\
\textbf{Case F23}\\
\\
\hline
\end{tabular}
\begin{tabular}{lc|ccccccccccccccccc}
\hline
& breaks & & $\gamma_{\mathrm{c}}$ & & $\widehat{\gamma_{\mathrm{m}}}$ & & $\widehat{\gamma_0}$ & & $\gamma_0$ & & \red{$\gamma_{\mathrm{m}}$} & & $\widehat{\widehat{\gamma_0}}$ & & $\widehat{\widehat{\gamma_{\mathrm{m}}}}$ & & $\widehat{\gamma_{\mathrm{c}}}$ &  \\
\hline
f & slopes & & | & $2$ & | & $3/2$ & | & $1$ & | & $2$ & \red{|} & \multicolumn{7}{c}{$p+1$} \\
\hline
h & slopes & \multicolumn{3}{c}{$0$} & | & $1/2$ & | & \multicolumn{5}{c}{$1$} & | & $3/4$ & | & $1/2$ & | & $4/3$ \\
\hline
\end{tabular}
\begin{tabular}{|l|}
\hline
\\
$\gamma_0 = \widehat{\gamma_{\mathrm{m}}}^{1/2} \widehat{\gamma_0}^{1/2} Y_{\mathrm{c}} = \widehat{\gamma_{\mathrm{m}}}^{1/4} \gamma_{\mathrm{self}}^{3/4} Y_{\mathrm{c}}^{1/2}$\\
\\
\hline
\end{tabular}


\begin{tabular}{|l|}
\hline
\\
\textbf{Case F26}\\
\\
\hline
\end{tabular}
\begin{tabular}{lc|ccccccccccc}
\hline
& breaks & & $\gamma_{\mathrm{c}}$ & & \red{$\gamma_{\mathrm{m}}$} & & $\widehat{\gamma_{\mathrm{m}}}$ & & $\gamma_0$ & & $\widehat{\gamma_{\mathrm{c}}}$ &  \\
\hline
f & slopes & & | & $2$ & \red{|} & $p+1$ & | & $p+1/2$ & | & \multicolumn{3}{c}{$p+1$} \\
\hline
h & slopes & \multicolumn{5}{c}{$0$} & | & \multicolumn{3}{c}{$1/2$} & | & $4/3$ \\
\hline
\end{tabular}
\begin{tabular}{|l|}
\hline
\\
$\gamma_0 = \widehat{\gamma_{\mathrm{m}}} Y_{\mathrm{c}}^2$\\
\\
\hline
\end{tabular}

\begin{tabular}{|l|}
\hline
\\
\textbf{Case F27}\\
\\
\hline
\end{tabular}
\begin{tabular}{lc|ccccccccccc}
\hline
& breaks & & $\gamma_{\mathrm{c}}$ & & \red{$\gamma_{\mathrm{m}}$} & & $\widehat{\gamma_{\mathrm{m}}}$ & & $\widehat{\gamma_{\mathrm{c}}}$ & & $\gamma_0$ &  \\
\hline
f & slopes & & | & $2$ & \red{|} & $p+1$ & | & $p+1/2$ & | & $p-1/3$ & | & $p+1$ \\
\hline
h & slopes & \multicolumn{5}{c}{$0$} & | & $1/2$ & | & \multicolumn{3}{c}{$4/3$} \\
\hline
\end{tabular}
\begin{tabular}{|l|}
\hline
\\
$\gamma_0 = \widehat{\gamma_{\mathrm{m}}}^{3/8} \widehat{\gamma_{\mathrm{c}}}^{5/8} Y_{\mathrm{c}}^{3/4}$\\
\\
\hline
\end{tabular}

\end{minipage}}
\end{center}
    \vspace*{-3ex}

    \tablefoot{
    For each case, the first row of the table gives the list of breaks appearing either in the normalised electron distribution $f(x)$ (i.e. the list of $x_i$), or in the normalised Compton parameter $h(x)$ (i.e. the list of relevant $\widehat{x}_i$), as well as all other necessary characteristic Lorentz factors to fully define the case in agreement with Fig.~\ref{fig:all_cases} (blue). The break corresponding to the peak of the synchrotron spectrum is highlighted in red. The second and third rows give the slopes $-\mathrm{d}\ln{f}/d\ln{x}$ and  $-\mathrm{d}\ln{g}/d\ln{x}$ for each branch. Finally, the last column gives the expression of $\gamma_0$. Cases F24 and F25 require a specific treatment: see text in appendix~\ref{sec:sol_fgh}.}
    \label{tab:cases_F}
\end{table*}

\begin{table*}
    \caption{Slow cooling: list of all possible radiative regimes and corresponding electron distribution.}
    \vspace*{-5.5ex}
    
\begin{center}
\resizebox{\linewidth}{!}{\begin{minipage}{26.5cm}
\begin{tabular}{|l|}
\hline
\\
\textbf{Case S1}\\
\\
\hline
\end{tabular}
\begin{tabular}{lc|ccccccccccc}
\hline
& breaks & & $\gamma_{\mathrm{m}}$ & & \red{$\gamma_{\mathrm{c}}$} & &  $\widehat{\gamma_{\mathrm{c}}}$ & & $\widehat{\gamma_{\mathrm{m}}}$ &  \\
\hline
f & slopes & & | & $p$ & \red{|} & \multicolumn{5}{c}{$p+1$}  \\
\hline
h & slopes & & \multicolumn{4}{c}{$0$} & | & $(3-p)/2$ & | & $4/3$ \\
\hline
\end{tabular}
\begin{tabular}{|l|}
\hline
\\
$\gamma_0$ undefined\\
\\
\hline
\end{tabular}

\begin{tabular}{|l|}
\hline
\\
\textbf{Case S2}\\
\\
\hline
\end{tabular}
\begin{tabular}{lc|ccccccccccc}
\hline
& breaks & & $\gamma_{\mathrm{m}}$ & & $\widehat{\gamma_{\mathrm{c}}}$ & & \blue{$\gamma_0$} & & \red{$\gamma_{\mathrm{c}}$} & & $\widehat{\gamma_{\mathrm{m}}}$ &  \\
\hline
f & slopes & & | & \multicolumn{5}{c}{$2$} & \red{|} & \multicolumn{3}{c}{$p+1$} \\
\hline
h & slopes & \multicolumn{3}{c}{$0$} & | & \multicolumn{5}{c}{$(3-p)/2$} & | & $4/3$ \\
\hline
\end{tabular}
\begin{tabular}{|l|}
\hline
\\
$\gamma_0 = \gamma_{\mathrm{c}} Y_{\mathrm{c}}^{2/(3-p)}$\\
\\
\hline
\end{tabular}

\begin{tabular}{|l|}
\hline
\\
\textbf{Case S3}\\
\\
\hline
\end{tabular}
\begin{tabular}{lc|ccccccccccccc}
\hline
& breaks & & $\widehat{\gamma_{\mathrm{c}}}$ & & \blue{$\gamma_0$} & & $\gamma_{\mathrm{m}}$ & & \blue{$\gamma_0$} & & \red{$\gamma_{\mathrm{c}}$} & & $\widehat{\gamma_{\mathrm{m}}}$ &  \\
\hline
f & slopes & & & & & & | & \multicolumn{3}{c}{$2$} & \red{|} & \multicolumn{3}{c}{$p+1$} \\
\hline
h & slopes & $0$ & | & \multicolumn{9}{c}{$(3-p)/2$} & | & $4/3$ \\
\hline
\end{tabular}
\begin{tabular}{|l|}
\hline
\\
$\gamma_0 = \gamma_{\mathrm{c}} Y_{\mathrm{c}}^{2/(3-p)}$\\
\\
\hline
\end{tabular}

\begin{tabular}{|l|}
\hline
\\
\textbf{Case S4}\\
\\
\hline
\end{tabular}
\begin{tabular}{lc|ccccccccccccc}
\hline
& breaks & & $\widehat{\gamma_{\mathrm{c}}}$ & & \blue{$\gamma_0$} & & $\gamma_{\mathrm{m}}$ & & \blue{$\gamma_0$} & & $\widehat{\gamma_{\mathrm{m}}}$ & & \red{$\gamma_{\mathrm{c}}$} &  \\
\hline
f & slopes & & & & & & | & \multicolumn{5}{c}{$2$} & \red{|} & $p+1$ \\
\hline
h & slopes & $0$ & | & \multicolumn{7}{c}{$(3-p)/2$} & | & \multicolumn{3}{c}{$4/3$} \\
\hline
\end{tabular}
\begin{tabular}{|l|}
\hline
\\
$\gamma_0 = \gamma_{\mathrm{c}}^{8/(9-3p)} \widehat{\gamma_{\mathrm{m}}}^{(1-3p)/(9-3p)} Y_{\mathrm{c}}^{2/(3-p)}$\\
\\
\hline
\end{tabular}

\begin{tabular}{|l|}
\hline
\\
\textbf{Case S5}\\
\\
\hline
\end{tabular}
\begin{tabular}{lc|ccccccccccc}
\hline
& breaks & & $\widehat{\gamma_{\mathrm{c}}}$ & & $\gamma_{\mathrm{m}}$ & & $\widehat{\gamma_{\mathrm{m}}}$ & & \blue{$\gamma_0$} & & \red{$\gamma_{\mathrm{c}}$} &  \\
\hline
f & slopes & & & & | & \multicolumn{5}{c}{$p$} & \red{|} & $p+1$ \\
\hline
h & slopes & $0$ & | & \multicolumn{3}{c}{$(3-p)/2$} & | & \multicolumn{5}{c}{$4/3$} \\
\hline
\end{tabular}
\begin{tabular}{|l|}
\hline
\\
$\gamma_0 = \gamma_{\mathrm{c}} Y_{\mathrm{c}}^{3/4}$\\
\\
\hline
\end{tabular}

\begin{tabular}{|l|}
\hline
\\
\textbf{Case S6}\\
\\
\hline
\end{tabular}
\begin{tabular}{lc|ccccccccccc}
\hline
& breaks & & $\widehat{\gamma_{\mathrm{c}}}$ & &  \blue{$\gamma_0$} & & $\widehat{\gamma_{\mathrm{m}}}$ &   & $\gamma_{\mathrm{m}}$ & &  \red{$\gamma_{\mathrm{c}}$} &  \\
\hline
f & slopes & \multicolumn{7}{c}{} & | & $p$ & \red{|} & $p+1$ \\
\hline
h & slopes & $0$ & | & \multicolumn{3}{c}{$(3-p)/2$} & | & \multicolumn{5}{c}{$4/3$} \\
\hline
\end{tabular}
\begin{tabular}{|l|}
\hline
\\
$\gamma_0 = \gamma_{\mathrm{c}}^{8/(9-3p)} \widehat{\gamma_{\mathrm{m}}}^{(1-3p)/(9-3p)} Y_{\mathrm{c}}^{2/(3-p)}$\\
\\
\hline
\end{tabular}

\begin{tabular}{|l|}
\hline
\\
\textbf{Case S7}\\
\\
\hline
\end{tabular}
\begin{tabular}{lc|ccccccccccccc}
\hline
& breaks & & $\widehat{\gamma_{\mathrm{c}}}$ & & $\widehat{\gamma_{\mathrm{m}}}$ & & \blue{$\gamma_0$} & &  $\gamma_{\mathrm{m}}$ & & \blue{$\gamma_0$} & & \red{$\gamma_{\mathrm{c}}$} &  \\
\hline
f & slopes & & & & & & & & | & \multicolumn{3}{c}{$p$} & \red{|} & $p+1$ \\
\hline
h & slopes & $0$ & | & $(3-p)/2$ & | & \multicolumn{9}{c}{$4/3$} \\
\hline
\end{tabular}
\begin{tabular}{|l|}
\hline
\\
$\gamma_0 = \gamma_{\mathrm{c}} Y_{\mathrm{c}}^{3/4}$\\
\\
\hline
\end{tabular}

\begin{tabular}{|l|}
\hline
\\
\textbf{Case S8}\\
\\
\hline
\end{tabular}
\begin{tabular}{lc|ccccccccccccc}
\hline
& breaks & & $\widehat{\gamma_0}$ & & $\widehat{\gamma_{\mathrm{c}}}$ & & $\widehat{\gamma_{\mathrm{m}}}$ & & $\gamma_{\mathrm{m}}$ & & $\gamma_{\mathrm{c}}$ & & \red{$\gamma_0$} &  \\
\hline
f & slopes & \multicolumn{7}{c}{} & | & $p$ & | & $p-1/3$ & \red{|} & $p+1$ \\
\hline
h & slopes & $0$ & | & $(10-3p)/6$ & | & $(3-p)/2$ & | & \multicolumn{7}{c}{$4/3$} \\
\hline
\end{tabular}
\begin{tabular}{|l|}
\hline
\\
$\gamma_0 = \gamma_{\mathrm{c}} Y_{\mathrm{c}}^{3/4}$\\
\\
\hline
\end{tabular}

\begin{tabular}{|l|}
\hline
\\
\textbf{Case S9}\\
\\
\hline
\end{tabular}
\begin{tabular}{lc|ccccccccccccc}
\hline
& breaks & & $\widehat{\gamma_0}$ & & $\widehat{\gamma_{\mathrm{c}}}$ & & $\gamma_{\mathrm{m}}$ & & $\widehat{\gamma_{\mathrm{m}}}$ & & $\gamma_{\mathrm{c}}$ & & \red{$\gamma_0$} &  \\
\hline
f & slopes & \multicolumn{5}{c}{} & | & \multicolumn{3}{c}{$p$} & | & $p-1/3$ & \red{|} & $p+1$ \\
\hline
h & slopes & $0$ & | & $(10-3p)/6$ & | & \multicolumn{3}{c}{$(3-p)/2$} & | & \multicolumn{5}{c}{$4/3$} \\
\hline
\end{tabular}
\begin{tabular}{|l|}
\hline
\\
$\gamma_0 = \gamma_{\mathrm{c}} Y_{\mathrm{c}}^{3/4}$\\
\\
\hline
\end{tabular}

\begin{tabular}{|l|}
\hline
\\
\textbf{Case S10}\\
($p<7/3$)\\
\hline
\end{tabular}
\begin{tabular}{lc|ccccccccccccc}
\hline
& breaks & & $\widehat{\gamma_0}$ & & $\widehat{\gamma_{\mathrm{c}}}$ & & $\gamma_{\mathrm{m}}$ & & $\gamma_{\mathrm{c}}$ & & \red{$\gamma_0$} & & $\widehat{\gamma_{\mathrm{m}}}$ &  \\
\hline
f & slopes & \multicolumn{5}{c}{} & | & $p$ & \violet{|} & $(3p-1)/2$ & \blue{|} & \multicolumn{3}{c}{$p+1$} \\
\hline
h & slopes & $0$ & | & $(7-3p)/4$ & | & \multicolumn{7}{c}{$(3-p)/2$} & | & $4/3$ \\
\hline
\end{tabular}
\begin{tabular}{|l|}
\hline
\\
$\gamma_0 = \gamma_{\mathrm{c}} Y_{\mathrm{c}}^{2/(3-p)}$\\
\\
\hline
\end{tabular}

\begin{tabular}{|l|}
\hline
\\
\textbf{Case S10}\\
($p>7/3$)\\
\hline
\end{tabular}
\begin{tabular}{lc|ccccccccccccc}
\hline
& breaks & & $\widehat{\gamma_0}$ & & $\widehat{\gamma_{\mathrm{c}}}$ & & $\gamma_{\mathrm{m}}$ & & \red{$\gamma_{\mathrm{c}}$} & & $\gamma_0$& & $\widehat{\gamma_{\mathrm{m}}}$ &  \\
\hline
f & slopes & \multicolumn{5}{c}{} & | & $p$ & \violet{|} & $(3p-1)/2$ & \blue{|} & \multicolumn{3}{c}{$p+1$} \\
\hline
h & slopes & \multicolumn{3}{c}{$0$} & | & \multicolumn{7}{c}{$(3-p)/2$} & | & $4/3$ \\
\hline
\end{tabular}
\begin{tabular}{|l|}
\hline
\\
$\gamma_0 = \gamma_{\mathrm{c}} Y_{\mathrm{c}}^{2/(3-p)}$\\
\\
\hline
\end{tabular}

\begin{tabular}{|l|}
\hline
\\
\textbf{Case S11}\\
($p<7/3$)\\
\hline
\end{tabular}
\begin{tabular}{lc|ccccccccccccccc}
\hline
& breaks & & $\widehat{\gamma_0}$ & & $\widehat{\widehat{\gamma_{\mathrm{m}}}}$ & & $\widehat{\gamma_{\mathrm{c}}}$ & & $\gamma_{\mathrm{m}}$ & & $\gamma_{\mathrm{c}}$ & & $\widehat{\gamma_{\mathrm{m}}}$ & & \red{$\gamma_0$} & \\
\hline
f & slopes & \multicolumn{7}{c}{} & | & $p$ & \violet{|} & $(3p-1)/2$ & | & $p-1/3$ & \blue{|} & $p+1$ \\
\hline
h & slopes & $0$ & | & $(10-3p)/6$ & | & $(7-3p)/4$ & | & \multicolumn{5}{c}{$(3-p)/2$} & | & \multicolumn{3}{c}{$4/3$} \\
\hline
\end{tabular}
\begin{tabular}{|l|}
\hline
\\
$\gamma_0 = \widehat{\gamma_{\mathrm{m}}}^{(3p-1)/8} \gamma_{\mathrm{c}}^{(9-3p)/8} Y_{\mathrm{c}}^{3/4}$\\
\\
\hline
\end{tabular}

\begin{tabular}{|l|}
\hline
\\
\textbf{Case S11}\\
($p>7/3$)\\
\hline
\end{tabular}
\begin{tabular}{lc|ccccccccccccccc}
\hline
& breaks & & $\widehat{\gamma_0}$ & & $\widehat{\widehat{\gamma_{\mathrm{m}}}}$ & & $\widehat{\gamma_{\mathrm{c}}}$ & & $\gamma_{\mathrm{m}}$ & & \red{$\gamma_{\mathrm{c}}$} & & $\widehat{\gamma_{\mathrm{m}}}$ & & $\gamma_0$ & \\
\hline
f & slopes & \multicolumn{7}{c}{} & | & $p$ & \violet{|} & $(3p-1)/2$ & | & $p-1/3$ & \blue{|} & $p+1$ \\
\hline
h & slopes & \multicolumn{5}{c}{$0$} & | & \multicolumn{5}{c}{$(3-p)/2$} & | & \multicolumn{3}{c}{$4/3$} \\
\hline
\end{tabular}
\begin{tabular}{|l|}
\hline
\\
$\gamma_0 = \widehat{\gamma_{\mathrm{m}}}^{(3p-1)/8} \gamma_{\mathrm{c}}^{(9-3p)/8} Y_{\mathrm{c}}^{3/4}$\\
\\
\hline
\end{tabular}

\begin{tabular}{|l|}
\hline
\\
\textbf{Case S12}\\
($p<7/3$) \\
\hline
\end{tabular}
\begin{tabular}{lc|ccccccccccccc}
\hline
& breaks & & $\gamma_{\mathrm{m}}$ & & $\widehat{\gamma_0}$ & & $\widehat{\gamma_{\mathrm{c}}}$ & & $\gamma_{\mathrm{c}}$ & & \red{$\gamma_0$} & & $\widehat{\gamma_{\mathrm{m}}}$ &  \\
\hline
f & breaks & & | & \multicolumn{5}{c}{$p$} & \violet{|} & $(3p-1)/2$ & \blue{|} & \multicolumn{3}{c}{$p+1$} \\
\hline
h & breaks & \multicolumn{3}{c}{$0$} & | & $(7-3p)/4$ & | & \multicolumn{5}{c}{$(3-p)/2$} & | & $4/3$ \\
\hline
\end{tabular}
\begin{tabular}{|l|}
\hline
\\
$\gamma_0 = \gamma_{\mathrm{c}} Y_{\mathrm{c}}^{2/(3-p)}$\\
\\
\hline
\end{tabular}

\begin{tabular}{|l|}
\hline
\\
\textbf{Case S12}\\
($p>7/3$) \\
\hline
\end{tabular}
\begin{tabular}{lc|ccccccccccccc}
\hline
& breaks & & $\gamma_{\mathrm{m}}$ & & $\widehat{\gamma_0}$ & & $\widehat{\gamma_{\mathrm{c}}}$ & & \red{$\gamma_{\mathrm{c}}$} & & $\gamma_0$ & & $\widehat{\gamma_{\mathrm{m}}}$ &  \\
\hline
f & breaks & & | & \multicolumn{5}{c}{$p$} & \violet{|} & $(3p-1)/2$ & \blue{|} & \multicolumn{3}{c}{$p+1$} \\
\hline
h & breaks & \multicolumn{5}{c}{$0$} & | & \multicolumn{5}{c}{$(3-p)/2$} & | & $4/3$ \\
\hline
\end{tabular}
\begin{tabular}{|l|}
\hline
\\
$\gamma_0 = \gamma_{\mathrm{c}} Y_{\mathrm{c}}^{2/(3-p)}$\\
\\
\hline
\end{tabular}

\begin{tabular}{|l|}
\hline
\\
\textbf{Case S13}\\
($p<7/3$) \\
\hline
\end{tabular}
\begin{tabular}{lc|ccccccccccccc}
\hline
& breaks & & $\widehat{\gamma_0}$ & & $\gamma_{\mathrm{m}}$ & & $\widehat{\gamma_{\mathrm{c}}}$ & & $\gamma_{\mathrm{c}}$ & & \red{$\gamma_0$} & & $\widehat{\gamma_{\mathrm{m}}}$ &  \\
\hline
f & xlope & \multicolumn{3}{c}{} & | & \multicolumn{3}{c}{$p$} & \violet{|} & $(3p-1)/2$ & \blue{|} & \multicolumn{3}{c}{$p+1$} \\
\hline
h& slope &  $0$ & | & \multicolumn{3}{c}{$(7-3p)/4$} & | & \multicolumn{5}{c}{$(3-p)/2$} & | & $4/3$ \\
\hline
\end{tabular}
\begin{tabular}{|l|}
\hline
\\
$\gamma_0 = \gamma_{\mathrm{c}} Y_{\mathrm{c}}^{2/(3-p)}$\\
\\
\hline
\end{tabular}

\begin{tabular}{|l|}
\hline
\\
\textbf{Case S13}\\
($p>7/3$)\\
\hline
\end{tabular}
\begin{tabular}{lc|ccccccccccccc}
\hline
& breaks & & $\widehat{\gamma_0}$ & & $\gamma_{\mathrm{m}}$ & & $\widehat{\gamma_{\mathrm{c}}}$ & & \red{$\gamma_{\mathrm{c}}$} & & $\gamma_0$ & & $\widehat{\gamma_{\mathrm{m}}}$ &  \\
\hline
f & xlope & \multicolumn{3}{c}{} & | & \multicolumn{3}{c}{$p$} & \violet{|} & $(3p-1)/2$ & \blue{|} & \multicolumn{3}{c}{$p+1$} \\
\hline
h & slope &  \multicolumn{5}{c}{$0$} & | & \multicolumn{5}{c}{$(3-p)/2$} & | & $4/3$ \\
\hline
\end{tabular}
\begin{tabular}{|l|}
\hline
\\
$\gamma_0 = \gamma_{\mathrm{c}} Y_{\mathrm{c}}^{2/(3-p)}$\\
\\
\hline
\end{tabular}

\begin{tabular}{|l|}
\hline
\\
\textbf{Case S14}\\
($p<7/3$) \\
\hline
\end{tabular}
\begin{tabular}{lc|ccccccccccccccc}
\hline
& breaks & & $\widehat{\gamma_0}$ & & $\widehat{\widehat{\gamma_{\mathrm{m}}}}$ & & $\gamma_{\mathrm{m}}$ & & $\widehat{\gamma_{\mathrm{c}}}$ & & $\gamma_{\mathrm{c}}$ & & $\widehat{\gamma_{\mathrm{m}}}$ & & \red{$\gamma_0$} & \\
\hline
f & slopes & \multicolumn{5}{c}{} & | & \multicolumn{3}{c}{$p$} & \violet{|} & $(3p-1)/2$ & | & $p-1/3$ & \blue{|} & $p+1$ \\
\hline
h& slopes & $0$ & | & $(10-3p)/6$ & | & \multicolumn{3}{c}{$(7-3p)/4$} & | & \multicolumn{3}{c}{$(3-p)/2$} & | & \multicolumn{3}{c}{$4/3$} \\
\hline
\end{tabular}
\begin{tabular}{|l|}
\hline
\\
$\gamma_0 = \widehat{\gamma_{\mathrm{m}}}^{(3p-1)/8} \gamma_{\mathrm{c}}^{(9-3p)/8} Y_{\mathrm{c}}^{3/4}$\\
\\
\hline
\end{tabular}

\begin{tabular}{|l|}
\hline
\\
\textbf{Case S14}\\
($p>7/3$)\\
\hline
\end{tabular}
\begin{tabular}{lc|ccccccccccccccc}
\hline
& breaks & & $\widehat{\gamma_0}$ & & $\widehat{\widehat{\gamma_{\mathrm{m}}}}$ & & $\gamma_{\mathrm{m}}$ & & $\widehat{\gamma_{\mathrm{c}}}$ & & \red{$\gamma_{\mathrm{c}}$} & & $\widehat{\gamma_{\mathrm{m}}}$ & & $\gamma_0$ & \\
\hline
f & slopes & \multicolumn{5}{c}{} & | & \multicolumn{3}{c}{$p$} & \violet{|} & $(3p-1)/2$ & | & $p-1/3$ & \blue{|} & $p+1$ \\
\hline
h & slopes & \multicolumn{7}{c}{$0$} & | & \multicolumn{3}{c}{$(3-p)/2$} & | & \multicolumn{3}{c}{$4/3$} \\
\hline
\end{tabular}
\begin{tabular}{|l|}
\hline
\\
$\gamma_0 = \widehat{\gamma_{\mathrm{m}}}^{(3p-1)/8} \gamma_{\mathrm{c}}^{(9-3p)/8} Y_{\mathrm{c}}^{3/4}$\\
\\
\hline
\end{tabular}

\begin{tabular}{|l|}
\hline
\\
\textbf{Case S15}\\
\\
\hline
\end{tabular}
\begin{tabular}{lc|ccccccccccc}
\hline
& breaks & & $\gamma_{\mathrm{m}}$ & & \red{$\gamma_{\mathrm{c}}$} & & $\widehat{\gamma_{\mathrm{c}}}$ & & $\gamma_0$ & & $\widehat{\gamma_{\mathrm{m}}}$ & \\
\hline
f & slopes & & | & $p$ & \red{|} & $p+1$ & | & $(3p-1)/2$ & | & \multicolumn{3}{c}{$p+1$} \\
\hline
h & slopes &  \multicolumn{5}{c}{$0$} & | & \multicolumn{3}{c}{$(3-p)/2$} & | & $4/3$ \\
\hline
\end{tabular}
\begin{tabular}{|l|}
\hline
\\
$\gamma_0 = \widehat{\gamma_{\mathrm{c}}} Y_{\mathrm{c}}^{2/(3-p)}$\\
\\
\hline
\end{tabular}

\begin{tabular}{|l|}
\hline
\\
\textbf{Case S16}\\
\\
\hline
\end{tabular}
\begin{tabular}{lc|ccccccccccc}
\hline
& breaks & & $\gamma_{\mathrm{m}}$ & & \red{$\gamma_{\mathrm{c}}$} & & $\widehat{\gamma_{\mathrm{c}}}$ & & $\widehat{\gamma_{\mathrm{m}}}$ & & $\gamma_0$ & \\
\hline
f & slopes & & | & $p$ & \red{|} & $p+1$ & | & $(3p-1)/2$ & | & $p-1/3$ & | & $p+1$ \\
\hline
h & slopes & \multicolumn{5}{c}{$0$} & | & $(3-p)/2$ & | & \multicolumn{3}{c}{$4/3$} \\
\hline
\end{tabular}
\begin{tabular}{|l|}
\hline
\\
$\gamma_0 = \widehat{\gamma_{\mathrm{m}}}^{(3p-1)/8} \widehat{\gamma_{\mathrm{c}}}^{(9-3p)/8} Y_{\mathrm{c}}^{3/4}$\\
\\
\hline
\end{tabular}

\end{minipage}}
\end{center}
    \tablefoot{Same convention as in table~\ref{tab:cases_F}. }
    \label{tab:cases_S}
\end{table*}

\begin{table*}
 \caption{Fast cooling cases F24 and F25.}
 \vspace*{-3ex}

\newlength{\mytablong}
\setlength{\mytablong}{39.5cm}
\resizebox{\linewidth}{!}{\rotatebox{90}{
{\begin{minipage}{40cm}
\noindent{\large \textbf{Column 1:} $(\gamma_\mathrm{m}^2 \gamma_\mathrm{c})^{1/3} < \gamma_\mathrm{self} < \gamma_\mathrm{self, cr, 4}=\left(\gamma_\mathrm{m}^8 \gamma_\mathrm{c}\right)^{1/9}$:}
\vspace*{-3ex}

\begin{center}
\resizebox{\linewidth}{!}{\begin{minipage}{\mytablong}
\begin{tabular}{|c|}
\hline
 \textbf{F24, subcase:}\\
$\gamma_\mathrm{m}<\gamma_0<\widetilde{\gamma}_\mathrm{c}$\\
\\
\hline
\end{tabular}
\begin{tabular}{lc|cccccccccccccccccc}
\hline
& breaks & & $\gamma_{\mathrm{c}}$ & & $\widehat{\gamma_0}$ & & $\widehat{\gamma_\mathrm{m}}$ & & $\gamma_{\mathrm{m}}$ & &  \red{$\gamma_0$} & & $^{(2)}\widehat{\gamma_\mathrm{m}}$ & & $^{(2)}\widehat{\gamma_0}$ & & $\widehat{\gamma_{\mathrm{c}}}$ &  \\
\hline
f & slopes & & | & $2$ & | & $(p+1)/2$ & | & $1$ & | & $p$ & \red{|} & \multicolumn{7}{c}{$p+1$}  \\
\hline
h & slopes & & \multicolumn{2}{c}{$0$} & | & $(3-p)/2$ & | & \multicolumn{5}{c}{$1$} & | & $(5-p)/4$ & | & $1/2$ & | & $4/3$ \\
\hline
\end{tabular}
\begin{tabular}{|l|}
\hline
\\
$\gamma_0 = \widehat{\gamma_\mathrm{m}}^{\frac{p-1}{2(4-p)}} \gamma_\mathrm{self}^{\frac{3(3-p)}{2(4-p)}} Y_\mathrm{c}^{\frac{1}{4-p}}$\\
\\
\hline
\end{tabular}
\end{minipage}}
\vspace*{1.5ex}

\resizebox{\linewidth}{!}{\begin{minipage}{\mytablong}
\begin{tabular}{|c|}
\hline
\textbf{F24, subcase:}\\
$\widetilde{\gamma}_\mathrm{c} < \gamma_0 < ^{(2)}\widehat{\gamma}_\mathrm{m}$\\
\\
\hline
\end{tabular}
\begin{tabular}{lc|cccccccccccccccc}
\hline
& breaks & & $\widehat{\gamma_0}$ & & $\gamma_{\mathrm{c}}$ & & $\widehat{\gamma_\mathrm{m}}$ & & $\gamma_{\mathrm{m}}$ & &  \red{$\gamma_0$} & & $^{(2)}\widehat{\gamma_\mathrm{m}}$ & & $\widehat{\gamma_{\mathrm{c}}}$ &  \\
\hline
f & slopes & & & & | & $(p+1)/2$ & | & $1$ & | & $p$ & \red{|} & \multicolumn{5}{c}{$p+1$}  \\
\hline
h & slopes & $0$ & | & \multicolumn{3}{c}{$(3-p)/2$} & | & \multicolumn{5}{c}{$1$} & | & $(5-p)/4$ & | & $4/3$ \\
\hline
\end{tabular}
\begin{tabular}{|l|}
\hline
\\
$\gamma_0 = \gamma_\mathrm{c}^{\frac{3-p}{2}} \widehat{\gamma_\mathrm{m}}^{\frac{p-1}{2}} Y_\mathrm{c}$\\
\\
\hline
\end{tabular}
\end{minipage}}
\vspace*{1.5ex}

\resizebox{\linewidth}{!}{\begin{minipage}{\mytablong}
\begin{tabular}{|c|}
\hline
\textbf{F24, subcase:}\\
$^{(2)}\widehat{\gamma}_\mathrm{m} < \gamma_0 < \widehat{\gamma}_\mathrm{c}$\\
($p < 13/5$)\\
\hline
\end{tabular}
\begin{tabular}{lc|cccccccccccccccccc}
\hline
& breaks & & $\widehat{\gamma_0}$ & & $^{(3)}\widehat{\gamma_\mathrm{m}}$ & & $\gamma_{\mathrm{c}}$ & & $\widehat{\gamma_\mathrm{m}}$ & & $\gamma_{\mathrm{m}}$ & & $^{(2)}\widehat{\gamma_\mathrm{m}}$  & & \red{$\gamma_0$} & & $\widehat{\gamma_{\mathrm{c}}}$ &  \\
\hline
f & slopes & & & & & & | & $(p+1)/2$ & | & $1$ & | & $p$ & | & $(5p-1)/4$ & \red{|} & \multicolumn{3}{c}{$p+1$}  \\
\hline
h & slopes & $0$ & | & $(13-5p)/8$ & | & \multicolumn{3}{c}{$(3-p)/2$} & | & \multicolumn{3}{c}{$1$} & | & \multicolumn{3}{c}{$(5-p)/4$} & | & $4/3$ \\
\hline
\end{tabular}
\begin{tabular}{|l|}
\hline
\\
$\gamma_0 = \gamma_\mathrm{c}^{\frac{2(3-p)}{5-p}} \widehat{\gamma_\mathrm{m}}^{\frac{2(p-1)}{5-p}}\, ^{(2)}\widehat{\gamma_\mathrm{m}}^{\frac{1-p}{5-p}} Y_\mathrm{c}^{\frac{4}{5-p}}$\\
\\
\hline
\end{tabular}
\end{minipage}}
\vspace*{1.5ex}

\resizebox{\linewidth}{!}{\begin{minipage}{\mytablong}
\begin{tabular}{|c|}
\hline
\textbf{F24, subcase:}\\
$^{(2)}\widehat{\gamma}_\mathrm{m} < \gamma_0 < \widehat{\gamma}_\mathrm{c}$\\
($p > 13/5$)\\
\hline
\end{tabular}
\begin{tabular}{lc|cccccccccccccccc}
\hline
& breaks & & $^{(3)}\widehat{\gamma_\mathrm{m}}$ & & $\gamma_{\mathrm{c}}$ & & $\widehat{\gamma_\mathrm{m}}$ & & $\gamma_{\mathrm{m}}$ & & \red{$^{(2)}\widehat{\gamma_\mathrm{m}}$} & & $\gamma_0$ & & $\widehat{\gamma_{\mathrm{c}}}$ &  \\
\hline
f & slopes & & & & | & $(p+1)/2$ & | & $1$ & | & $p$ & \red{|} & $(5p-1)/4$ & | & \multicolumn{3}{c}{$p+1$}  \\
\hline
h & slopes & $0$ & | & \multicolumn{3}{c}{$(3-p)/2$} & | & \multicolumn{3}{c}{$1$} & | & \multicolumn{3}{c}{$(5-p)/4$} & | & $4/3$ \\
\hline
\end{tabular}
\begin{tabular}{|l|}
\hline
\\
$\gamma_0 = \gamma_\mathrm{c}^{\frac{2(3-p)}{5-p}} \widehat{\gamma_\mathrm{m}}^{\frac{2(p-1)}{5-p}}\, ^{(2)}\widehat{\gamma_\mathrm{m}}^{\frac{1-p}{5-p}} Y_\mathrm{c}^{\frac{4}{5-p}}$\\
\\
\hline
\end{tabular}
\end{minipage}}
\vspace*{1.5ex}

\resizebox{\linewidth}{!}{\begin{minipage}{\mytablong}
\begin{tabular}{|c|}
\hline
\textbf{F25, subcase:}\\
$\widehat{\gamma}_\mathrm{c} < \gamma_0$\\
($p < 13/5$)\\
\hline
\end{tabular}
\begin{tabular}{lc|cccccccccccccccccccc}
\hline
& breaks & & $\widehat{\gamma_0}$ & & $^{(2)}\widehat{\gamma_\mathrm{c}}$ & & $^{(3)}\widehat{\gamma_\mathrm{m}}$ & & $\gamma_{\mathrm{c}}$ & & $\widehat{\gamma_\mathrm{m}}$ & & $\gamma_{\mathrm{m}}$ & & $^{(2)}\widehat{\gamma_\mathrm{m}}$  & & $\widehat{\gamma_{\mathrm{c}}}$ & & \red{$\gamma_0$} &  \\
\hline
f & slopes & & & & & & & & | & $(p+1)/2$ & | & $1$ & | & $p$ & | & $(5p-1)/4$ & | & $p - 1/3$ & \red{|} & $p+1$  \\
\hline
h & slopes & $0$ & | & $(10 - 3p)/6$ & | & $(13-5p)/8$ & | & \multicolumn{3}{c}{$(3-p)/2$} & | & \multicolumn{3}{c}{$1$} & | & $(5-p)/4$ & | & \multicolumn{3}{c}{$4/3$} \\
\hline
\end{tabular}
\begin{tabular}{|l|}
\hline
\\
$\gamma_0 = \gamma_\mathrm{c}^{\frac{3(3-p)}{8}} \widehat{\gamma_\mathrm{m}}^{\frac{3(p-1)}{8}}$\\
\hspace*{0.75cm} $\times ^{(2)}\widehat{\gamma_\mathrm{m}}^{\frac{3(1-p)}{16}} \widehat{\gamma_\mathrm{c}}^{\frac{3(1-p)}{16}} Y_\mathrm{c}^{\frac{3}{4}}$\\
\hline
\end{tabular}
\end{minipage}}
\vspace*{1.5ex}

\resizebox{\linewidth}{!}{\begin{minipage}{\mytablong}
\begin{tabular}{|c|}
\hline
\textbf{F25, subcase:}\\
$\widehat{\gamma}_\mathrm{c} < \gamma_0$\\
($p > 13/5$)\\
\hline
\end{tabular}
\begin{tabular}{lc|cccccccccccccccc}
\hline
& breaks & & $^{(3)}\widehat{\gamma_\mathrm{m}}$ & & $\gamma_{\mathrm{c}}$ & & $\widehat{\gamma_\mathrm{m}}$ & & $\gamma_{\mathrm{m}}$ & & \red{$^{(2)}\widehat{\gamma_\mathrm{m}}$}  & & $\widehat{\gamma_{\mathrm{c}}}$ & & $\gamma_0$ &  \\
\hline
f & slopes & & & & | & $(p+1)/2$ & | & $1$ & | & $p$ & \red{|} & $(5p-1)/4$ & | & $p - 1/3$ & | & $p+1$  \\
\hline
h & slopes & $0$ & | & \multicolumn{3}{c}{$(3-p)/2$} & | & \multicolumn{3}{c}{$1$} & | & $(5-p)/4$ & | & \multicolumn{3}{c}{$4/3$} \\
\hline
\end{tabular}
\begin{tabular}{|l|}
\hline
\\
$\gamma_0 = \gamma_\mathrm{c}^{\frac{3(3-p)}{8}} \widehat{\gamma_\mathrm{m}}^{\frac{3(p-1)}{8}}\, ^{(2)}\widehat{\gamma_\mathrm{m}}^{\frac{3(1-p)}{16}} \widehat{\gamma_\mathrm{c}}^{\frac{3(1-p)}{16}} Y_\mathrm{c}^{\frac{3}{4}}$\\
\\
\hline
\end{tabular}
\end{minipage}}
\end{center}

\noindent{\large \textbf{Column 2:} $\gamma_\mathrm{self, cr, 4}=\left(\gamma_\mathrm{m}^8 \gamma_\mathrm{c}\right)^{1/9} < \gamma_\mathrm{self} < \gamma_\mathrm{self, cr, 6}=\left(\gamma_\mathrm{m}^{32} \gamma_\mathrm{c}\right)^{1/33}$:}
\vspace*{-3ex}

\begin{center}
\resizebox{\linewidth}{!}{\begin{minipage}{\mytablong}
\begin{tabular}{|c|}
\hline
 \textbf{F24, subcase}\\
$\gamma_\mathrm{m}<\gamma_0<^{(2)}\widehat{\gamma}_\mathrm{m}$\\
\\
\hline
\end{tabular}
\begin{tabular}{lc|cccccccccccccccccc}
\hline
& breaks & & $\gamma_{\mathrm{c}}$ & & $\widehat{\gamma_0}$ & & $\widehat{\gamma_\mathrm{m}}$ & & $\gamma_{\mathrm{m}}$ & &  \red{$\gamma_0$} & & $^{(2)}\widehat{\gamma_\mathrm{m}}$ & & $^{(2)}\widehat{\gamma_0}$ & & $\widehat{\gamma_{\mathrm{c}}}$ &  \\
\hline
f & slopes & & | & $2$ & | & $(p+1)/2$ & | & $1$ & | & $p$ & \red{|} & \multicolumn{7}{c}{$p+1$}  \\
\hline
h & slopes & & \multicolumn{2}{c}{$0$} & | & $(3-p)/2$ & | & \multicolumn{5}{c}{$1$} & | & $(5-p)/4$ & | & $1/2$ & | & $4/3$ \\
\hline
\end{tabular}
\begin{tabular}{|l|}
\hline
\\
$\gamma_0 = \widehat{\gamma_\mathrm{m}}^{\frac{p-1}{2(4-p)}} \gamma_\mathrm{self}^{\frac{3(3-p)}{2(4-p)}} Y_\mathrm{c}^{\frac{1}{4-p}}$\\
\\
\hline
\end{tabular}
\end{minipage}}
\vspace*{1.5ex}

\resizebox{\linewidth}{!}{\begin{minipage}{\mytablong}
\begin{tabular}{|c|}
\hline
\textbf{F24, subcase:}\\
$^{(2)}\widehat{\gamma}_\mathrm{m} < \gamma_0 < \widetilde{\gamma}_\mathrm{c}$\\
($p < 13/5$)\\
\hline
\end{tabular}
\begin{tabular}{lc|cccccccccccccccccccccc}
\hline
& breaks & & $\gamma_\mathrm{c}$ & & $\widehat{\gamma_0}$ & & $^{(3)}\widehat{\gamma_\mathrm{m}}$ & & $\widehat{\gamma_\mathrm{m}}$ & & $\gamma_{\mathrm{m}}$ & & $^{(2)}\widehat{\gamma_\mathrm{m}}$ & & \red{$\gamma_0$} & & $^{(4)}\widehat{\gamma_\mathrm{m}}$ & & $^{(2)}\widehat{\gamma_0}$ & & $\widehat{\gamma_{\mathrm{c}}}$ &  \\
\hline
f & slopes & & | & $2$ & | & $(3+5p)/8$ & | & $(p+1)/2$ & | & $1$ & | & $p$ & | & $(5p-1)/4$ & \red{|} & \multicolumn{5}{c}{$p+1$}  \\
\hline
h & slopes & \multicolumn{3}{c}{$0$} & | & $(13-5p)/8$ & | & $(3-p)/2$ & | & \multicolumn{3}{c}{$1$} & | & \multicolumn{3}{c}{$(5-p)/4$} & | & $(21 - 5p)/16$ & | & $1/2$ & | & $4/3$ \\
\hline
\end{tabular}
\begin{tabular}{|l|}
\hline
\\
$\gamma_0 = \widehat{\gamma_\mathrm{m}}^{\frac{p-1}{3(3-p)}}\, ^{(2)}\widehat{\gamma_\mathrm{m}}^{\frac{p-1}{6(3-p)}}\, ^{(3)}\widehat{\gamma}_\mathrm{m}^{\frac{p-1}{12(3-p)}} \gamma_\mathrm{self}^{\frac{13-5p}{4(3-p)}} Y_\mathrm{c}^{\frac{2}{3(3-p)}}$\\
\\
\hline
\end{tabular}
\end{minipage}}
\vspace*{1.5ex}

\resizebox{\linewidth}{!}{\begin{minipage}{\mytablong}
\begin{tabular}{|c|}
\hline
\textbf{F24, subcase:}\\
$^{(2)}\widehat{\gamma}_\mathrm{m} < \gamma_0 < \widetilde{\gamma}_\mathrm{c}$\\
($p > 13/5$)\\
\hline
\end{tabular}
\begin{tabular}{lc|cccccccccccccccccc}
\hline
& breaks & & $\gamma_\mathrm{c}$ & & $^{(3)}\widehat{\gamma_\mathrm{m}}$ & & $\widehat{\gamma_\mathrm{m}}$ & & $\gamma_{\mathrm{m}}$ & & \red{$^{(2)}\widehat{\gamma_\mathrm{m}}$} & & $\gamma_0$ & & $^{(4)}\widehat{\gamma_\mathrm{m}}$ & & $\widehat{\gamma_{\mathrm{c}}}$ &  \\
\hline
f & slopes & & | & $2$ & | & $(p+1)/2$ & | & $1$ & | & $p$ & \red{|} & $(5p-1)/4$ & | & \multicolumn{3}{c}{$p+1$}  \\
\hline
h & slopes & \multicolumn{3}{c}{$0$} & | & $(3-p)/2$ & | & \multicolumn{3}{c}{$1$} & | & \multicolumn{3}{c}{$(5-p)/4$} & | & $1/2$ & | & $4/3$ \\
\hline
\end{tabular}
\begin{tabular}{|l|}
\hline
\\
$\gamma_0 = \widehat{\gamma_\mathrm{m}}^{\frac{2(p-1)}{5-p}}\, ^{(2)}\widehat{\gamma_\mathrm{m}}^{\frac{1-p}{5-p}}\, ^{(3)}\widehat{\gamma}_\mathrm{m}^{\frac{2(3-p)}{5-p}} Y_\mathrm{c}^{\frac{4}{5-p}}$\\
\\
\hline
\end{tabular}
\end{minipage}}
\vspace*{1.5ex}

\resizebox{\linewidth}{!}{\begin{minipage}{\mytablong}
\begin{tabular}{|c|}
\hline
\textbf{F24, subcase:}\\
$\widetilde{\gamma}_\mathrm{c} < \gamma_0 < ^{(4)}\widehat{\gamma}_\mathrm{m}$\\
($p < 13/5$)\\
\hline
\end{tabular}
\begin{tabular}{lc|cccccccccccccccccccc}
\hline
& breaks & & $\widehat{\gamma_0}$ & & $\gamma_\mathrm{c}$ & & $^{(3)}\widehat{\gamma_\mathrm{m}}$ & & $\widehat{\gamma_\mathrm{m}}$ & & $\gamma_{\mathrm{m}}$ & & $^{(2)}\widehat{\gamma_\mathrm{m}}$ & & \red{$\gamma_0$} & & $^{(4)}\widehat{\gamma_\mathrm{m}}$ & & $\widehat{\gamma_{\mathrm{c}}}$ &  \\
\hline
f & slopes & & & & | & $(3+5p)/8$ & | & $(p+1)/2$ & | & $1$ & | & $p$ & | & $(5p-1)/4$ & \red{|} & \multicolumn{3}{c}{$p+1$}  \\
\hline
h & slopes & $0$ & | & \multicolumn{3}{c}{$(13-5p)/8$} & | & $(3-p)/2$ & | & \multicolumn{3}{c}{$1$} & | & \multicolumn{3}{c}{$(5-p)/4$} & | & $(21 - 5p)/16$ & | & $4/3$ \\
\hline
\end{tabular}
\begin{tabular}{|l|}
\hline
\\
$\gamma_0 = \gamma_\mathrm{c}^{\frac{13-5p}{2(5-p)}} \widehat{\gamma_\mathrm{m}}^{\frac{2(p-1)}{5-p}}\, ^{(2)}\widehat{\gamma_\mathrm{m}}^{\frac{1-p}{5-p}}\, ^{(3)}\widehat{\gamma}_\mathrm{m}^{\frac{25-9p}{2(5-p)}} Y_\mathrm{c}^{\frac{4}{5-p}}$\\
\\
\hline
\end{tabular}
\end{minipage}}
\vspace*{1.5ex}

\resizebox{\linewidth}{!}{\begin{minipage}{\mytablong}
\begin{tabular}{|c|}
\hline
\textbf{F24, subcase}\\
$\widetilde{\gamma}_\mathrm{c} < \gamma_0 < ^{(4)}\widehat{\gamma}_\mathrm{m}$\\
($p > 13/5$)\\
\hline
\end{tabular}
\begin{tabular}{lc|cccccccccccccccccc}
\hline
& breaks & & $\gamma_\mathrm{c}$ & & $^{(3)}\widehat{\gamma_\mathrm{m}}$ & & $\widehat{\gamma_\mathrm{m}}$ & & $\gamma_{\mathrm{m}}$ & & \red{$^{(2)}\widehat{\gamma_\mathrm{m}}$} & & $\gamma_0$ & & $^{(4)}\widehat{\gamma_\mathrm{m}}$ & & $\widehat{\gamma_{\mathrm{c}}}$ &  \\
\hline
f & slopes & & | & $2$ & | & $(p+1)/2$ & | & $1$ & | & $p$ & \red{|} & $(5p-1)/4$ & | & \multicolumn{3}{c}{$p+1$}  \\
\hline
h & slopes & \multicolumn{3}{c}{$0$} & | & $(3-p)/2$ & | & \multicolumn{3}{c}{$1$} & | & \multicolumn{3}{c}{$(5-p)/4$} & | & $1/2$ & | & $4/3$ \\
\hline
\end{tabular}
\begin{tabular}{|l|}
\hline
\\
$\gamma_0 = \widehat{\gamma_\mathrm{m}}^{\frac{2(p-1)}{5-p}}\, ^{(2)}\widehat{\gamma_\mathrm{m}}^{\frac{1-p}{5-p}}\, ^{(3)}\widehat{\gamma}_\mathrm{m}^{\frac{2(3-p)}{5-p}} Y_\mathrm{c}^{\frac{4}{5-p}}$\\
\\
\hline
\end{tabular}
\end{minipage}}
\vspace*{1.5ex}

\resizebox{\linewidth}{!}{\begin{minipage}{\mytablong}
\begin{tabular}{|c|}
\hline
\textbf{F24, subcase:}\\
$^{(4)}\widehat{\gamma}_\mathrm{m} < \gamma_0 < \widehat{\gamma}_\mathrm{c}$\\
($p < 53/21$)\\
\hline
\end{tabular}
\begin{tabular}{lc|cccccccccccccccccccccc}
\hline
& breaks & & $\widehat{\gamma_0}$ & & $^{(5)}\widehat{\gamma_\mathrm{m}}$ & & $\gamma_\mathrm{c}$ & & $^{(3)}\widehat{\gamma_\mathrm{m}}$ & & $\widehat{\gamma_\mathrm{m}}$ & & $\gamma_{\mathrm{m}}$ & & $^{(2)}\widehat{\gamma_\mathrm{m}}$ & & $^{(4)}\widehat{\gamma_\mathrm{m}}$ & & \red{$\gamma_0$} & & $\widehat{\gamma_{\mathrm{c}}}$ &  \\
\hline
f & slopes & & & & & & | & $(3+5p)/8$ & | & $(p+1)/2$ & | & $1$ & | & $p$ & | & $(5p-1)/4$ & | & $(21p-5)/16$ & \red{|} & \multicolumn{3}{c}{$p+1$}  \\
\hline
h & slopes & $0$ & | & $(53-21p)/32$ & | & \multicolumn{3}{c}{$(13-5p)/8$} & | & $(3-p)/2$ & | & \multicolumn{3}{c}{$1$} & | & $(5-p)/4$ & | & \multicolumn{3}{c}{$(21 - 5p)/16$} & | & $4/3$ \\
\hline
\end{tabular}
\begin{tabular}{|l|}
\hline
\\
$\gamma_0 = \gamma_\mathrm{c}^{\frac{26-10p}{21-5p}} \widehat{\gamma_\mathrm{m}}^{\frac{8(p-1)}{21-5p}}\, ^{(2)}\widehat{\gamma_\mathrm{m}}^{\frac{4(1-p)}{21-5p}}\, ^{(3)}\widehat{\gamma}_\mathrm{m}^{\frac{2(p-1)}{21-5p}}\, ^{(4)}\widehat{\gamma}_\mathrm{m}^{\frac{1-p}{21-5p}} Y_\mathrm{c}^{\frac{16}{21-5p}}$\\
\\
\hline
\end{tabular}
\end{minipage}}
\vspace*{1.5ex}

\resizebox{\linewidth}{!}{\begin{minipage}{\mytablong}
\begin{tabular}{|c|}
\hline
\textbf{F24, subcase:}\\
$^{(4)}\widehat{\gamma}_\mathrm{m} < \gamma_0 < \widehat{\gamma}_\mathrm{c}$\\
($53/21 < p < 13/5$)\\
\hline
\end{tabular}
\begin{tabular}{lc|cccccccccccccccccccc}
\hline
& breaks & & $^{(5)}\widehat{\gamma_\mathrm{m}}$ & & $\gamma_\mathrm{c}$ & & $^{(3)}\widehat{\gamma_\mathrm{m}}$ & & $\widehat{\gamma_\mathrm{m}}$ & & $\gamma_{\mathrm{m}}$ & & $^{(2)}\widehat{\gamma_\mathrm{m}}$ & & \red{$^{(4)}\widehat{\gamma_\mathrm{m}}$} & & $\gamma_0$ & & $\widehat{\gamma_{\mathrm{c}}}$ &  \\
\hline
f & slopes & & & & | & $(3+5p)/8$ & | & $(p+1)/2$ & | & $1$ & | & $p$ & | & $(5p-1)/4$ & \red{|} & $(21p-5)/16$ & | & \multicolumn{3}{c}{$p+1$}  \\
\hline
h & slopes & $0$ & | & \multicolumn{3}{c}{$(13-5p)/8$} & | & $(3-p)/2$ & | & \multicolumn{3}{c}{$1$} & | & $(5-p)/4$ & | & \multicolumn{3}{c}{$(21 - 5p)/16$} & | & $4/3$ \\
\hline
\end{tabular}
\begin{tabular}{|l|}
\hline
\\
$\gamma_0 = \gamma_\mathrm{c}^{\frac{26-10p}{21-5p}} \widehat{\gamma_\mathrm{m}}^{\frac{8(p-1)}{21-5p}}\, ^{(2)}\widehat{\gamma_\mathrm{m}}^{\frac{4(1-p)}{21-5p}}\, ^{(3)}\widehat{\gamma}_\mathrm{m}^{\frac{2(p-1)}{21-5p}}\, ^{(4)}\widehat{\gamma}_\mathrm{m}^{\frac{1-p}{21-5p}} Y_\mathrm{c}^{\frac{16}{21-5p}}$\\
\\
\hline
\end{tabular}
\end{minipage}}
\vspace*{1.5ex}

\resizebox{\linewidth}{!}{\begin{minipage}{\mytablong}
\begin{tabular}{|c|}
\hline
\textbf{F24, subcase}\\
$^{(4)}\widehat{\gamma}_\mathrm{m} < \gamma_0 < \widehat{\gamma}_\mathrm{c}$\\
($p > 13/5$)\\
\hline
\end{tabular}
\begin{tabular}{lc|cccccccccccccccccc}
\hline
& breaks & & $\gamma_\mathrm{c}$ & & $^{(3)}\widehat{\gamma_\mathrm{m}}$ & & $\widehat{\gamma_\mathrm{m}}$ & & $\gamma_{\mathrm{m}}$ & & \red{$^{(2)}\widehat{\gamma_\mathrm{m}}$} & & $^{(4)}\widehat{\gamma_\mathrm{m}}$ & & $\gamma_0$ & & $\widehat{\gamma_{\mathrm{c}}}$ &  \\
\hline
f & slopes & & | & $2$ & | & $(p+1)/2$ & | & $1$ & | & $p$ & \red{|} & $(5p-1)/4$ & | & $p + 1/2$ & | & \multicolumn{3}{c}{$p+1$}  \\
\hline
h & slopes & \multicolumn{3}{c}{$0$} & | & $(3-p)/2$ & | & \multicolumn{3}{c}{$1$} & | & $(5-p)/4$ & | & \multicolumn{3}{c}{$1/2$} & | & $4/3$ \\
\hline
\end{tabular}
\begin{tabular}{|l|}
\hline
\\
$\gamma_0 = \widehat{\gamma_\mathrm{m}}^{p-1}\, ^{(2)}\widehat{\gamma_\mathrm{m}}^{\frac{(1-p)}{2}}\, ^{(3)}\widehat{\gamma}_\mathrm{m}^{3-p}\, ^{(4)}\widehat{\gamma}_\mathrm{m}^{\frac{p-3}{2}} Y_\mathrm{c}^{2}$\\
\\
\hline
\end{tabular}
\end{minipage}}
\vspace*{1.5ex}

\resizebox{\linewidth}{!}{\begin{minipage}{\mytablong}
\begin{tabular}{|c|}
\hline
\textbf{F25, subcase:}\\
$\widehat{\gamma}_\mathrm{c} < \gamma_0$\\
($p < 53/21$)\\
\hline
\end{tabular}
\begin{tabular}{lc|cccccccccccccccccccccccc}
\hline
& breaks & & $\widehat{\gamma_0}$ & & $^{(2)}\widehat{\gamma_\mathrm{c}}$ & & $^{(5)}\widehat{\gamma_\mathrm{m}}$ & & $\gamma_\mathrm{c}$ & & $^{(3)}\widehat{\gamma_\mathrm{m}}$ & & $\widehat{\gamma_\mathrm{m}}$ & & $\gamma_{\mathrm{m}}$ & & $^{(2)}\widehat{\gamma_\mathrm{m}}$ & & $^{(4)}\widehat{\gamma_\mathrm{m}}$ & & $\widehat{\gamma_{\mathrm{c}}}$ & & \red{$\gamma_0$} &  \\
\hline
f & slopes & & & & & & & & | & $(3+5p)/8$ & | & $(p+1)/2$ & | & $1$ & | & $p$ & | & $(5p-1)/4$ & | & $(21p-5)/16$ & | & $p - 1/3$ & \red{|} & $p+1$  \\
\hline
h & slopes & $0$ & | & $(10 - 3p)/6$ & | & $(53-21p)/32$ & | & \multicolumn{3}{c}{$(13-5p)/8$} & | & $(3-p)/2$ & | & \multicolumn{3}{c}{$1$} & | & $(5-p)/4$ & | & $(21 - 5p)/16$ & | & \multicolumn{3}{c}{$4/3$} \\
\hline
\end{tabular}
\begin{tabular}{|l|}
\hline
\\
$\gamma_0 = \gamma_\mathrm{c}^{\frac{39-15p}{32}} \widehat{\gamma_\mathrm{m}}^{\frac{3(p-1)}{8}}\, ^{(2)}\widehat{\gamma_\mathrm{m}}^{\frac{3(1-p)}{16}}\,$\\
\hspace*{0.75cm} $\times ^{(3)}\widehat{\gamma}_\mathrm{m}^{\frac{3(p-1)}{32}}\, ^{(4)}\widehat{\gamma}_\mathrm{m}^{\frac{3(1-p)}{64}} \widehat{\gamma_\mathrm{c}}^{\frac{1+15p}{64}} Y_\mathrm{c}^{\frac{3}{4}}$\\
\hline
\end{tabular}
\end{minipage}}
\vspace*{1.5ex}

\resizebox{\linewidth}{!}{\begin{minipage}{\mytablong}
\begin{tabular}{|c|}
\hline
\textbf{F25, subcase:}\\
$\widehat{\gamma}_\mathrm{c} < \gamma_0$\\
($53/21 < p < 13/5$)\\
\hline
\end{tabular}
\begin{tabular}{lc|cccccccccccccccccccc}
\hline
& breaks & & $^{(5)}\widehat{\gamma_\mathrm{m}}$ & & $\gamma_\mathrm{c}$ & & $^{(3)}\widehat{\gamma_\mathrm{m}}$ & & $\widehat{\gamma_\mathrm{m}}$ & & $\gamma_{\mathrm{m}}$ & & $^{(2)}\widehat{\gamma_\mathrm{m}}$ & & \red{$^{(4)}\widehat{\gamma_\mathrm{m}}$} & & $\widehat{\gamma_{\mathrm{c}}}$ & & $\gamma_0$ &  \\
\hline
f & slopes & & & & | & $(3+5p)/8$ & | & $(p+1)/2$ & | & $1$ & | & $p$ & | & $(5p-1)/4$ & \red{|} & $(21p-5)/16$ & | & $p - 1/3$ & | & $p+1$  \\
\hline
h & slopes & $0$ & | & \multicolumn{3}{c}{$(13-5p)/8$} & | & $(3-p)/2$ & | & \multicolumn{3}{c}{$1$} & | & $(5-p)/4$ & | & $(21 - 5p)/16$ & | & \multicolumn{3}{c}{$4/3$} \\
\hline
\end{tabular}
\begin{tabular}{|l|}
\hline
\\
$\gamma_0 = \gamma_\mathrm{c}^{\frac{39-15p}{32}} \widehat{\gamma_\mathrm{m}}^{\frac{3(p-1)}{8}}\, ^{(2)}\widehat{\gamma_\mathrm{m}}^{\frac{3(1-p)}{16}}\,$\\
\hspace*{0.75cm} $\times ^{(3)}\widehat{\gamma}_\mathrm{m}^{\frac{3(p-1)}{32}}\, ^{(4)}\widehat{\gamma}_\mathrm{m}^{\frac{3(1-p)}{64}} \widehat{\gamma_\mathrm{c}}^{\frac{1+15p}{64}} Y_\mathrm{c}^{\frac{3}{4}}$\\
\hline
\end{tabular}
\end{minipage}}
\vspace*{1.5ex}

\resizebox{\linewidth}{!}{\begin{minipage}{\mytablong}
\begin{tabular}{|c|}
\hline
\textbf{F25, subcase:}\\
$\widehat{\gamma}_\mathrm{c} < \gamma_0$\\
($p > 13/5$)\\
\hline
\end{tabular}
\begin{tabular}{lc|cccccccccccccccccc}
\hline
& breaks & & $\gamma_\mathrm{c}$ & & $^{(3)}\widehat{\gamma_\mathrm{m}}$ & & $\widehat{\gamma_\mathrm{m}}$ & & $\gamma_{\mathrm{m}}$ & & \red{$^{(2)}\widehat{\gamma_\mathrm{m}}$} & & $^{(4)}\widehat{\gamma_\mathrm{m}}$ & & $\widehat{\gamma_{\mathrm{c}}}$ & & $\gamma_0$ &  \\
\hline
f & slopes & & | & $2$ & | & $(p+1)/2$ & | & $1$ & | & $p$ & \red{|} & $(5p-1)/4$ & | & $p + 1/2$ & | & $p - 1/3$ & | & $p+1$  \\
\hline
h & slopes & \multicolumn{3}{c}{$0$} & | & $(3-p)/2$ & | & \multicolumn{3}{c}{$1$} & | & $(5-p)/4$ & | & $1/2$ & | & \multicolumn{3}{c}{$4/3$} \\
\hline
\end{tabular}
\begin{tabular}{|l|}
\hline
\\
$\gamma_0 = \widehat{\gamma_\mathrm{m}}^{\frac{3(p-1)}{8}}\, ^{(2)}\widehat{\gamma_\mathrm{m}}^{\frac{3(1-p)}{16}}\, ^{(3)}\widehat{\gamma}_\mathrm{m}^{\frac{3(3-p)}{8}}\, ^{(4)}\widehat{\gamma}_\mathrm{m}^{\frac{-3(p+3)}{16}} \widehat{\gamma_\mathrm{c}}^{\frac{5}{8}} Y_\mathrm{c}^{\frac{3}{4}}$\\
\\
\hline
\end{tabular}
\end{minipage}}
\end{center}
\end{minipage}}}}
 \vspace*{-3ex}

    \tablefoot{
    The electron distribution is given in the first subcases of the two first columns (see text in Appendix~\ref{sec:sol_fgh}). Same convention as in Table~\ref{tab:cases_F}.}
    \label{tab:cases_F2425_12}
\end{table*}

\section{Posterior distributions 
of the three fits of the afterglow of GW~170817}
\label{ap:posterior_3_models}

\begin{figure*}[h]
    \centering
    \includegraphics[width=\textwidth]{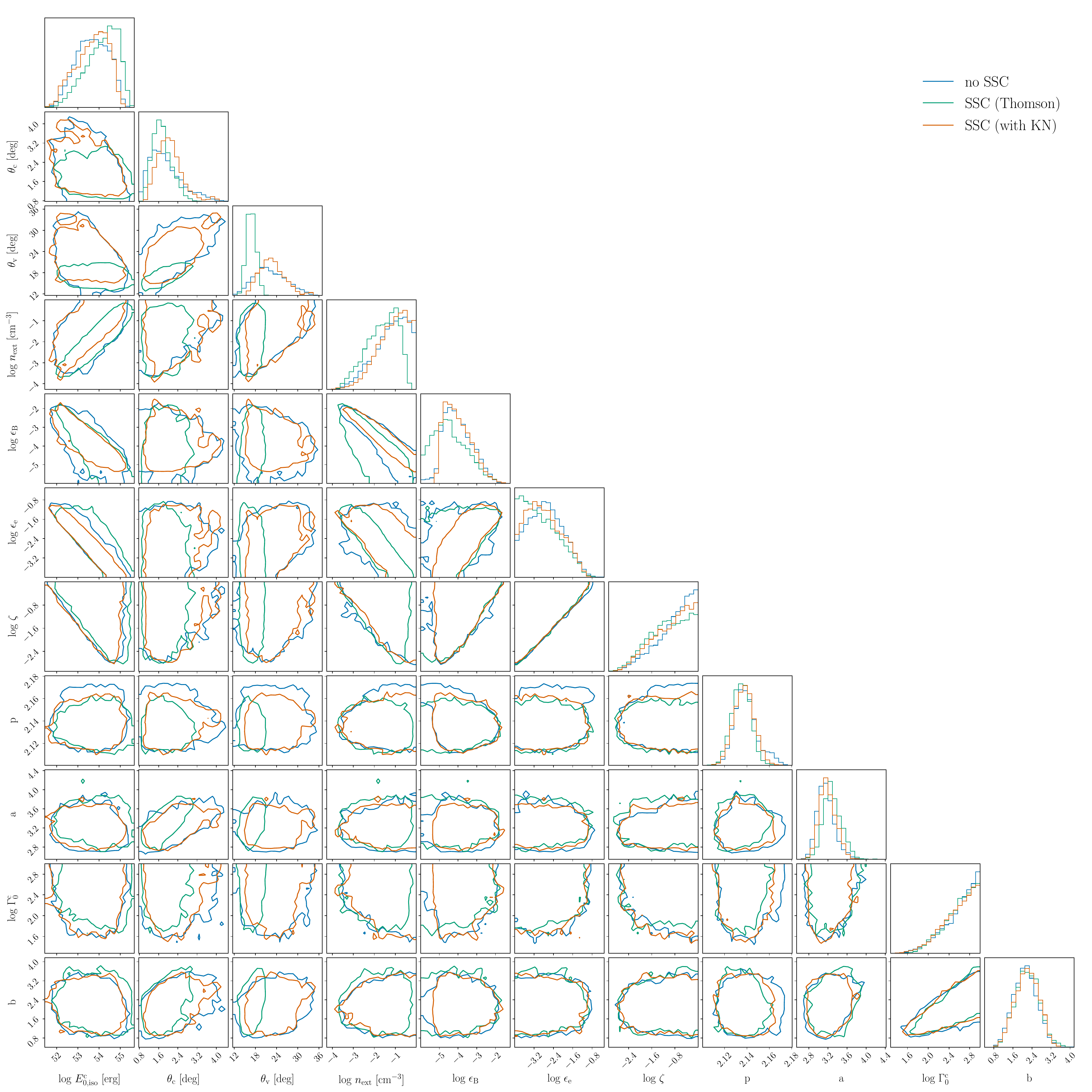}
    \caption{Posterior joint and marginalised distributions of the 
    free model parameters, $E_\mathrm{0,iso}^\mathrm{c}$, $\theta_{\mathrm{c}}$, $\theta_{\mathrm{v}}$, $n_{\mathrm{ext}}$, $\epsilon_{\mathrm{B}}$, $\epsilon_{\mathrm{e}}$, $\zeta$, $p$ $a$, $\Gamma_0^\mathrm{c}$, $b$ for the three fits of the afterglow of GW~170817 presented in Sect.~\ref{sec:170817}, which differ only by the treatment of the radiative processes. In blue, no SSC scatterings are taken into account ("no SSC"); in green, SSC scatterings are assumed to be only in the Thomson regime ("SSC (Thomson)"); in orange, the SSC scattering is depleted at high energy in the KN regime ("SSC (with KN)"). The coloured contours correspond to the $3\sigma$ confidence intervals for each model and each parameter.}
    \label{fig:MCMC_3_models}
\end{figure*}

Fig.~\ref{fig:MCMC_3_models} gives the posterior joint and marginalised distributions of the model parameters for the three fits of the multi-wavelength afterglow of GW~170817 described in Sect.~\ref{sec:170817}. The three fits use the same observational constraints and the same priors on the free parameters, and differ only by the radiative treatment: synchrotron only ("no SSC"), synchrotron and SSC assuming that all IC scatterings occur in Thomson regime ("SSC (Thomson)"), and self-consistent calculation of synchrotron and SSC processes taking into account the KN attenuation ("SSC (with KN)"). It appears clearly that the "no SSC" and "SSC (with KN)" fits lead to similar results, whereas the parameters inferred by the "SSC (Thomson)" fit are slighltly different (see also Table~\ref{table:best_fit_values}). This especially true for the viewing  and core jet opening angles, but also for the kinetic energy and the external density. Note the correlations between these quantities. This difference is due to the fact that the SSC component in the afterglow of GW~170817 is very weak due to the KN attenuation, and is therefore overestimated in the "SSC (Thomson)" case: see discussion in Sect.~\ref{sec:discussion_170817}.

\end{document}